**Detecting Structural breakpoints in natural gas and electricity wholesale prices via Bayesian ensemble approach and assessing their interaction-causality relationships via partial mutual information, in the era of energy prices turmoil of 2022 period: the cases of ten European markets**

**Panayotis G. Papaioannou [1†], George P. Papaioannou [2, †,] George Evangelidis [3, †,] George Gavalakis[4]**

1 Eurobank SA, FX & Derivatives Trading Desk-Global Markets & Assets Management, Greece
2 Center for Research and Applications in Nonlinear Systems (CRANS), Department of Mathematics, University of Patras, Patras 26 500, Greece
3 Department of Transmission System Operations & Control, Independent Power Transmission Operator (IPTO) S.A., Athens 104 43, Greece
4 University of West Attica, Athens, Greece (giorgosgavalakis@gmail.com)
***** Correspondence gpthespies@gmail.com; Tel.: +30-6972420181
† These authors contributed equally to this work.

## Abstract

We investigate the impact of several critical events associated with the Russo-Ukrainian war, started officially on 24 February 2022 with the Russian invasion of Ukraine, on ten European electricity markets, two natural gas markets (the European reference trading hub TTF and N.Y.'s NGNMX market) and how these markets interact to each other and with USD/RUB exchange rate, a 'financial market'. We analyze the reactions of these markets, manifested as breakpoints attributed to these critical events, and their interaction, by using a set of three tools that can shed light on different aspects of this complex situation. We combine the concepts of market efficiency, measured by quantifying the ***Efficient market hypothesis (EMH)*** via rolling ***Hurst exponent***, with ***structural breakpoints*** occurred in the time series of gas, electricity and financial markets, the detection of which is possible by using a ***Bayesian ensemble approach, the Bayesian Estimator of Abrupt change, Seasonal change and Trend (BEAST), a powerful tool that can effectively detect*** structural breakpoints, trends, seasonalities and sudden abrupt changes in time series. *We perform also* ***causality analysis*** *using* the ***Partial Mutual Information with Mixed Embedding (PMIME) and rolling Mutual information (rMI)*** approaches, to analyze the direction of flow of information between the markets to understand the nature of their interaction, especially during the period of crisis and intense – turmoiled economic and geopolitical conditions. *The results show that the analyzed markets have exhibited different modes of reactions to the critical events, both in respect of number, nature, and time of occurrence (leading, lagging, concurrent with dates of critical events) of breakpoints as well as of the dynamic behavior of their trend components.* ***The most critical event, in respect of causing a strong structural breakpoint, for most of the markets, is not that of 24 February 2022, the day of the Russian invasion, but other critical events before this date,*** because of each market's 'idiosyncrasy' and 'readiness'. Also, the interaction between TTF, NGNMX and USD/RUB markets is found to be strongly mutual i.e. bidirectional, the financial market (USD/RUB) affects both gas markets, that in turn affects, to a different degree, the electricity markets. These findings support the results of similar works in literature. The three tools of analysis provide consistent results, linking rationally the concepts of market efficiency ('readiness' and degree of independence from Russia gas inflows), number of breakpoints, dynamic profile of trend and seasonality curves and the direction of 'causalities' in the complex interaction of the markets during the Russo-Ukrainian crisis.

**Keywords: Russo-Ukrainian war;** rolling Partial Mutual Information PMIME; BEAST Bayessian ensemble tool; wholesale Electricity prices; natural gas prices; energy and financial markets interaction.

## 1. Introduction and literature review

The geopolitical conflicts occurred during 2021, enhanced by the Russian invasion of Ukraine, February 2022, resulted in dramatic jumps in gas prices and a drastic reduction of the gas inflows from Russian to Europe, therefore changing the characterization of natural gas, as a commodity, from almost a neglected to a crucial one, especially for the Electricity generation, as well as a key issue on the media and in policy debates. The security of supply of this vital commodity for the European economies has turned now to be one of a hottest issues in macroeconomic and energy research and analyses.

Nat gas, unlike oil, also, is regarded as an unknown factor, from a macroeconomic perspective. Scant evidence exists on how gas prices affect inflation and economic activity. More importantly, the natural gas market has several idiosyncrasies that influence heavily the dynamic evolution of the shocks. To mention a few, the ***trading*** of this commodity is mainly via long-term contracts in fragmented markets, thus the associated structures and regulations weaken the interaction between wholesale and retail prices. Also, the strong seasonality in its consumption induces the maintaining of large storage capacities by buyers. The natural gas price at the European reference trading hub TTF peaked at 339 Euro/MWh on August 26, 2022, a huge increase since the historical average is around 20 Euro/MWh (see see https://tradingeconomics.com/commodity/eu-natural-gas). Two days ago, on February 24, 2022, Russia invaded Ukraine and western states responded almost directly.

In this study we address the issue of the impact of the natural gas price of the Netherlands fund Title Transfer Facility, TTF), as well as some other 'related' to natural gas influencing factors, on the wholesale electricity prices of some selective European countries, with an extra focus on the energy turmoil of 2022. More specifically, we examine the very interesting and 'hot' period from 1$^{st}$ Jan. 2020 to Q2-2023 during which the risen outburst on natural gas prices, as a result of the invasion of Russia to Ukraine, is assumed to have a very strong impact on the evolutionary dynamics of wholesale prices and their volatility of the examined European Electricity (EE) Day-Ahead (DA) Markets, however with a different response behavior, due to their different inherent idiosyncrasies stemming out of their particular structural forms. The study of impact of TTF on DA electricity prices is implementing via adopting the ***Bayesian Estimator of Abrupt change, Seasonal change, and Trend (BEAST),*** a strong tool for the investigation of how TTF and other 'related to NG prices' exogenous factors shape the dynamics of prices and their volatility in each separate market. We combine BEAST with other supplementary approaches, the *Partial Mutual Information with Mixed Embedding (PMIME) and the rolling Mutual Information (rMI),* which are assumed also to shed light on our research target with a different perspective.

In the models we implemented, the other two 'Nat gas related' exogenous factors (variables), that we consider crucial in shaping the dynamics of electricity prices are the Natural Gas Futures Index at NYMEX (NGFI), the USD/RUB foreign exchange pair, taking also into consideration a number of major state, geopolitical and regulatory events that also are assumed to have a crucial impact on the electricity prices, as described in the report of (ACER-CEER, 2023).

The rest of this paper is organized as follows: in the present section 1, the introduction and literature review are provided, i.e. a short information on the conceptual connection of efficient markets hypothesis, Hurst exponent and structural breakpoints, a justification of choosing the Bayesian approach and finally a statement of the research questions of this paper. In section 2 we state the research questions and main targets of our work. In section 3 we describe shortly the ten European electricity and the TTF markets, with emphasis on the drivers leading to the price surge that occurred in the summer 2022, and the inefficiency of the natural gas markets in Europe. Also, a list of major events (geopolitical, economic, policy, regulations etc.) deemed to have a direct or indirect significant influence on the dynamics of the factors affecting the formation of prices. A description of the data sets, their necessary tests (for stationarity, normality etc.) as well as related graphs and descriptive statistics of all variables analyzed, is given in section 4. Section 5 describes the methodologies used (Hurst exponent, PMIME, rMI, BEAST), and the empirical results are provided in section 6, followed by a discussion and conclusions in section 7.

### 1.2. Structural breakpoints of a financial time series and the efficient market hypothesis, EMH

There is a strong conceptual connection between the *Structural breakpoints* of a financial market's time series and the efficiency of this market, as shown by the ***efficient market hypothesis, EMH***, perspective (Fama, E. F., 1960; Fama, E. F., 1970; Fama, E. F., 1991). This connection, to understandable, requires a grasp of both concepts: ***Structural breakpoints*** refer to points in time at which the statistical properties of a time series change (Bai, J., and Perron, P., 2003; Andrews, D. W. K., 1993). These changes could be in the mean, variance, correlation structure, or any other statistical characteristic of the series. In the context of financial markets, structural breaks might be caused by ***significant geopolitical,*** economic events, policy changes, technological innovations, or sudden shifts in investor behavior. Identifying these breakpoints is crucial for accurate modeling and forecasting, as they indicate periods where historical patterns no longer apply. The work of (Bai, J., and Perron, P.,2003), expands their earlier paper, providing more sophisticated methods for detecting and analyzing structural breaks in econometric models.

The Efficient Market Hypothesis suggests that at any given time, asset prices fully reflect all available information. According to EMH, it is impossible to consistently achieve higher returns on a risk-adjusted basis than the market average because asset prices always incorporate and reflect all relevant information. The hypothesis is categorized into three forms based on the extent of information reflected in prices:

- Weak form: All past market prices and data are fully reflected in asset prices.
- Semi-strong form: All publicly available information is reflected in asset prices.
- Strong form: All information, public and private, is fully reflected in asset prices.

As far as the ***connection between structural breakpoints*** in financial time series and ***market efficiency*** is concerned, it revolves around how markets respond to new information and how that response is reflected in asset prices. The paper of (Zivot, E., et al., 1992) provide the foundations for how the concepts of market efficiency and structural breaks integrate. The way they are connected is as follows: a) ***information incorporation:*** Structural breaks often occur due to the release of new, unexpected information or *sudden changes in economic conditions*. According to EMH, the market quickly and efficiently incorporates this information into asset prices. *The occurrence of a structural break could be seen as a test of how efficiently the market responds to new information,* b) ***predictability and arbitrage:*** if markets are truly efficient (especially in the semi-strong and strong forms), it should be difficult to predict future price movements based on past information, as all known information is already reflected in current prices. **However, *the identification of structural breaks* can sometimes lead to *profitable*** ***trading strategies,*** *challenging the EMH.* ***If investors can systematically identify and exploit these breaks before the market fully adjusts, it suggests a form of market inefficiency***, and finally c) ***adaptive markets hypothesis :*** the concept of *structural breaks* and their identification might be better explained by the ***Adaptive Markets Hypothesis (AMH)*** (Lo, A. W., 1991; Lo, A. W., 2004) , which proposes that *market efficiency is not a static condition but evolves over time as participants and conditions change, i.e. a dynamic process influenced by the evolution of the structure of the market and external factors*. ***From this perspective, structural breaks could be seen as points where the market is adjusting to a new equilibrium, reflecting a dynamic form of efficiency.***

*As a general conclusion, the relationship between structural breakpoints and the EMH hinges on how these breakpoints reflect the market's ability to process and incorporate new information into asset prices. The presence of significant structural breaks and the ability of market participants to exploit them for profit can challenge the traditional view of market efficiency, suggesting a more nuanced, dynamic understanding of how financial markets operate.*

### 1.3. Connection of Hurst Exponent, EMH and structural breakpoints

The ***Hurst Exponent (HE)*** (Hurst H.E., 1951; Hurst H.E. 1965; Mandelbrot, B.B., et al., 1969; Peters, E.E, 1994; Cajueiro, D.O., et al., 2004, Bui Q., et al., 2022) is a statistical measure that can help to understand two key aspects of financial time series: the degree of *market efficiency* (as proposed by the Efficient Market Hypothesis) and the presence of *structural breakpoints* in time series data. The work of Cajueiro et. al. 2004, above, applies the Hurst exponent to analyze the efficiency of emerging financial markets over time, providing empirical evidence linked to the concept of EMH. Hurst exponent is used as a tool to select stocks for algorithmic investment strategy, in a very recent work of how HE is applied in financial markets. More specifically HE is used as a tool to generate a *trading strategy that can beat the market, i.e. it challenges the EMH* (Bui Q., et al., 2022). The application of HE in assessing the efficiency of electricity markets is described in the work of (Papaioannou G., et al, 2019), where extensive literature on this issue is also given. In a similar work to ours (Keharan C.C., et al., 2024), the authors have applied a rolling Hurst exponent, among other tools, to study the *efficiency* of several European electricity markets, including most of the markets used in our study. They sampled the time of examination according to the *breakpoints* in each market that correspond to the *dates of coupling-decoupling events* of the markets with other markets. Their results support our results in this paper, as described in section 6 of the results below.

We describe here shortly how the Hurst Exponent connects to each of these concepts: the Efficient Market Hypothesis (EMH) suggests that asset prices fully reflect all available information, making it impossible to consistently achieve higher returns than the market average through any analysis of available information. The Hurst Exponent, denoted as H, is a measure used to analyze *the long-term memory* of time series data, including financial markets data. It can take on values between 0 and 1, with the following interpretation:

- H=0.5 suggests a completely *random walk,* supporting the *weak form of the EMH,* indicating that future price movements are completely *independent of past movements* (uncorrelated process). **Market Implication:** This suggests a ***neutral stability*** where the market is neither highly stable nor highly unstable. The price movements are mostly random, possibly driven by a mix of balanced and fluctuating forces.
- H>0.5 indicates a *persistent behavior*, suggesting a *trend-following pattern* where future price increases are likely to follow past increases (and vice versa for decreases), *challenging the EMH* by implying *predictability* in price movements. This could imply a certain degree of ***stability, as trends (whether increasing or decreasing) are more predictable.*** **Market Implication:** A persistent Hurst exponent indicates a stable market where supply and demand are relatively balanced, and price movements are smoother and more predictable. This reflects lower market risk and higher confidence among market participants.
- H<0.5 indicates an *anti-persistent behavior*, suggesting a *mean-reverting* pattern where an *increase is likely to be followed by a decrease (and vice versa),* which also challenges the EMH by suggesting prices are *predictably moving away from trends.* This could suggest **high volatility** and potentially ***lower stability***. **Market Implication:** An anti-persistent Hurst exponent reflects ***instability*** in the electricity market, characterized by frequent price reversals and ***higher volatility***. This could lead to increased risk for market participants and challenges in ensuring a reliable electricity supply.

The estimation of Hurst exponent is described in section 5.1. In the context of **electricity markets**, the Hurst exponent can provide insights into the ***market's stability*** by analyzing the price dynamics over time.

***Hurst Exponent and Structural Breakpoints.*** *Structural breakpoints* in a financial time series indicate points at which the statistical properties of the series (such as mean, variance, or correlation structure) change significantly. *These breakpoints can significantly impact the long-term memory and dependencies within the series, which are precisely what the Hurst Exponent measures. H can be used for* ***detecting changes in dependence****:* A change in the Hurst Exponent value before and after a certain point in time could suggest a structural break in the time series. For instance, a shift from $H > 0.5$ *to* $H < 0.5$ might indicate a *transition from trending behavior to mean-reverting behavior*, potentially due to a significant market event or change in market dynamics. H is also used to monitor the ***evolving market efficienc***y: If the *market's efficiency level changes over time* (possibly due to changes in regulation, technology, or market participants), such shifts may also manifest as *changes in the Hurst Exponent*. This could provide a quantitative measure of how market efficiency evolves, with implications for *strategies based on trend following or mean reversion.* An investigation of stock return predictability and the adaptive nature of markets, with implications for both structural breaks and the HE's role in the analysis of financial and energy time markets, is described in the works of (Phillips, P.C.B., et al., 2006; Kim, J.H., et al., 2011; Balcilar, M., et al., 2015, Kaharan C.C., 2024).

As a conclusion, the Hurst Exponent offers a nuanced tool for examining financial markets through the lenses of the *Efficient Market Hypothesis* and the presence of *structural breakpoints*. By measuring the degree of long-term memory in price changes, the Hurst Exponent can provide insights into market efficiency, potential predictability of price movements, and the ***impact of structural changes in the market's behavior***. However, interpreting the Hurst Exponent requires careful consideration of the broader market context, including economic conditions, market regulations, technological changes and more importantly, after considering the impact of *Russian invasion in Ukraine, geopolitical conflicts, that might influence market dynamics.*

### 1.4. 'Mainstream' and Bayesian approach in detecting structural breakpoints

Despite existing numerous successful applications in the time series analysis in finance and economics, challenges remain. For example, various inconsistencies in the results that occur when applying different models on the same time series, is still a challenging issue in the time series analysis. The inconsistencies are indeed attributed partly to different approaches and perspectives adopted for time series analysis and prediction. A preponderance of financial time series analysis takes usually a 'best-model' seeking, statistical modeling perspective in the scene that out of many candidates models a 'best' model is selected to reliably decompose a given time series in its dynamic components i.e. trends, seasonality and abrupt changes. In a variety of scientific fields, not only in finance, this single-best-model paradigm is broadly adopted, however, the usefulness for extracting information on the dynamics of the underlying systems, is not necessarily safe. This is so because the choices of statistical models adopted for the analysis have a crucial impact on the way of interpreting the dynamics 'hidden' in the data. It is completely different, for example, when fitting a simple linear model to a given time series, from the fitting of a piece-wise linear model with one break point (structural break), on the same series. In addition, the fitting of piece-wise models with multiple breakpoints may generate alternative specifications regarding the generative factors (drivers) responsible for the 'observed' changes in the dynamics of the time series.

As mentioned earlier, the adoption of the single-best-model methodology, in analyzing time series, can possibly create inconsistent or contradictory insights. It is known that optimization criteria as Akaike's information criterion (AIC) or the Bayesian information criterion (BIC) used in the process of selecting the 'best model', play a crucial role in this approach, together with the optimization method used. This means that choosing one model and ignoring other ones is indeed a strategically not efficient approach because in such a way possibly interesting characteristics of the alternative models ignored are lost while at the same time ignores the possible uncertainties associated with the chosen 'best' model. The inability of a single model, respectively of how appropriate or 'good' is thought to be, in capturing the complexity of the underlying dynamics in the data, as well as the preferences of the modeler in specifying the model, may complicate the modeling process. In fact, the larger the number of model parameters and the more its structural complexity is, the greater the chance of overfitting and miss specifying the model, even though the model chosen is likely to capture changes in time series at multiple time scales, based on its high structural complexity.

When analyzing time series assumed to be recorded from a high dimensional, underlying dynamical system, as in our case energy and financial systems, the inability of main-stream or conventional methodologies in describing such systems becomes directly obvious. This inability however can be overcome by adopting Bayesian statistics, an inferential paradigm that can probabilistically treat model parameters as well as structural specifications within a unified context, taking into consideration all possible uncertainties encountered in modeling process. In the section that follows, the adoption of the Bayesian approach is justified.

In this point we provide a short look at the field of structural break testing, in general. Hartley (1950) and Page (1954) and Page (1955), are the pioneers in the field. Hartley (1950) introduced a test that can identify *variance differences* between data groups while Page (1954) and Page (1955) offered a method to find *changes that affect model parameters* over time. Generalization and extension of the above tests followed the next decades, producing tests with enhanced abilities to detect structural changes. For example, while Hartley's test was design to detect variance heterogeneity in groups of data that are of equal size, and normally distributed, the later modern breakpoint tests can detect differences and changes in a variety of parameter types, in models with not so restrictive specifications. Possible formulations of the breakpoint test include the *a-posteriori type* in which breaks can be detected retrospectively in a complete data and the *monitoring type* in which the data is sequentially updated. Direct estimation of model parameters or accepting the results of an optimization method for the parameters are also two approaches that differentiate parametric structural break tests. Nonparametric tests, on the other hand, are adopted in more general applications. However, because in this approach the properties of the underlying data distribution is unknown (i.e. the generating mechanism of the data is a stochastic dynamical system), its enhanced applicability bears the cost of a low testing power, as is described in Andrews (1993). An example of the monitoring test type is the work of (Page 1954), which is the base for the development, some years later, of the standardized CUSUM test of Brown et al. (1975). In this method, recursive residuals are added up to form the cumulative sum (CUSUM) statistic, from predictions that come only from preceding data. CUSUM is used broadly in detecting structural instabilities in more general model parameters (Lee et al. (2003). In detecting changes in several Autoregressive (AR) parameters, Gombay and Serban (2009) provided critical values for a generalized application of CUSUM test in this type of time series models. A further development of the CUSUM test is presented in Groen et al. (2011) in which the main finding is that the testing power of an adjusted CUSUM test is further enhanced via using multivariate data systems, an outcome of using the maximum absolute value of individual CUSUM statistics, at different stages.

As far as the branch of *a-posteriori methods*, the work of in Chow (1960) is considered the pioneer one, in which the significance of a structural break at a given date can be determined. Chow (1960) examined also the concurrent structural breaks of two interacting time series so he searches to see whether two groups of economic units can be assumed to share the same regression parameters. Towards this target he developed a method for testing the equality of linear regression coefficients. In Chow's test, sums of squared residuals and sums of squared differences in fitted values are used, by estimating the regression coefficients of full and smaller samples of the time series. The test is used to infer a structural break in a time series, since this break divides a given sample in two segments having statistically significant differences in the parameter values. However, the candidate date at which the breakpoint occurs and separates the sample is considered as an exogenous event. In the Bayesian approach we follow this exogenously provided fact is not necessary, as we will be described later. Chow's test exhibited, after some years of its application, some limitations. An example of an attempt to address these limitations is the test of Goldfeld and Quandt (1965), another seminal publication in the field of break testing. It describes a test for the equality of error term variances and compares sums of squared residuals between two equal-sized samples. Chow's and Goldfeld-Quandt's test assume normal linear regression models. The test of structural break of Ploberger et al. (1989) has a more extensive applicability as in dynamic regression models with stationary and ergodic errors.

A worth mentioned development, regarding the *a-posteriori structural break testing*, is the detection of structural *breaks endogenously*, instead of testing for a structural break at a exogenously given (preselected) event, as described in the work of Hansen (1992). An *endogenously* structural *break* is defined as the most likely break located at a date that causes the maximum data fit improvement among all possible candidates break dates. The detection of endogenous structural break is possible by using a likelihood ratio (LR) statistic, as presented in Hinkley (1970), assuming linear regression models with normally distributed errors and mean parameter change. For i.i.d (independent and identically distributed) errors and a change in all model parameters at the break date, Hawkins (1987) provides a Wald statistic. A natural and very significant extension of the previous test is the one developed by Andrews (1993). More specifically he expands the applicability of endogenous break detection to *non-linear, non-stationary models* that may contain temporally correlated (dependent) data. A great percentage of the of the subsequent literature in testing for structural breaks has enhanced the generalization and applicability the Andrews's test. The detection of *multiple structural breaks* in a sample *at once,* is described in the paper of Bai and Perron (2003), however in the BEAST approach used in this work, this detection is much easier and more reliable than the Bai and Perron's test, as described below. In the work of Qu and Perron (2007), the asymptotic distribution of the maximum LR statistic, in the case that the system contains more than one dependent variable is described, and the detection of multiple structural breaks is done via a test in any segment of the sample of model parameters, including covariance terms of error distributions.

Various break tests have also been developed for specific conditions. Bataa et al. (2013) examines conditional testing, according to which various types of parameters must be tested that are nuisance parameters to each other. Another such specific case is the structural breaks that are not asymptotically distinct but locally ordered, as analyzed by Perron and Oka (2011). The examination of a more general parameter changes than the usual *abrupt parameter change* considered in 'mainstream' structural break testing, is given in Lin and Teräsvirta (1994), in which a test is described in the context of smooth transition, where a stability of a parameter is rejected in favor of a smooth transition alternative hypothesis. We must emphasize here that the testing for structural break detection belongs to the field of model diagnostics, for assessing the fitting of a model to a time series data, so referring to seminal texts as Engle R.F (1982a;1982b) and Brockwell and Davis (1996; 2016), is considered a must.

### 1.5 Justification in adopting a Bayesian approach in breakpoint detection.

Instead of following a conventional, AIC or BIC criterion-based approaches for choosing a single best model, we adopt the Bayessian approach that combines all possible candidate models to produce an average model, evaluating concurrently how much reliable (how close to truth) this model can be. The method followed here is called Bayessian Model Average (BMA), that is a member of the multi-model method, broadly known as ensemble learning (Zhao et al., 2019).

Considering several models instead of a single 'best' one, enhances the ability of the average model to capture any uncertainties, reduces the chance of model misspecification, while at the same time enhances model's flexibility and generalizability, two crucial properties needed in modeling complex time series. BMA has been applied in a variety of fields (Banner, et al., 2017; Zhang et al., 2012; Steel, Mark F. J, 2018; Nonejad N., 2021; Fragoso T. et al., 2015; Chua C. L, et al., 2013).

Despite the advantages of BMA, its application in energy time series analysis remains rather limited, with enormous potential to tap. In this study we seek to reliably detect breakpoints in time series of European Electricity and Natural gas data, adopting the BMA approach. We use the ***BEAST- a Bayessian Estimator of Abrupt change, Seasonal Change, and Trend***, developed by (Zhao et al., 2019), an approach that exhibits many advantages over 'main-stream', conventional, non-Bayessian one. BEAST gives up the single best-model approach and uses the BMA concept to combine several potential alternative models to generate an average model incorporating a rich amount of information on the underlying dynamics of the time series, revealing complex nonlinear structures, thus shedding light on the various sources of uncertainties and reliably detecting breakpoints and abrupt changes in time series.

## 2. Related research questions and the main targets of the work

Drawing from the literature review, and 'exploiting' the insights gained from studies related to this work, we state below the research questions:

During the examined period of the Russo-Ukrainian conflict ('long' before, at the onset and 'long' after the invasion:

1. What are the significant (occurring with the largest probability) breakpoints of the time series of the price of specific variables of the energy and financial markets analyzed? Do these breakpoints correspond only to the critical events E1-E13 linked to the mentioned conflict in the time examined or there are also other 'hidden' ones?
2. How does the Hurst Exponent, a measure to quantify the weak-, semi-, and strong-form of the efficiency of markets analyzed, vary across them, an information that can help us to assess their adaptability, especially in the context of drastic geopolitical and economic changes occurred due to the invasion?

3. Are the detected structural breakpoints, by using BEAST approach, linked to the market efficiency measured via the rolling Hurst exponent over the period of our analysis?
4. How the (war related) critical events, seen us the sources of exogenous shocks in the markets, contribute to the development of the structural breakpoints as well as on the deviations of the markets analyzed from the EMH limit?
5. How the energy markets (electricity and gas) analyzed, interacted with the USD/RUB financial market and what is the direction of the interaction (or 'causality')? Is there any mutual (bidirected) interaction present?

## 3. European Electricity Markets analyzed: a short description on their structure and responsive behavior to the natural gas price surges

### *3.1. The European Electricity Markets*

Electricity has become the most basic form of energy for everyday use, with natural gas, renewable and other energy resources mainly used for its production. With its flexibility, easy control, immediate availability and pure form, electricity immediately became a much needed and multipurpose form of energy. Modern societies have become very dependent on the supply of electricity. European electricity markets operating under the new regime of liberalization, energy transactions between participants take place at various time levels, from the long term to the real time. Depending on the regulatory framework of each country, we have the following basic categories of electricity markets (Simoglou, C.K., 2011). The Mid-Long-term Forward market and the Future market, the Day-ahead market, intraday or adjustment market, the Balancing, Real-time market and Ancillary Services and Regulation market.

The day-ahead market is a semi-compulsory market, where orders cover the availability and are compatible with Price Coupling of Regions algorithm (PCR EUPHEMIA) standards, and the biddings. There are also provisions for exchange-based futures and over-the-counter (OTC) contract limits on the volumes. Single Intraday Coupling (SIDC) creates a single EU cross-zonal intraday electricity market. In simple terms, buyers and sellers of energy (market participants) can work together across Europe to trade electricity continuously on the day the energy is needed.

### *3.2. Critical events are assumed to have shaped the price of European Electricity Markets.*

Figure 1.1 presents the time series of all European wholesale electricity prices analyzed in this work. The price surge is pronounced after the end of 2021 and more evident during 2022. The large increase and fluctuation of electricity prices dominated the discussion in Europe during 2022. The wide volatility of prices daily creates confusion and requires their analysis over longer time periods. As shown in Figure 1.2, Greece with €279.90 per MWh is the 2$^{nd}$ most expensive country of our study, after Italy (€298.90/MWh). In neighboring Bulgaria, the cost was €233.32/MWh and Spain has the lowest cost with €167.52/MWh. The European average was €234.19/MWh. It should be noted that Day ahead prices, although they mainly shape the prices of the wholesale energy market, are not the only ones, since the final prices also include quantities supplied in the futures market through bilateral contracts, as well as in the deviations market. These markets are developed to varying degrees in different countries. Comparing the average annual prices for 2020, 2021 and 2022 shows the very large increase in day-ahead market prices between these three years. In the chart above, countries are ranked in increasing order by percentage increase in average annual DAM price between 2020 and 2022.

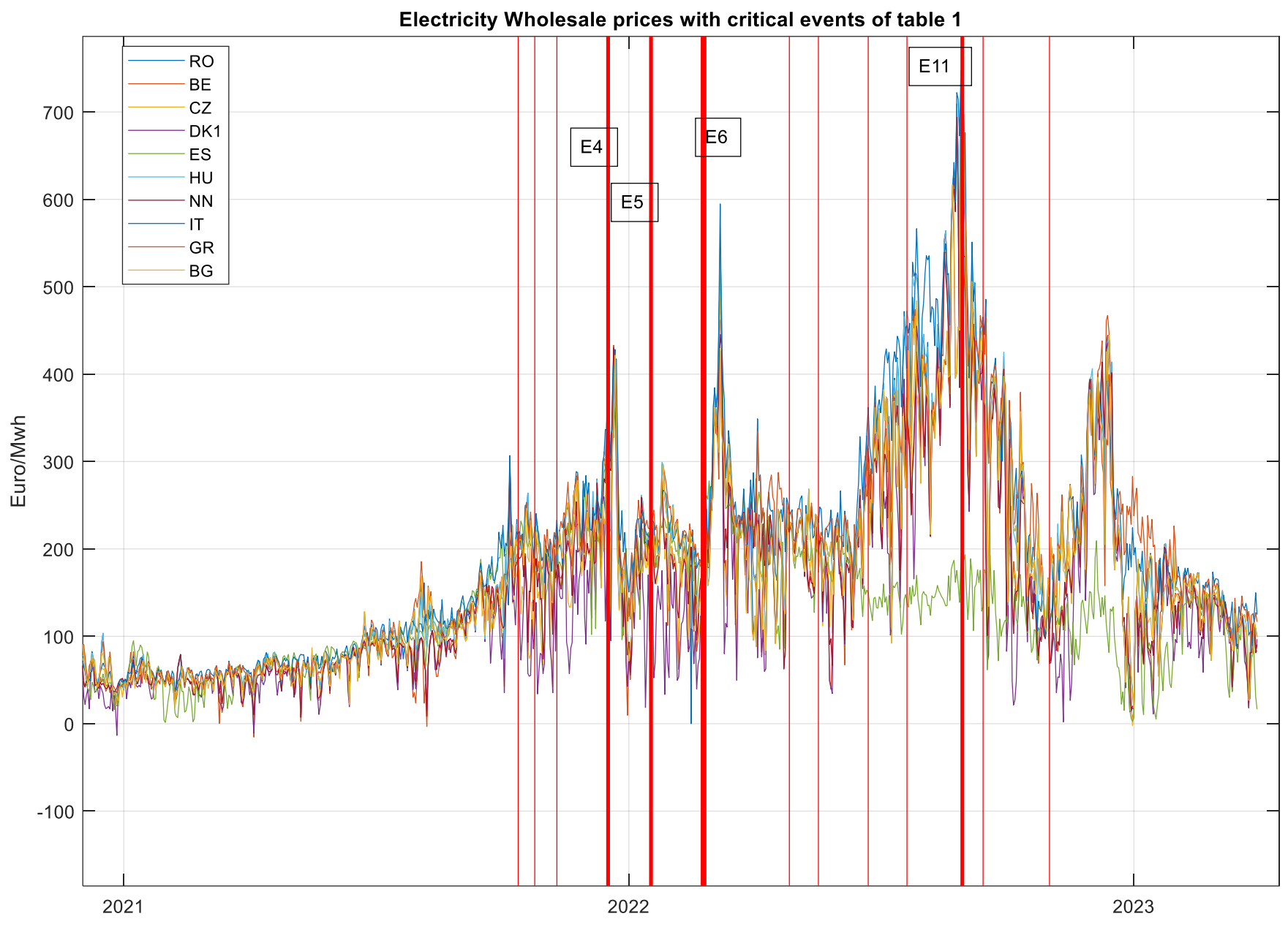

Figure 1.1: Time series of European wholesale electricity prices, analyzed in this work. Red vertical lines are dates of critical events E1-E13 as described at table 1.

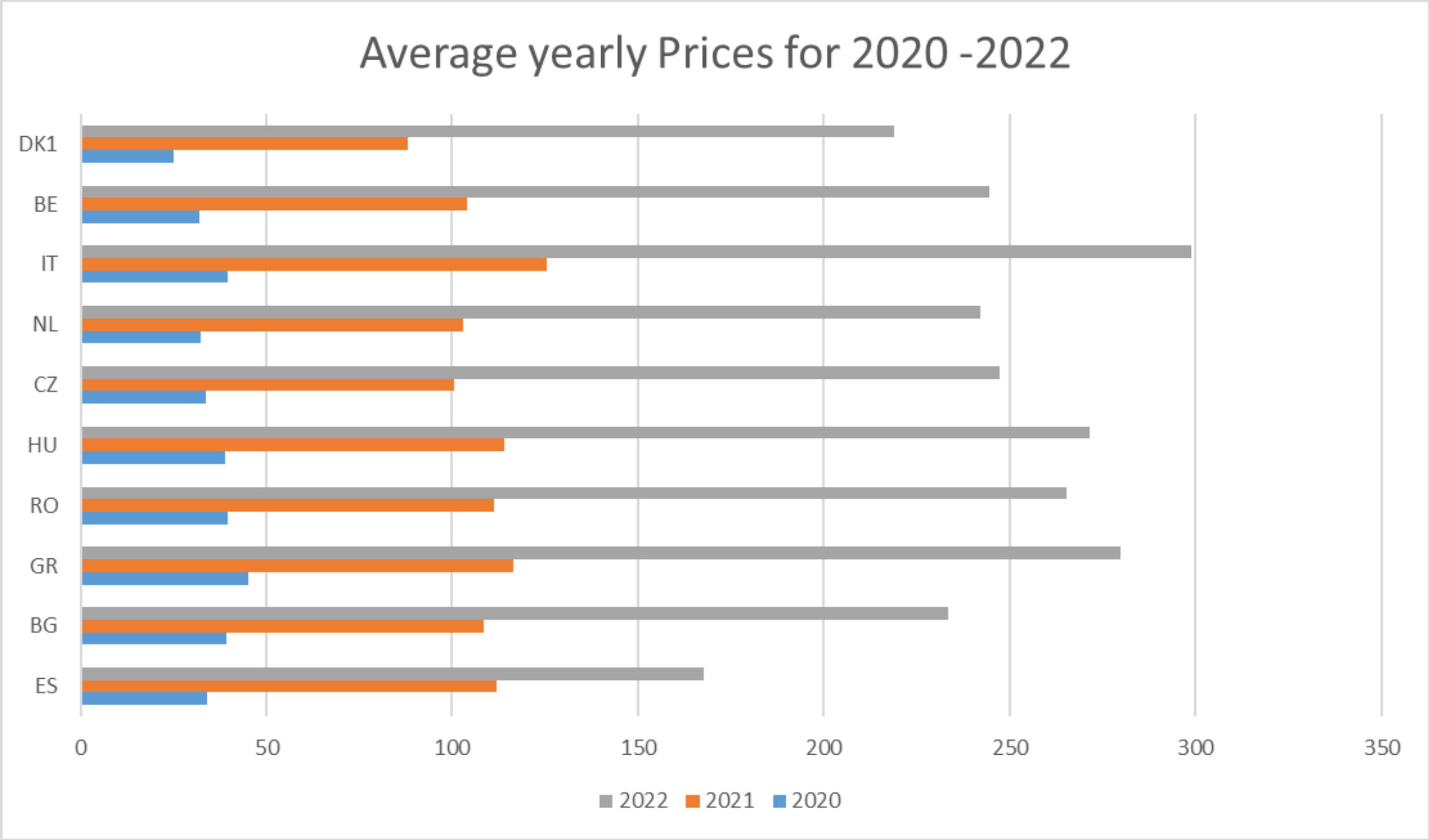

Figure 1.2. Average annual prices in day-ahead markets (DAM) in the European countries under analysis, for 2020, 2021, and 2022 years.

Spain has the smallest increase (493%). Followed by Bulgaria with an increase of 595%, and Greece with 620%. Finally, the countries with the biggest increases are Belgium with 877% and Denmark with a price increase of 767%. In 2022, the spot price averaged 279.90 EUR/MWh compared to just 45.09 EUR/MWh in 2020.

In figure 2 we plot the graph of the evolution of the TTF gas price as well as several events (E1-E13, red vertical lines) that are very critical in shaping the European gas price, during period Q1-2021 and Q2-2023, based on an extensive literature review. For events E3, E5 and E19, an information textbox is also provided. In **table 1** we list these events, associated with a short description, which are also discussed in the next sections. The ***E1 critical event***, refers to the date ***13 October 2021***, on which the Commission adopted a Communication on Energy Prices, to tackle the exceptional rise in global energy prices, which was projected to last through the winter of 2021, and help Europe's people and businesses. A toolbox of measures is included in the communication, that the EU and its Member States could use to **address the immediate impact of current prices surge**, and in addition to enhance **resilience against future shocks**. Emergency income to support households' income, aid for businesses from the Member States (MS) and focused tax reliefs, consist of the main short-term national measures. Also, via this communication, would also support investments in renewable energy and energy efficiency and adopted possible measures regarding energy storage and purchasing of gas reserves. The reassessment of the current European electricity market design was also a topic in this communication. On mid-October 2021 (critical **event E2**), as well, Gazprom started to cease selling volumes at EU gas hubs, with only limited long-term pipeline contracts remaining. Disruptions in gas flows via the routes of Yamal and Nord Stream occurred, resulting in the stopping of Hub trading, from May to September 2022. Critical **event E3 (10**

**November 2021)**, is associated with the unusual movement of Russian troops near borders of Ukraine, a fact reported early by USA authorities.

Table 1: List of critical events E1 to E13, with dates and quarters of the year, based on accumulated information from various sources, shown also as vertical red lines in figure 2.

| | Date | Symbol | Event Description |
|---|---|---|---|
| 1 | 13 October 2021 (Q4) | E1 | The European Commission presents a “toolbox” of measures to tackle exceptional situation and its impacts (Communication on Energy Prices) |
| 2 | Mid-October 2021 (Q4) | E2 | Gazprom ceased selling volumes at EU gas hubs since mid-October 2021, with only limited long-term pipeline contracts remaining. Hub trading has stopped, and disruptions in gas flows occurred via the Yamal and Nord Stream routes from May to September 2022. Meanwhile, Russian LNG deliveries to the EU increased in 2022 and 2023, despite discussions about a potential future ban. |
| 3 | 10 November 2021 (Q4) | E3 | The US reports unusual movement of Russian troops near borders of Ukraine. |
| 4 | 17 December 2021 (Q4) | E4 | Putin proposes a prohibition on Ukraine joining NATO |
| 5 | 17 January 2022 (Q1) | E5 | Russian troops begin arriving in Russia's ally Belarus, "for military exercises". |
| 6 | **24 February 2022 (**Q1) | **E6** | **Russia invades Ukraine.** Economic sanctions against Russia began, including the **removal of elected Russian banks from the SWIFT interbank system**, and prohibition of the Central Bank of Russia from access to foreign exchange reserves. |
| 7 | 27 April 2022 (Q2) | E7 | Gazprom cuts off gas supplies to Bulgaria and Poland, |
| 8 | 18 May 2022 (Q2) | E8 | the European Commission presents its €300 billion REPowerEU plan to eliminate Russian energy imports by 2027 |
| 9 | 23 June 2022 (Q2) | E9 | Germany moves closer to rationing gas, raising alert level to the 2nd of 3 stages. |
| 10 | 21 July 2022 (Q3) | E10 | New package of measures in response to Russia's invasion of Ukraine |
| 11 | 30 August 2022 (Q3) | E11 | Nordstream out of operation |
| 12 | 14 September 2022 (Q3) | E12 | EU announces tax energy companies |
| 13 | 1 November 2022 (Q4) | E13 | Obligation of member states to achive a minimum filling target of 80% of their gas storage capacity (Driver 3 of ACER-CEER’s report) |

On 17 December 2021 (**critical event E4**), President Putin proposed a prohibition on Ukraine joining NATO. This event is also concurrent with the conflict between Gazprom and Ukraine's Naftogaz (see next section).

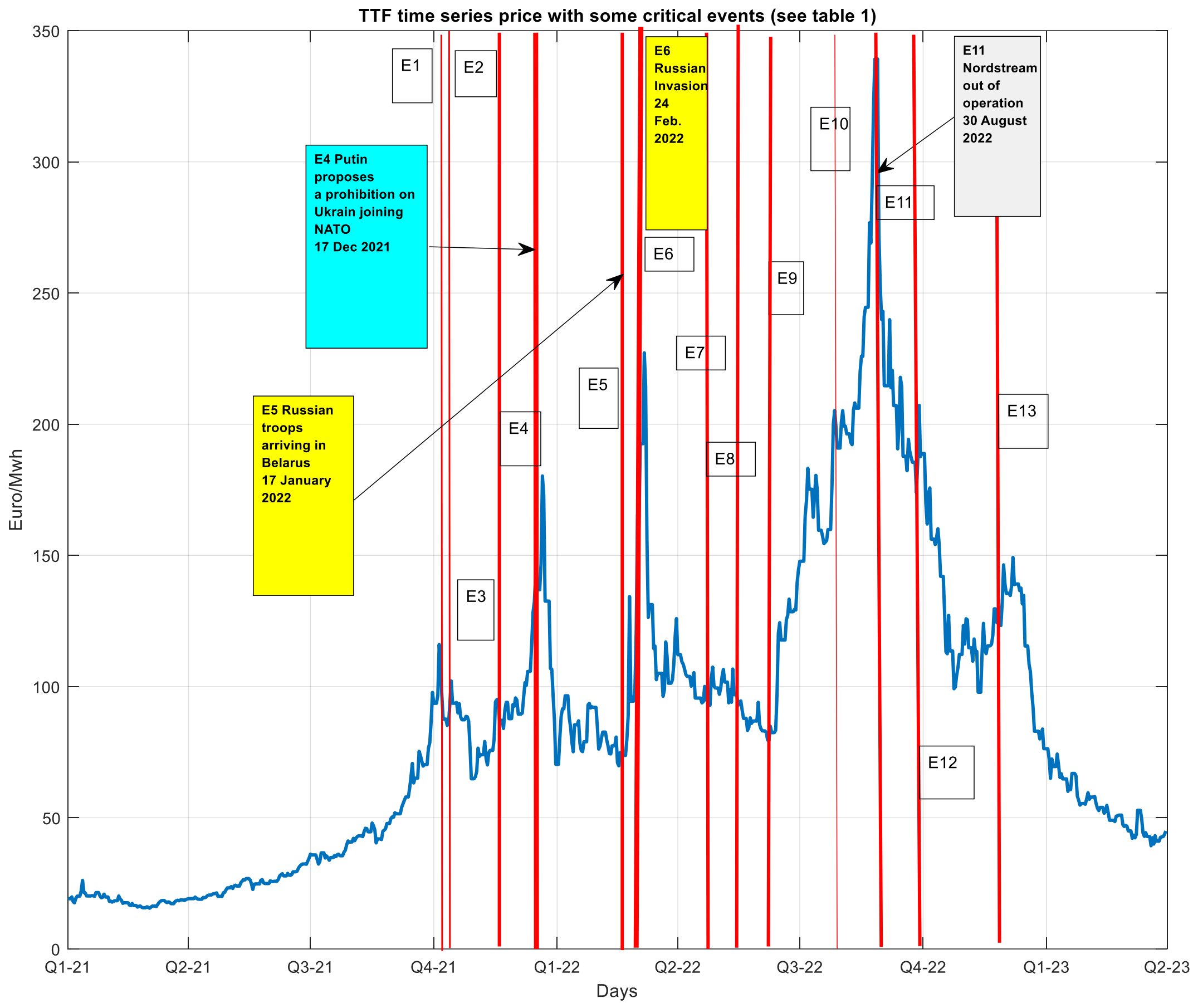

Figure 2: Evolution of the TTF gas price and critical events (E1-E13) (red vertical lines) during period Q1-2021 and Q2-2023, shown also in table 1.

### 3.2.1 Recent research works related to the effects of Russo-Ukrainian conflict on European economy and markets. Interaction of USD/RUB with energy markets

Our research is related to several strands of recent emerging literature on the effects of the *onset of the Russo-Ukrainian war on the Russian* as well as European economy and markets by analyzing the period from the 1st of December 2021 to the end of the 7th of March 2022. Mamonov and Pestova (2021) have analyzed how significant are the macroeconomic effects of financial sanctions imposed in Russia, using the Bayesian (S)VAR model. The authors focused on the Western financial sanctions that were imposed on the Russian economy in 2014 and 2017. The main finding was that the effects of these sanctions were negative and non-negligible. The most important finding, related to our work is that the imposed sanctions have a moderate impact on **the *ruble real exchange rate USD/RUB*** and on output, consumption, investment, and trade balance, *but a* significant impact on the real interest rate and external corporate debt. Ozili (2022) has examined the global economic consequences of the Russian invasion of Ukraine, the effect of imposed sanctions on Russia as well as possible spillover effects related to the global economy. An estimation of the possible economic costs of the Russia-Ukraine conflict was recently made by Liadze et al. (2022). Halouskova et al. (2022) have recently showed that the amount of attention paid to the Russo-Ukrainian crisis period, could exploited to predict the next day's price fluctuations in stock markets, globally, and they also found a negative correlation between geographical and economic distance with attention and price fluctuations. Finally, using a vector autoregression framework, Polyzos (2022) provides evidence that shocks in the number of tweets were related to the depreciation of the Russian Ruble. The author used Twitter to extract number of tweets (an *attention measures*) and a sentiment index to study their impact on intraday returns of several stock market indices, commodities, the U.S. Treasury bill index, Bitcoin and three currencies, including Ruble.

After Russia's invasion of Ukraine, the USD to RUB value lost significant ground, reaching a low of 135 rubles in March 2022, just after **critical event E6**, (the invasion of *24 February 2022*). A chart (not provided in this work) comparing the monthly average of RUB against USD and Euro, shows also this significant devaluation, since 2008. An interesting fact for which no specific reason was given for its timing, is a decline that started in November 2020, and continued into 2021. More specifically, the USD/RUB exchange rate reached its highest point earlier in 2020, as one U.S. dollar could buy nearly 80 rubles in March 2020. Years later, values were significantly different. Figure 2.1 shows the time series of USD/RUB exchange rate from June 2020 to April 2024. As a result of the Russian invasion, a very crucial event occurred on 26 February, 2 days after invasion, that is related to USD/RUB rate: *the removal of selected Russian banks from the **SWIFT interbank system**, and the prohibition of the Central Bank of Russia from access to foreign exchange reserves* (Lyocsa S., et al., 2022). According to the previous paper, on 27 February 2022 restrictive measures from EU on exports sectors of Belarus and Russia were launched, and the Russian nuclear armed forces were set on high alert. These events considered together with the drastic increase in TTF and NGNMX prices due to invasion, suggest that these markets might be driving factors of the USD/RUB dynamics, and this is the case as it is shown in the results section 6: the rolling mutual information and the PMIME causality analysis indicate a strong information flow from TTF and NGNMX markets to the USD/RUB. A similar chart (not provided in this work) comparing the monthly average of RUB against USD and Euro, shows also this significant devaluation, since 2008. An interesting fact for which no specific reason was given for its timing, is a decline that started in November 2020, and continued into 2021. More specifically, as figure 2.1 shows, the USD/RUB exchange rate reached its highest point earlier in 2020, as one U.S. dollar could buy nearly 80 rubles in March 2020. A work very related to this paper is the research of (Lyocsa S. et al., 2022) that models the intraday price fluctuations of USD/RUB and EUR/RUB exchange rates from 1st of December 2021 to the 7th of March 2022. They used several specifications of Heterogeneous Autoregressive (HAR) type of models of Corsi (2009), to analyze the variations of implied volatility of the rates above (dependent variable) using as independent variables two sets of variables: ***attention measures*** (google searches as a proxy of investor's attention) and lagged values of ***implied volatility*** of the rate (a proxy of investor's expectations).

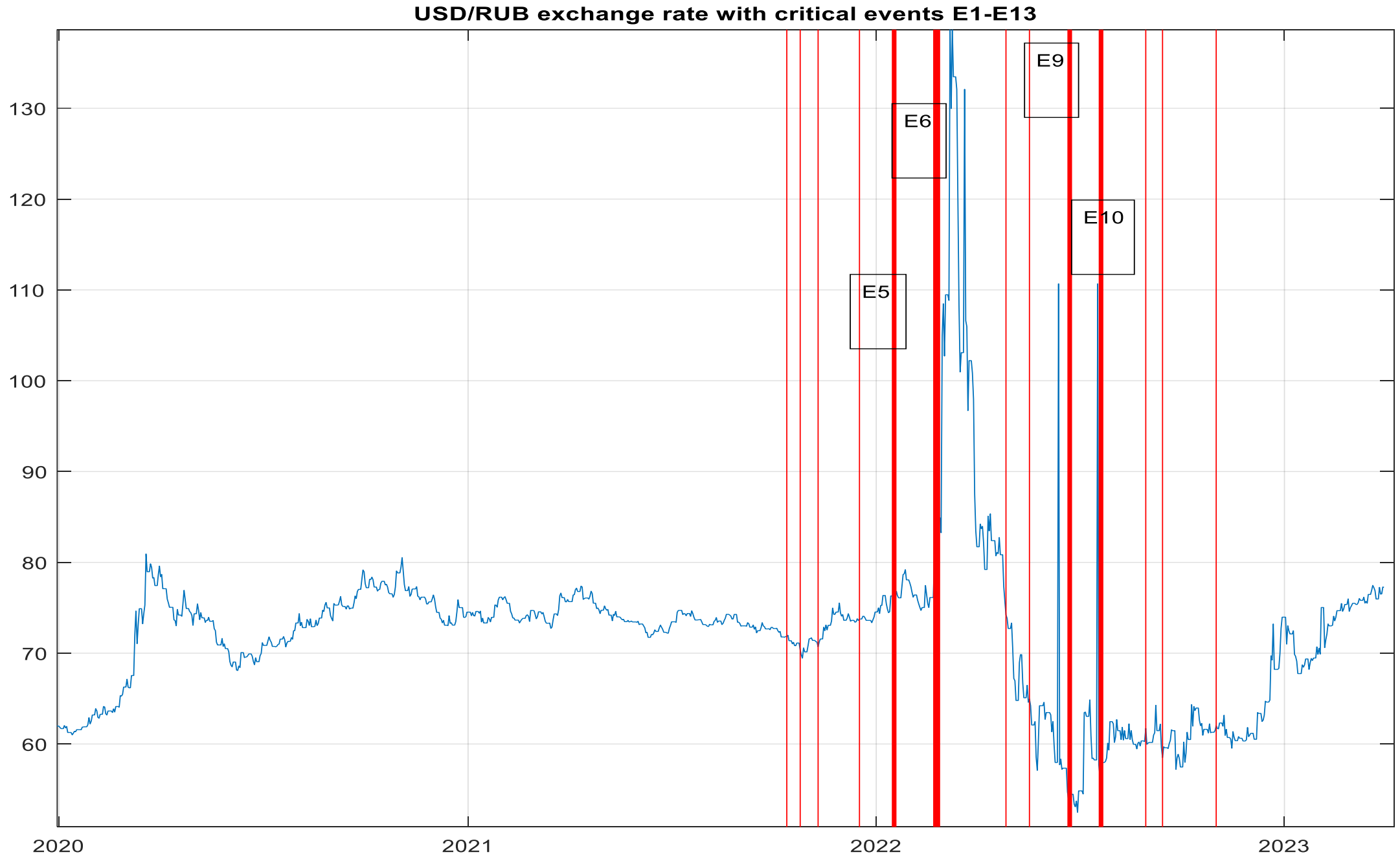

Figure 2.1: Time series of USD/RUB exchange rate from January 2020 to Q1 2023. Red vertical lines mark the critical events E1-E13 of table 1.

The attention measures, based on limited attention theory (Barber and Odean,2008), are exogenous variables collected as *Google Trend data* used to predict future price variations (Dimpfl and Jank, 2016; Lyocsa et al., 2020). The attention measures consist of three sets of variables: 1) general financial market (e.g., SP500, VIX, FX market etc.), 2) Ruble (e.g., Ruble, *USD/RUB*, Russian interest rate), and 3) Russian economy related (e.g., *economics sanctions to Russia*, *asset freeze*, *Nord Stream 2*, *export controls*, *fx reserves*, *SWIFT Russia*, etc.). From this list we see that some of the crucial variables in the 3$^{rd}$ set above are the ones that are considered also in our paper to form the list of critical events E1 to E13 (see Table 1 and section 3). The findings of (Lyocsa S. and Plihal T, 2022) are consistent and further supported by our findings, described in section 6.3.1.

Some recent works related closely to this work, especially in how the efficiency of financial compared to that of energy markets has been affected during the Russo-Ukrainian war, are: (Aslam F, et al., 2022; Ahmed S., et al., 2023; Manelli A., et al., 2024; Sari E.L, et al., 2023; Mishra A.K., et al., 2024; Michiyuki Y., et al., 2023; Marwan I., et al., 2023; Hansen J., et al., 2023. The work of Choi S.Y, (2021), provides information on the EMH of financial markets during a crisis, in general.

In the very recent work of (Manelli, A., et al., 2024), the authors analyzed the nature of interaction between a financial market (the Euro index Eurostoxx50) and Dutch TTF gas market. This resembles, partially, our work since we both examine the interaction of TTF gas market with a financial, in our case

USD/RUB rate, instead of the Euro index. They used a Quantile VAR (QVAR) model and found that the two markets interact, as exhibited by the asymmetries of TTF coefficients, that cannot be detected via ordinary OLS. They showed how the TTF market has affected the left and right tail of the of the Eurostoxx50, and how the effects of the shock are limited and positive for upper quantile and negative for the lower, for the TTF, using 3-D Impulse Response Functions (IRF**)**. ***Their most important finding related to our work is that the interaction of the markets is mutual, bidirected, i.e. the TTF market affects Euro index and vice versa. This result supports our finding that there is a mutual interaction between USD/RUB rate and TTF price, as shown by applying the tool PMIME, described in section 6.3.1. Thus, our work contributes further in the growing recent research of detecting 'causalities' between financial and energy markets during the Russo-Ukrainian crisis.***

Similarly, the work of (Aslam F., et al., 2022) resembles our work, in respect to the *efficiency* in several energy markets they analyzed (TTF included), during the mentioned crisis. They used multifractal detrended fluctuation analysis (MFDFA), to intra-day data, from the period October 21, 2021, to May 20, 2022. The main findings is that during the crisis, TTF market exhibited multifractality that varied drastically accompanied also by a strong decline in all other energy markets, but surprisingly, not in the TTF market. *More specifically, the TTF market, being the least efficient before the invasion, turned out to be more efficient afterwards, suggesting that the investors in TTF market are likely to exhibit herding, more prominently after invasion. The improved efficiency observed in the TTF market, after invasion, highlights its unique characteristic and underlying complexity. The above results support our findings* related ***to the increased in the rolling Hurst values of the TTF (as well as of all electricity markets analyzed), during the same critical period, as described in the results section.***

Also, the issue of how war in Ukraine has affected the efficiency of a financial market, is the subject of the paper of (Sari E.L., et al., 2023), in which they examined the interplay between geopolitical events and the responses of the Indonesian Stock Exchange (IDX), during three distinct periods focused around the Russian invasion : *a) the announcement of Russia's invasion of Ukraine on the* ***24th of February****, b) the announcement of the oil import embargo on Russia by the EU* ***on the 31st of May****, and c) the announcement of the first wheat export ship's departure from the harbor of Odesa on the* ***1st August 2022***. They found that events involving several countries or institutions have exerted stronger impacts on the energy sector within IDX, resulting in more pronounced market responses. During the critical events IDX ha exhibited a semi-strong form of efficiency. From comparative tests results, using various methods, the authors have concluded that the IDX exhibited a level of semi-strong form of efficiency (see sections 3.1 and 6.1), and that this market has reacted more strongly following event c), compared to the other two events, because this event is a significant diplomatic breakthrough, involving multiple countries globally. Several countries intervened to enable Russia's resolving agreements with Ukraine of exporting grain. Thus, *the involvement of multiple countries in the conflict has a stronger impact on the prices of energy stocks in the IDX, manifested via their significant differences before and after the war, an indication that this financial market has adjusted quickly and effectively to the 'entire' new available information generated by these critical events. Our result, in section 6, supports the above finding since the rolling Hurst of USD/RUB exchange rate has exhibited an increased trend towards the EMH limit at the critical events examined in our case.*

A paper related to our study, published very recently, is the that of (Kostaridou E., et al., 2024), in which they have analyzed the impact of the Russia-Ukraine unrest, reflected as geopolitical shocks on the prices of eight agricultural commodities, including the EUR/RUB daily exchange rate (in our paper we used the USD/RUB), by first detecting the *breakpoints* induced by the critical events associated with this conflict. They enhanced the Granger causality method to be able to conduct the Granger tests across different periods, emphasizing the changes of the price volatility, due two critical events that 'shaped' the formation of the corresponding breakpoints: the Russian occupation of Crimes in 2014 and the Russian invasion in Ukraine in 2022. The breakpoints were identified by using the Vogelsang and Perron method (optimizing the Dickey-Fuller t-statistic) (Vogelsang and Perron, 1998). They choose wheat as the primary indicator for assessing the impact of military actions, and *its identified breakpoint found to be aligned precisely with the onset of war in 2022 as well as with the 2014 events in Crimea*. Also, the EUR/RUB's breakpoint demonstrated similar behavior. Their results showed significant changes in the interlinkages among the variables during the crisis periods, compared to stable periods.

The role of information during the Russo-Ukraine war and how it affected the price volatility of stocks in forty European countries, is examined by the work of (Cataline Gheorghe, et al., 2023). Using a short time series but focusing exactly on the period of the Russian invasion, 24 Feb. 2022, and the Granger causality test they found that some markets proximate to Ukraine, notably Hungary, Czech Republic (both countries are included in our analysis), Poland, Servia etc., reacted in anticipation of the conflict, days prior to February 24, *a result that supports our findings, since the dates of their identified breakpoints by BEAST method used in our study, are leading the date of invasion* (see results section below).

In the paper of (Yousafet , I., et al., 2022) they examine the impact of the breakout of the conflict between Russia and Ukraine on the G20 and other selected stock markets, employing the method of event study. The found that the Russian invasion on 24 Feb.2022 revealed a *strong negative impact of this military action on most of the stock markets, especially on the Russian market.* On the event day and post event period, the Russian invasion impacted the markets negatively but with different strength and temporal characteristics (leading, concurrent and lagging response behaviors). More specifically, their country-wise analysis demonstrated that the stock markets of ***Hungary,*** Russia, Poland, and Slovakia were the ones that almost 'concurrently' reacted to the military events, exhibiting ***negative returns the period before the critical event,*** while the stock markets of ***Italy***, ***Spain, Romania***, Japan, Australia, France, Germany, India, South Africa, **,** and Turkey were adversely affected in the after invasion period. The countries with bold letters are included also in our preset study.

### 3.3 Explaining the inefficiency of TTF market via examining the crucial factors shaping its price. The impact of the disruption of Russian gas supply.

In this section we tried to identify the critical events of table 1 with specific events described in the report of ACER-CEER (ACER-CEER, 2023) and used extensively in the results of the BEAST decomposition. According to the report, the European gas market is ***not an efficient market***, a 'statement' that is not quantitatively supported in the report. This lack of information in the report gave us the incentive to quantitatively examine the inefficiency of the market and how this finding is connected to the structural breakpoints. Therefore, another significant contribution of our paper is to enhance the above argument, by assessing the efficiency of the TTF market as well as of all electricity markets via the Hurst exponent, and further examine if these efficiency results can be related (via correlation analysis) to the detected structural breakpoints of each separate electricity market. The results of the Hurst analysis are reported in section 6.

For this purpose, we first considered the main factors that are responsible for the inefficiency of the TTF market, as described in a recent report (ACER-CEER, 2023), due to complexities associated with gas production, transport and trading, gas market access and restricted options in the supply of gas, various barriers have been elevated in the market.

Six primary conclusions on the gas market developments, during the summer 2022, were presented in ACER-CEER report, a summary of which is described below:

1. The primer driver affecting EU gas prices (and subsequently, to a larger or lower extend the spot electricity prices) is the disruption of Russian gas supply.
2. In 2022, a reduction in EU gas consumption, by over 50 billion cubic meters, occurred. However, due to larger storage injections and increased gas-fired power generation, an additional demand took place during the summer months of 2022, resulting in record-high gas prices.
3. Substantial gas volumes were attracted, ahead of winter 2022/2023 because of implemented storage measures, but unfortunately in some instances associated with high injection costs.
4. Although LNG played a main role in securing EU gas supply, expensive spot LNG imports finally drove hub prices up. As a response, this caused a fast development of LNG infrastructure that was proved to be effective.
5. The integrated gas system of the European Union revealed a significant resilience during the period of our analysis. Furthermore, LNG terminals and pipelines were heavily congested, due to an extremely intensive supply stock, resulting in price disparities and trading disruptions.
6. Due to record-high gas prices, a surge in trading margins was observed, but despite this event, hub trading volumes remained robust. The trading climate was more disturbed.

The set of the above six conclusions, constitute the ***six critical drivers*** that are deemed to be responsible for the gas prices surge: *driver 1: Disruption of Russian supply, driver 2: demand developments, driver 3: storage analyses, driver 5: LNG price developments, driver 5: Transmission infrastructure congestion and finally driver 6: Trading developments*. The associated to these drivers' crucial events happened to specific dates which are compared to the dates detected by the BEAST tool, during the period of our analysis.

### 3.3.1 Extracting the dates of crucial events associated with the above six critical drivers.

Using a tool called 'market health', ACER-CEER institutions have assessed the structural competition 'environment' of the European gas markets, providing an evaluation of the available suppliers as well as their possible concentration. In the report above, an analysis presented for the gas supply diversification of EU member states in 2023, compared with 2021. An informed estimate of the market's gas supply contractual origins over the year, per Member states, is presented. In respect with the markets, we analyze here in this work, we have isolated specific information from the report, in an effort to shed light on the factors that have influenced TTF prices. In ***2023, Bulgaria (BG)***, imported gas from Greece and Turkey primarily via LNG flows, reaching their interconnectors and then imported to Bulgaria. Regarding ***Czech (CZ)*** market, all imports during this period are from Germany, after a cancellation of a long-term supply contract with Russia. The % of actual volumes purchased and the number of supply sources (in terms of the contractual origin of gas in EU Member States, MSs), as given in fig.20 of (ACER-CEER, 2023), for BG market in 2023 is: from Greece and Turkey (EL+TK) 67% and 32% from Azerbaijan (AZ). In 2021, the situation was 80% from Russia and 20% from Greece and Azerbaijan (EL+AZ), i.e. during 2021 the dependence of BG on the Russian gas was extremely high. Similar dependency on Russian gas supply is shown by other EU countries, as shown in the same report. The conclusion is that EU's dependence on external gas imports and especially from Russia (the largest supplier of natural gas until 2022), was very high. The decrease of 2/3 of the EU internal production, since 2010, has extremely enhanced this dependence. During 2010-2022(Q1), the consumption of gas has been marginally reduced, until the market shock in 2022, during which the EU gas demand was drastically reduced.

***The main driver responsible for the unprecedented EU gas price rises, happened during the 2021 and across 2022, is the significant reduction in Russian gas supplied volumes, as well as the total uncertainties associated with this supply decline expected forward***. This can be seen from the abrupt and gradual escalation of the TTF front-month prices shown in **figure 3**. This up-trending dynamic evolution is correlated with the gradual reduction of the aggregated Russian supply into the EU per supply corridor (Baltics, North Stream, Turk stream, Ukraine, Yamal), evident during Jan2020 to Jul2022. Just a little before July 2022 (i.e. before critical ***event E10***), the aggregated supply reaches the lowest value, forcing TTF price to increase rapidly to a historical high record value (>300 EUR/MWh) (see fig. 21 in ACER-CEER, 2023). According to (ACER-CEER, 2023), most of the Russian gas supplies were procured via long-term supply contracts between European byers and Gazprom (175 bcm annual nominal contractual capacities, in 2021). To reduce the dependency of EU on Russia energy supplies, due to Russia invasion of Ukraine, EU undertook several actions, so from June 2022, the Russian dependence dropped below 20% and moved gradually lower, reaching in September 2023 to a 7% level. From May 2022 (***critical event E8***) to September 2022, gas flows completely disrupted via Yamal and Nord Stream

corridors. The above report, presents also a list of specific events that were responsible for the gradual decrease in Russian supplies in EU, since mid-2021:

- ***A gradual intensified political conflict, regarding the entry into operation of the Nord Stream 2***, accompanied also by US commercial sanctions. As a response, Russia reduced the supply to **Germany**, Slovakia and **Hungary,** as an indication of Russia's pressures on EU and its member states. Gazprom stopped selling volumes at EU gas hubs, and from *mid-October 2021* (***critical event E2)*** completely disrupted the operation of its own trading and sales information system.
- ***The conflict between Russia's Gazprom and Ukraine's Naftogaz, creating frictions in their mutual long-term contract.*** In December 2021 (***critical event E4***), Naftogaz send a complaint to EC accusing Gazprom of market power in the EU gas market. This conflict created further risks regarding the continuation of Russian gas supplies in EU, because of its conflict with Ukraine and gas transit via this country.
- ***The achievement of EU's climate targets has made necessary several legislations and laws, for the promotion of RES, Hydrogen and Biomethane.*** These actions are considered to have put a pressure to Russia to maximize its revenues as 'soon as possible', before the expected reduction in gas imports, due to reduced need for gas from gas power plants. More specifically, EU proposed a Hydrogen and biomethane package in December 2021 (***event E4***), to accelerate the decoupling of EU power generation from Russian gas.
- ***The actions of Russia to reduce the inflow of gas in its underground gas storage facilities, controlled by Russia***. These actions took place before winter 2021/2022, and concerned the very low storage stocks in the markets of Germany, Austria and Netherland (NNL)

All the above events (conflicts and frictions) have been escalated during 2022, especially from January 2022 ***(event E5***), when Russian troops began arriving in Belarus, and more crucially after Russian invasion in Ukraine, on ***24 February 2022 (critical event E6)***. The consequent EU economic sanctions and worsening of EU (and its member states) and Russia diplomatic relations, led to Russia's response to use its continuation of gas supplies to EU as a tool for political and financial pressures to EU's member states. Due to the Russian invasion, there was a significant price shock in the EU gas market, at the end of February 2022, and the beginning of March 2022. TTF front-month prices increased drastically from 71 Euro/MWh on 21 February 2022, to 212 Euro/MWh on 8 March 2022, indicating the high risks in gas supply due to this military fact. In the second week of March 2022, TTF prices partially declined, stabilizing at a range of 95-105 Euro/MWh. The long-term supply commitments of Gazprom during this period were initially fulfilled, thus the gas flows to EU, up to the end of April 2022, ranged from 250 t0 340 mcm/day, an amount 25% lower than the previous year's one. From the end-April 2022, supply disruptions gradually occurred, raising concerns about the normal continuation and reliability of the Russian gas inflows in the future. The inflows to **Bulgaria** (3 bcm/year) first curtailed on ***27 April 2022*** **(Event E7**, table 1), due to contract denomination in rubles, imposed by Russia. In summary, examples of major disruptions of gas flows from Russia include (ACER-CEER, 2023): 1) to Germany, in mid-May 2022, 2) to Poland before end of 2022, 3) to Finnish Gasum, in May 2022, 4) to **Denmark** in early June 2022. On 14 June 2022, a date very close to event E9, presenting as an excuse technical problem, Gazprom decided to reduce gas flows to Germany via Nord Steam 1, creating a substantial drop in gas flows to **Italy**, Austria, France and Germany, causing TTF day-ahead prices to increase rapidly to 40 Euro/MWh, in just two days. After these events, the deterioration of the gas disruptions intensified. On ***11 Jully 2022*** **(Event E10)** Nord Stream 1 flows ceased completely, an event that in combination with the need to replenish storage stocks with expensive LNG, as well as the badly delayed maintenance in the Norwegian fields, exerted a large impact on the TTF prices, which reached their peak value above 300 Euro/MWh, on 26 August 2022, in anticipation that Nord stream would be out of operation (***event E11***). Things became much worse when an explosion happened in both Nord Stream 1 & 2, on 26 September 2022, causing TTF day-ahead prices to increase by circa 30 Euro/MWh in the next two days. On **September 14, 2022**, ***(event E12),*** the problem of energy supply in EU is resolved*:* the EU announces the taxation of energy companies, send the message that the problem of energy supply in EU is resolved, and prices began to fall, Europe had found alternative ways to supply but had also reduced its electricity needs, and this is what the BEAST tool manages to identify successfully. with the most significant relative decreases occurring in October and November 2022. The Council regulation adopted a European Commission proposal originally presented in March 2022. *Finally,* ***event E13 (1 November 2022)*** *market* a final sudden *price* increase took. The regulation introduced an obligation for member states to achieve a minimum filling target of 80% of their storage capacity by *this date*, to be extended to 90% in the following years, together with measures to determine the filling path. The delay in filling stocks for Greece and most other electricity markets is reflected in another upward trend change in November to return to a downward trend in December after securing stocks.

In summary, regarding this crucial period of summer 2022, ***Russian gas supply disruptions are considered as the most important driver in causing a surge in EU gas hub prices***. The Byers were concerned about future shortages in gas that could affect badly their needs, especially the coming winter season. In 2022, monthly average TTF prices reached a level >130 EUR/Mwh, which in comparison with the average value between 2016-2021 is seven times higher. Also, during the storage filling period between March 2022 and October 2022, the average TTF prices reached 160 Euro/Mwh. **Figure 3** is a pictorial presentation of the how the Restriction of Russian gas supply has affected severely the TTF gas prices and consequently the average wholesale electricity prices of the European markets analyzed in this work.

Considering all above, the main conclusion is that the dramatic increases on EU energy prices (gas and electricity) are due to the over-reliance or strong dependence of EU on its historically major gas supplier, which was harshly revealed by the gas supply shock occurred during 2022. All previous attempts of EC member states (using a plethora of tools) to diversify its gas supplies, were proved to be inefficient, as a new dimension, ignored so far in the process of designing the gas market, has emerged: the geopolitical implications. If this dimension had been considered in the gas market design, then the market level of diversification would have been different, enhanced, therefore the impacts of the Russian gas flow interruptions would have been softer and mitigated. But in 2022, the reality of the EU gas market was completely different. As an adaptive response of EU, the EC sanctioned its *REPowerEU flagship* *plan*, in early May 2022 (***event E8***), a plan that encompasses a set of measures targeting in transforming Europe to a continent that is strongly independent from Russian gas supplies.

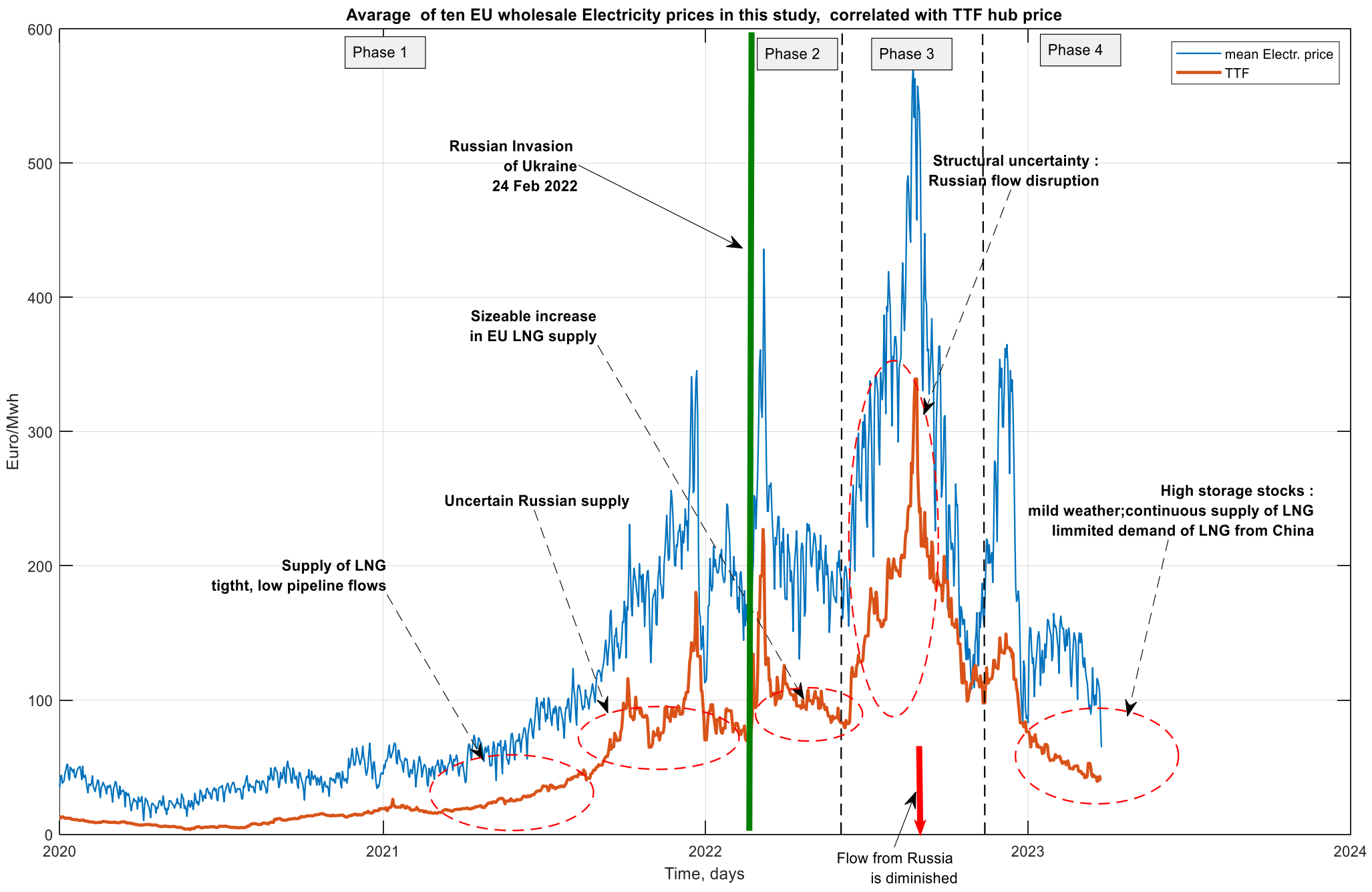

**Figure 3:** A pictorial presentation of how the Restriction of Russian gas supply has affected severely the TTF gas prices and consequently the average wholesale electricity prices of the European markets analyzed in this work.

### 3.3.2 Focusing on the onset of Russo-Ukrainian war and escalation of economic sanctions: November 2021- March 2022.

We focus our analysis of the impacts of the Russo-Ukrainian conflict on the electricity, gas and financial markets, *on the period November 21 (considered as the onset of the escalation) to March 2022*. The escalating presence of Russian troops near Ukrainian borders (also in Belarus and Crimea), pseudo-characterized as 'regular military exercises', started during November and December 2021. Russian President Vladimir Putin announced on February 21, 2022, that Russia recognized the independence of two pro-Russian regions, namely, Donetsk and Luhansk, in eastern Ukraine, triggering the first round of economic sanctions from NATO countries, almost immediately a full-edged war, shocking all countries globally, escalated on February 24, 2022, when Russia began a complete invasion of Ukraine. Therefore, European and other western countries imposed economic sanctions again Russia, as a) restrictions to Russian imports and exports, b) the removal of selected Russian banks from the SWIFT interbank system, and c) the prohibition of the Central Bank of Russia from access to foreign exchange reserves, to name a few. Thus, in the results of section 6, emphasis will be given to the period of November 2021 to March 2022.

## 4.0 Data Sets used, descriptive statistics and tests for normality and stationarity.

We use data from *ENTSO-E* for our set of 11 electricity markets, including Austria (AT), Belgium (BE), the Czech Republic (CH), Denmark (bidding zone 1) (DK1), Germany with Luxemburg (DELU), Spain (ES), France (FR), Greece (GR), Hungary (HU) and Italy (South bidding zone) (ITsouth), from January 2020 to April 20, 2022. The data are daily (average of 24-hourly data). All prices are in Euro/Mwh. In our specific study of the combined impacts of natural gas and RES generation on the Greek prices, we use the wind and solar generation in Greece (in MWh), for the same period considered.

As independent variables in our modeling, we use time series of the natural gas TTF futures price, the Natural Gas Futures Index at NYM and the USD/RUB exchange rate. TTF is the virtual trading point of the *Netherlands Transfer Fund* (title transfer facility, TTF), which is used as a reference gas market at European level and trades as Euro/Mwh. The futures prices of natural gas products traded on the TTF has become now the widely used gas pricing indicator in the majority of the European gas markets, including Greece, and this is the reason we consider this variable as the most crucial factor of influencing the whole sale electricity prices. We use the TTF futures monthly prices (May 22 futures). The Natural Gas Futures at New York Mercantile Exchange (NYM) series reflects the dynamics of the Natural Gas prices which are based at the Henry Hub in Louisiana (USA) and is considered one of the most liquid futures contracts in the Commodities Trading and Investing asset class in the world. Our intention on incorporating this information in the overall analysis is to possibly capture any cross-market effects between the TTF prices and the dynamics of the futures contracts in NYM, towards potentially explaining any lagging/leading behavior between the two and the SMP – i.e. which one of the two markets managed (or not) to adequately explain the SMP’s dynamics in the short-term. As per Statista (https://www.statista.com/statistics/217856/leading-gas-exporters-worldwide/), Russia was the biggest Natural Gas exporter for 2021 and also traditionally one of the top in the previous years. For that reason and since payments for exporting Natural Gas are made mostly in Rubbles, we also include the *USD/RUB foreign exchange pair* in our analysis to possible capture pre-signaling behaviors between the FX and the rest of the commodities markets’ related instruments in our dataset. We select the USD as the reference currency for the conversion as the one against which the largest volume is being traded in the spot and in the Derivatives FX Markets, as well. For the period considered, we observe that in the markets of Belgium (BE), Germany and Luxemburg (DELU), and Denmark (DK1), minimum values are negative. For these markets, in the modeling procedure, we follow the works of (Sewalt M., et al., 2003; Hui Q. et al., 2021; Naimoli Aet al., 2021; Atanasoae et al., 2020) and choose to just ignore the rare dates in which prices are negative, since taking differences of log prices is not permitted for negative prices. The non-normality of the electricity prices is confirmed through the rejection of the null hypothesis of the Jarque-Bera tests) (see Supplementary material, S1).

# 5. Methodology.

## *5.1 Hurst Exponent and the Efficient Market Hypothesis (EMH)*

Hurst exponent is a statistical tool (measure) for analyzing the scaling properties of a time series, which, in the case of financial asset prices, corresponds to patterns that are repeated at different time scales. The most popular model to study such patterns is the Brownian motion of Bachelier (Bachelier, L., 1900), and its modifications as fractional Brownian motion (Mandelbrot B.B., 1997; Clark P.K., 1973) and Levy motion (Mandelbrot B.B., 1967; Fama E.F, 1965; Fama E.F, 1970). The re-scaled **range (R/S)** for **distinguishing completely random time series from correlated time series** (Hurst, E., 1965), is the main concept in the seminal work of Hurst. In his approach, the procedure consists of several steps, the first is dividing a time series of length *L* into *d* subseries of length *n*, the second is finding the mean $(E_k)$ and standard deviation $(S_k)$, for each sub-series $k = 1, \dots d;$ a), then normalizing the data $(Z_{i,k})$ by subtracting the sample mean $X_{ik} = Z_{i,k} - E_k$ for $i = 1, \dots, n;$ . Finaly, creating a cumulative time series
$Y_{i,k} = \sum_{j=1}^{i} X_{j,k}$ for $i = 1, \dots, n;$ and then finding the range $R_k = max\{Y_{1,k}, \dots Y_{n,k}\} - min\{Y_{1,k}, \dots Y_{n,k}\}$; and rescaling the range $R_k/S_k$. The mean value of the rescaled range for subseries of length *n* is finally computed as $(R/S)_n = (1/d)\sum_{k=1}^{d} R_k/S_k$. We can also plot, the $(R/S)_n$ statistics against *n* on a double-logarithmic paper. *If the process of time series returns is white noise, then we get roughly a straight line with slope 0.5., while if the process is persistent then the slope is > 0.5. Finally, if the process is anti-persistent, the slope is <0.5.* The level of significance level is usually chosen to be $\sqrt{1/N}$ – the standard deviation of a Gaussian white noise. However, as (Weron, R. et al. 2000) has shown in applications of HE in energy markets, in the case that *n* is small, there may be a significant deviation from the 0.5 slope. For this reason, as he proposed, the theoretical (i.e. for white noise) values of the $(R/S)$ statistics can be better are approximated by

$$E(R/S)_n = \begin{cases} \frac{n-\frac{1}{2}}{n}\frac{\Gamma((n-1/2))}{\sqrt{\pi}\Gamma(n/2)}\sum_{i=1}^{n-1}\sqrt{\frac{n-i}{i}} & for\, n \le 340, \\ \frac{n-\frac{1}{2}}{n}\frac{1}{\sqrt{n}(\pi/2)}\sum_{i=1}^{n-1}\sqrt{\frac{n-i}{i}} & for\, n > 340\,. \end{cases} \tag{1}$$

$R/S$ is shown to follow asymptomatically the relation $(R/S)_n \sim cn^H$ thus by taking logs of both sides we have $log(R/S)_n = logc + Hlogn$. Thus, the value of $H$ can be estimated by simple regression.

*The Hurst exponent takes values in the range* $0 \le H \le 1$ *. As we have already mentioned H=0.5 indicates that the process is a pure random walk, while for H<0.5 the process is mean-reverting (anti-correlated or anti-persistent) and for H > 0.5, a persistent, positively correlated process.* A direct connection exists between HE the Hurst exponent and the "fractal dimension", which gives a measure of the roughness of a surface. The relationship between the fractal dimension, *D* and the Hurst exponent, *H*, is (Peters E.E, 1994):

$$D = 2 - H \tag{2}$$

Hurst exponents *quantify the correlation of a fractional Brownian motion***.** A fractional Brownian motion (fBm) is a *random walk with a Hurst exponent different from 0.5 and thus with a memory.* The decaying of spectral density of an fBm has a relationship with the Hurst exponent as follows:

$$P(f) = spectral\ density\ \propto \frac{1}{f^{\beta}} \qquad (3)$$

$$\text{where}\ \ \beta = 2H + 1 \qquad (4)$$

is the power spectrum exponent.

In a mean-reverting series, an increase in values will most likely be followed by a decrease or vice versa (i.e., values will tend to revert to a mean), so future values tend to return to a long-term mean. On the other hand, in a persistent time series an increase in values will most likely be followed by an increase in the short term and a decrease in values will most likely be followed by an increase in the short term. We can also interpret $H$ as a measure of the bias in the fractional Brownian motion (fBm). Thus, in our analysis of the electricity and gas markets, the deviation from random walk provides interesting information on the inherent *volatility* and *risk.* A relatively high $H$ underpins the relatively *smooth trend* and *less or controlled volatility*, and the persistent effect ($H > 0.5$) is stronger than the mean-reverting effect (when $H < 0.5$). These arguments are related to market efficiency, as we have mentioned previously. More specifically, *higher H values correspond to emerging or developing markets* in which the *EMH is not satisfied* (we remind that in competitive markets *all* information is instantaneously reflected in prices and it is not possible to systematically beat the market, so prices should follow a random walk ($H = 0.5$) (an unpredictable process), thus prices in such a market cannot be predicted. Therefore, the longer $H$ deviates from 0.5, the less noise in system or equivalently the mean-reverting (or anti-correlated) is the price time series, indicating a model's larger capability in making good forecasts. In other words, the closer the H value is to 0.00, the stronger is the tendency for the time series to revert to its long-term mean value.

As an efficient or perfectly competitive market is the market in which adequate numbers of buyers and sellers compete on equal terms and under full information, without any barriers or market power exerted by a single player that can significantly influence prices. The equilibrium or market-clearing price (MCP) is determined by matching supply and demand, i.e. suppliers sell gas only when they see that the price in the market exceeds the marginal cost of production, and buyers will buy gas if they benefit from the purchasing price. A very reliable indicator for assessing the efficiency of a market, based on the *EMH* (Fama, 1970; Fama, 1991), which is extensively used in the field of finance, is the ***Hurst Exponent (HE).*** The Hurst component takes values in the interval $0 \leq H \leq 1$ , and H= 0.5 for random walk time series, *H < 0.5 for anti-correlated (anti-persistent) or mean-reverting series, and H > 0.5 for positively correlated (persistent) series. In the case of market prices time series, if the Hurst exponent H > 0.5*, then it reveals that the price process is correlated or that a price increase in the past is more likely to be followed by an additional increase than a price decrease. Hence, daily price movements are persistent and subject to trends. Thus, the deviation from random walk additionally provides interesting information on the volatility and risk inherent in electricity market. An extensive application of Hurst exponent in the context of EMH has been implemented for several European electricity markets and described in the work of (Papaioannou et al., 2019), in which the reader is referred for a detail technical information and adequate literature review. The mean reversion behavior is one of the most characteristic stylized facts of the electricity and other energy commodities prices, thus these markets are in fact *incomplete* **(**due to real-world market frictions) since there may exist infinitely many risk-neutral probability measures (Oum Y., et al., 2006). Since ‘physical commodity’ electricity, as such, is not a tradable asset (like a stock), wholesale (spot) electricity markets are inherently incomplete, thus not efficient in the way the efficient market hypothesis requires (Vehvilainen, 2004; Karatzas and Shreve, 1998). Financial time series have been found to exhibit some universal characteristics that resemble the scaling laws, typical of natural systems, in which a very *large number of particles or units interacts.* Peters, E.E, (1996) has shown that HE, applied to a variety of ‘mature’ capital markets, has indicated that in these markets exist persistent memory, thus challenging the EMH.

### *5.1.2. Generalized Hurst Exponent (GHE)*

In our study we employ a ‘new version’ of HE, the **generalized Hurst exponent** (GHE), a generalization of the original method of Hurst (Hurst, H.E., 1951), a popular technique to study directly the scaling properties of our data via the qth-order moments of the distribution of the increments (Mandelbrot B.B, 1997; Barabasi A.L et al., 1991; Weron R., 2006)**.** The GHE is related to the long-term statistical dependence of a certain time series $X(t)$, with $t = (1,2, \dots, k, \dots, \Delta t)$, defined over a time-window $\Delta t$, with time-steps of one-unit. GHE quantifies the correlation persistence therefore there is a need for considering some fundamental statistical quantities, more specifically the *qth-order moments* of the distribution of the increments of the time series, defined as (Di Matteo T., 2007, Barabasi A.L et al., 1991)

$$K_q(\tau) = \frac{\langle |X(t+\tau) - X(t)|^q \rangle}{\langle |X(t)|^q \rangle} \qquad (5)$$

where $1 < \tau < \tau_{max}$ and and $\langle \cdot \rangle$ is the sample average over the time-window**.** We emphasize that for $q = 2$, $K_q(\tau)$ is analogous to the autocorrelation function: $C(t, \tau) = \langle X(t+\tau)X(t) \rangle$. From the scaling behavior of $K_q(\tau)$, the *generalized Hurst exponent* is then defined as follows:

$$K_q(\tau) \propto \tau^{qH(q)} \qquad (6)$$

Processes exhibiting this scaling behavior can be categorized into two groups: (a) processes having $H(q) = H$, i.e. independent of $q$, (they are uni-scaling or uni-fractal) and their behavior is uniquely determined by the constant $H$ (Hurst exponent or self-affine index)**,** (Di Matteo T., 2007**)**; (b) processes with varying $H(q)$, called multi-scaling **(or multi-fractal)** and each moment scales with a different exponent**.** (Di Matteo, T., (2007); Di Matteo, T. et al., 2005) have pointed out how financial data exhibit *multi-fractal* scaling behaviors. The GHE is calculated from an average over a set of values corresponding to different values of $\tau_{max}$ in equation (5) (Barabasi, A.L et al.,1991; Di Matteo, T., 2007). The analysis based on GHE approach is quite simple, since all the information about the scaling properties of a time series is contained in the scaling exponent $H(q)$**.** In our paper, we choose q =1 to compute the GHE, as we are not concerned about its multifractal feature. **Table 1** provides information on the time evolution of GHE, for all markets under study. The anti-persistent, mean reverting behavior, a deviation from the ‘reference’ value H=0.5 of the Brownian motion is clear, for both all series for every period.

### *5.2 Mutual Information and PMIME approach for detecting directed causality between TTF and electricity prices.*

Mutual information is a measure of the mutual dependence or the amount of information that one random variable contains about another. In mathematical terms, it is a measure of the reduction in entropy (uncertainty) of one random variable given the knowledge of another. In time series forecasting and causal analysis, mutual information can be used as a measure of association between two variables. If two variables are highly associated, it means that knowledge of one variable can be used to predict the other. This is useful in time series forecasting, where mutual information can be used to select the best predictor variables from a set of candidate predictors. In causal analysis, mutual information can be used to determine if there is a causal relationship between two variables. If there is a causal relationship, then changes in one variable should cause changes in the other. Mutual information can be used to quantify the strength of this relationship and determine the direction of causality. This is especially useful in cases where it is difficult to establish causality based on a simple linear relationship between two variables. In summary, mutual information is a useful statistic in both time series forecasting and causal analysis, as it measures the dependence and association between variables, and can provide insights into the direction and strength of the relationship. Several research papers have been exploring the strenghts and weaknesses of Mutual Information and its application in time series modelling and forecasting, like (Cover T.M., et al., 2012) ; McAllester , 1999; Kullback and Leibler, 1951; Kraskov et al., 2004; Palus M., 2001)
On a more formal mathematical context the Mutual Information is closely to the notions of entropy, joint entropy and conditional entropy. The entropy of a random variable X with probability distribution p(x) is given by:

$$H(X) = -\sum_z p(x) log p(x) \qquad (7)$$

while the joint entropy of two random variables X and Y with joint probability distribution p($x, y$) by:

$$H(X,Y) = -\sum_z p(x,y) log p(x,y) \qquad (8)$$

The conditional entropy of X given Y with conditional probability distribution p(x|y) is given by:

$$H(X|Y) = -\sum_{x,y} p(x,y) \log p\,(x|y) \qquad (9)$$

The ***mutual information*** of X and Y is given by:

$$I(X;Y) = \sum_{x,y} p(x,y) log \frac{p(x,y)}{p(x)p(y)} \qquad (10)$$

which can be expressed in terms of ***entropy*** as: $I(X;Y) = H(X) - H(X|Y)$ (11).

*Partial Mutual Information from Mixed Embedding (PMIME)* is a method for causality analysis. In (Kugiumtzis D., 2013) a new method for progressively building embedding vectors for multivariate analysis is introduced, addressing scaling, relevance, and redundancy issues in observed variables. The paper discusses the significance of Takens' embedding theorem in nonlinear time series analysis and proposes a progressive embedding scheme to capture system dynamics effectively. PMIME involves the progressive building of embedding vectors for different variables and delays to analyze information transfer in coupled systems, with considerations for redundancy, computational complexity, and scaling challenges.

**The Granger-oriented PMIME measure**

We describe here the main concepts of PMIME given in (Kugiumtzis D., 2013):

- The transfer entropy, TE quantifies c***ausality from a driving*** variable $X_1$ to a response variable $X_2$, of a two-dimensional (bivariate) time series $\{\chi_{1,t}, \chi_{2,t}\}$, $t=1,\dots,n$
- TE is defined from the conditional mutual information, as:

$$TE_{X_1 \to X_2} = I(x_{2,t+1}; \boldsymbol{x_{1,t}} | \boldsymbol{x_{2,t}}) \quad (12)$$

*Where the embedding vector* $\boldsymbol{x_{1,t}}$ contains the information of $X_1$ from present to the past, defined simply as the maximum delay L,

$$\boldsymbol{x_{1,t}} = [x_{1,t},\ x_{1,t-\tau}, \dots,\ x_{1,t-(L-1)\tau}] \qquad (13)$$

(for signals of discrete time, τ=1).

- The partial transfer entropy, PTE, extends the TE in the case that other observables variables are present $Z = [X_3, \dots, X_\kappa]$ , and measures the ***direct causality of*** $X_1$ to $X_2$

$$PTE_{X_1 \to X_2} = I(x_{2,t+1}; \boldsymbol{x_{1,t}} | \boldsymbol{x_{2,t}}, Z_t) \quad (14)$$

- PTE quantifies the present and past information in of $X_1$, *while explains the future value of* $X_2$ *which is not contained already in the present and past of each other variable* ($X_2$ and the rest κ-2 variables).
- Although conceptually PTE is suitable for measuring the ***direct causality***, in practice is not so useful due to the difficulty in estimating the CMI for large K or L or both.
- PMIME treats the dimensionality problem, due to using K embedding vectors, each one with L components (delay variables). PMIME creates gradually the ***mixed embedding vector***, $\boldsymbol{w}_1$ , that contains the most predictive delay variables, for the future value of $X_2$ ($x_{2,t+1}$), in respect of a small subset of the set of all KL delay variables.

- The $\boldsymbol{w}_1$ can contain variables of the driving X*1*, the response $X_2$ and the rest of the variables Z, i.e. of $w_t^{x_1}$, $w_t^{x_2}$, $w_t^{z}$ . The ***predictive information*** of $X_2$ exclusively from X*1,* is quantified by

$$\boldsymbol{I}(x_{2,t+1};\ w_t^{x_1} \mid w_t^{x_2}, w_t^{z}) \quad (15)$$

standardized by the mutual information $x_{2,t}$ and $\boldsymbol{w}_t$ .
Finally PMIME is defined as follows:

$$PMIME_{x_1 \to x_2} = \frac{\boldsymbol{I}(x_{2,t+1};\ w_t^{x_1} \mid w_t^{x_2}, w_t^{z})}{I(x_{2,t+1};\ \boldsymbol{w}_t)} \quad (16)$$

- PMIME (estimated for each directed pair of K variables) , provides the ***adjacency matrix*** , if positive values are set to one, and thus the corresponding ***Causality network of*** weighted or binary connections.

Application in investigating information flow across brain areas in scalp epileptic EEG records is also addressed (Papana A., et al., 2012). In (Papapetrou M., et al., 2022) a work very related to our present paper on market couplings is outlined. More specifically, the paper introduces the DPMIME method for causality analysis of discrete valued multivariate time series, comparing its performance with PMIME and applying it to study the causality network of capital markets pre and post-global financial crisis. Main findings include the application of DPMIME in analyzing the crisis's impact on financial market causality, comparison between parametric and resampling tests in DPMIME, and superior performance of DPMIME on *multivariate integer autoregressive* (MINAR) sequences over *mixture transition distribution model* (MTD) sequences. It introduces DPMIME for estimating direct causality in such time series, while the methodology involves developing and applying the DPMIME algorithm and investigating the *crisis's impact on financial market causality*. DPMIME successfully captured the *financial world market's causality network* before and after the crisis using DPMIME. Finally, the work most related to our work, is the paper of (Fotiadis A., et al., 2023), based on PMIME, in which the authors presented a novel scheme for detecting ***structural breaks***, through the occurrence or vanishing of nonlinear causal relationships in a complex system. The above scheme was applied to different records of financial indices regarding the global financial crisis of 2008, the two commodity crises of 2014 and 2020, the Brexit referendum of 2016, and the outbreak of COVID-19, ***accurately identifying the structural breaks at the identified times.***

### 5.3 Bayesian Estimator of Abrupt Change, Seasonal Change, and Trend (BEAST) approach.

We present here a short description of the BEAST (the Bayesian Ensemble Algorithm) approach, referring the readers to the paper (Zhao, K., et al., 2019) for details. Initially developed in the field of satellite time series recording and analysis, the method can be applied to any time series, as in climate temperature, biological system or even in socioeconomic, financial systems, however on the condition that certain assumptions must be satisfied, described in the paper above. Thus, the problem at hand and the nature of the involved time series influences crucially the possible interpretations.

It is well known from a typical theory in time series analysis (Brockwell and Davis, 2016; Hamilton,1994) that a given time series can be decomposed in four constituents : the **seasonal** one (modeled by a harmonic function), a **trend** or background component (modeled via piecewise linear regression), a possible number of **breakpoints** related to both seasonal and trend components), and finally an amount of **random noise.**

Let $G = \{y_i, t_i\}$ be a combination of a time series $y_i$, where $i = 1,2,\ldots n$, are $n$ time points at which data are recorded, then the afore mentioned (statistical) decomposition is expressed as a model $\tilde{y}(t)=f(t)$, which assumes that the time series is a composition of *seasonal* $S(.)$ and *trend* $T(.)$ components, breakpoints or abrupt changes, and noise. The model is written as:

$$y_i = S(t_i, \Theta_S) + T(t_i \Theta_T) + \varepsilon_i \qquad (17)$$

The noise $\varepsilon_i$ captures all the (related to the problem) variables (data) that are not explained by these time series, and is assumed to follow Gaussian distribution with variance $\sigma$. In this work he general linear models are adopted to parameterize signals $S(.)$ and $T(.)$, while the breakpoints (abrupt changes) in the time series are encapsulated in the *parameters* $\Theta_S$ and $\Theta_T$, respectively.

The seasonal signal *S(t)* is approximated by a piecewise harmonic model with respect to p knots, which divide the time series with starting time $\xi_0 = t_0$ and ending time $\xi_{p+1} = t_n$ into $p$ +1 intervals, such as $[\xi_0, \xi_1]$, $[\xi_1, \xi_2]$,…, $[\xi_p, \xi_{p+1}]$ . Thus, for each of the $p$+1 intervals, expressed as $[\xi_k, \xi_{k+1}]$ , $k = 0, \ldots, p$, the model $\tilde{y}(t)=f(t)$ is formulated as:

$$S(t) = \sum_{l=1}^{L_k} [a_{k,l} \sin\left(\frac{2\pi l t}{P}\right) + b_{k,l} \cos\left(\frac{2\pi l t}{P}\right)] \text{ , for } \xi_k \le t \le \xi_{k+1}, k = 0, \ldots p \qquad (18)$$

where $P$ expresses the period of the seasonal component, $L_k$ the harmonic order for the $k$-th segment, $a_{k,l}$ the parameter for the sine function, and $b_{k,l}$ the parameter for the cosine function.
The set of parameters $\Theta_S = \{p\} \cup \{\xi_k\}_{k-1,\ldots,p} \cup \{L_S\}_{k=0,\ldots,p} \cup \{k_{k,l}, b_{k,l}\}_{k=0,\ldots,p; l=1,\ldots,L_k}$ specify the curve of the *seasonal* harmonic component.

The trend signal $T$(t) is approximated by a piecewise linear function with respect to m knots, which divide the time series with starting time $\tau_0$=$t_0$ and ending time $\tau_{m+1}$=$t_n$ into $m$ +1 intervals, such as $[\tau_0, \tau_1]$, $[\tau_1, \tau_2]$,..., $[\tau_m, \tau_{m+1}]$ . Thus, for each of the m+1 intervals, expressed as $[t_k, t_{k+1}]$ , $k$ = 0, . . . , m, the *line segment* model is formulated as:

$$T(t) = a_j + b_j\,,\ \ \tau_j \ \leq t \leq \tau_{j+1},\ j = 0, \dots m \qquad (19)$$

where $a_j$ and$b_j$ are the coefficients. Therefore, the curve of the *linear trend* is written, in respect with two sets of parameters as:

$$\Theta_T = \{m\} \cup \left\{\tau_j\right\}_{j=1,\dots,m} \cup \ \{a_j, b_j\}_{j=0,\dots,m} \qquad (20)$$

The above two parameters $\Theta_S$ and $\Theta_T$ are now re-expressed in respect of two groups, $M$ and $\beta_M$, where $M$ refers to the structure of the model incorporating the number and timings of the seasonal and trend breakpoints as well as the seasonal harmonic order.

$$M = \{m\} \cup \left\{\tau_j\right\}_{j=1,\dots,m} \cup \ \{p\} \cup \ \{\xi_\kappa\}_{\kappa=1,\dots,p} \cup \{L\}_{\kappa=0,\dots,p} \qquad (21)$$

The exact shapes of the two components, T(.) and S(.) are determined via the group $\beta_M$ which comprises the segment specific coefficient parameters. Therefore, equation (17) is reformulated as

$$y(t_i) = x_M(t_i)\beta_M + \varepsilon_t \qquad (22)$$

where $x_M$ and $\beta_M$ constitute respectively the dependent variables and associated coefficients.
We now place all above within the Bayesian modeling framework thus for the time series $= \{y_i, t_i\}$ , $i$ = 1, 2, . . . , $n$, the main target is to find the posterior probability distribution

$$p(\beta_M, \sigma^2, M|G) \qquad (23)$$

We can use the Baye’s theorem, so the *posterior* is the product of the likelihood and a *prior* model:

$$p(\beta_M, \sigma^2, M|G) \propto (G|\beta_M, \sigma^2, M)\pi(\beta_M, \sigma^2, M) \qquad (24)$$

The Gauussian likelihood is written as $p(G|\beta_M, \sigma^2, M) = \prod_{i=1}^{n} N(y_i; x_M(t_i)\beta_M, \sigma^2)$ and the prior distribution is $\pi(\beta_M, \sigma^2, M) = \pi(\beta_M, \sigma^2|M)\pi(M)$. We consider, firstly, a normal-inverse Gamma distribution for the term $\pi(\beta_M, \sigma^2|M)$, and in order to 'capture' the vague knowledge of the seasonal harmonic order $\beta_M$, an extra parameter $v$ introduced. Furthermore, for the posterior $\pi(M)$, the breakpoints are considered as non-negative numbers, which is equally probably a prior, thus (24), the posterior model, becomes

$$p(\beta_M, \sigma^2, \nu, M|G) \propto \prod_{i=1}^{n} N(y_i; x_M(t_i)\beta_M, \sigma^2).\ \pi_\beta(\beta_M, \sigma^2, \nu|M).\pi(M) \qquad (25)$$

where the term $p(\beta_M, \sigma^2, \nu, M|G)$ in (19) captures all the essential information needed for interpreting the dynamic evolution of the time series. However, writing an analytical expression for this term is intractable, so for this purpose a Markov Chain Monte Carlo (MCMC) sampling is used, in order to generate a realization of random samples, for the posterior inference. As described in section 3.3 of the paper by (Zhao et al., 2019), the Monte Carlo-based inference is based on a hybrid sampler that embeds a reverse jump (RJ) MCMC sampling into a Gibbs sampling framework. A pictorial description of the BEAST model of (Zhao et al., 2019), is shown in figure 4.

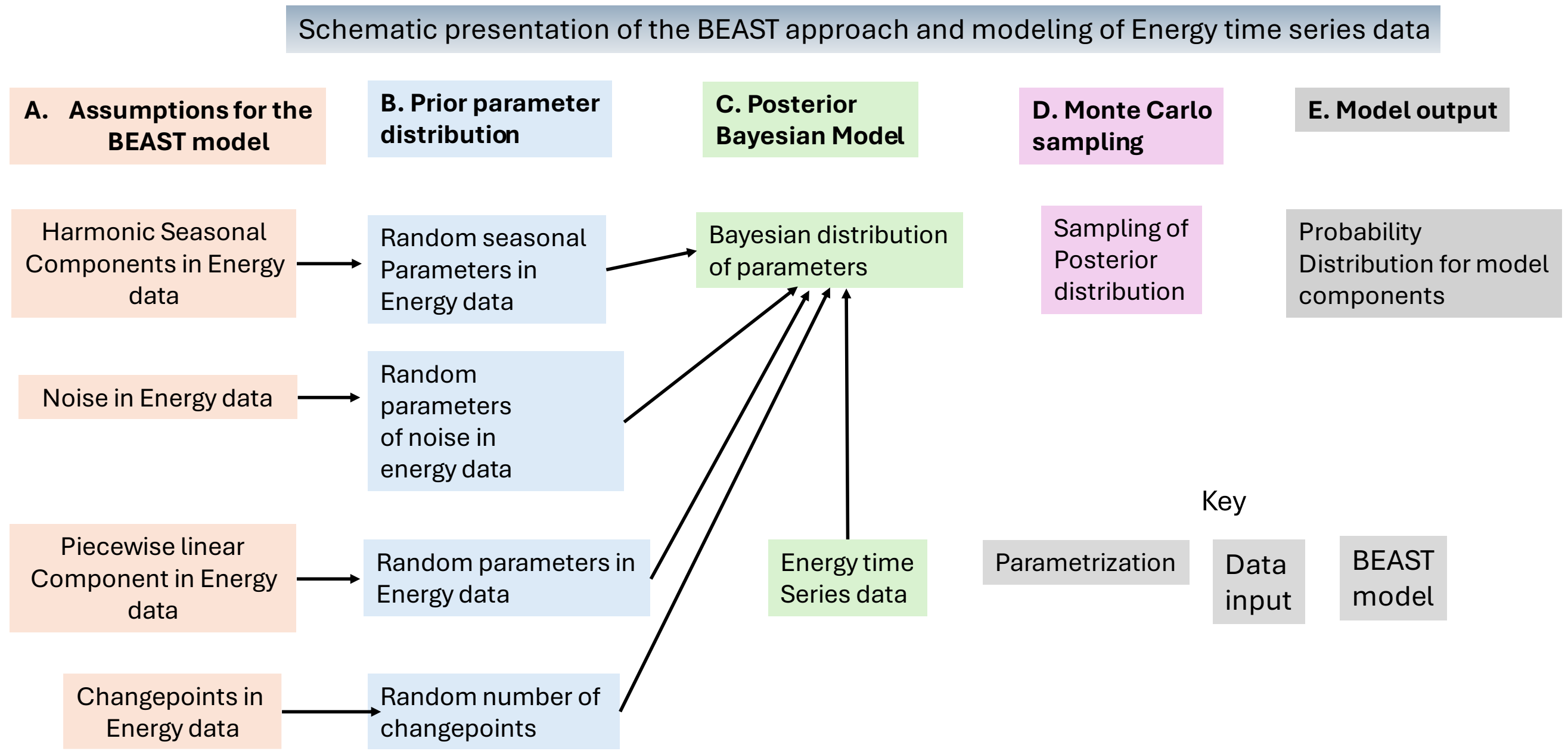

Figure 4: A schematic representation of the BEAST model, used in modeling the energy data in this work.

## 6. Empirical results

### 6.1 Hurst analysis and testing of EMH.

In this section, we assure quantitatively the above statement, by using the Hurst exponent as a tool for testing the ***EMH, efficient market hypothesis***, described in section 5.1. **Table 2a** shows the annual mean Hurst exponents, as well as the average value for the whole period 2019-2023, for each market. With bold we present the H values that are close to the H=0.5 value which corresponds to a perfectly competitive market, according to the EMH. The NGNMX and USD/RUB are the most efficient in 2021 and 2022, respectively. We observe that *all electricity markets* are *anti-persistent (mean reverting),* with H<0.5, in all separate years, as well as during the entire period 2018-2023, while both gas (except TTF in 2023) and USD/RUB are persistent (except in 2023). TTF, NGNMX and USD/RUB show $H \approx 0.5$ , i.e. are almost efficient in 2019, 2021, 2022 and the entire period, respectively (USD/RUB is very close to 'efficient' in 2018-2023 than any other 'market'). We also observe, from table 2a, that all electricity markets during 2022, exhibit the largest H values compared to H values in other years, and although they remain not efficient (H<0.5), *they tend to 'move' closer to the EMH limit during the crisis of 2022.*
**Table 2b** contains *specific dates* in the dynamics of rolling Hurst curves of the ten electricity market prices as compared with the dynamics of TTF gas price, that lie within period November 2021 to March 2023 which includes the most critical dates linked to the onset to Russo-Ukrainian war. These specific dates, shown as red vertical lines in figures 5 and 6 of the Greek and Spanish markets (similar figures are provided in **Supplementary material A,** for the rest of electricity markets)**,** set the limits of discrete dynamical behaviors of the markets mentioned.

**Table 2a:** Generalized Hurst Exponents (GHE) of markets as a measure for testing their efficiency deviations from EMH (H=0.5 indicates an ideal, perfectly competitive –efficient- market)

| Market | 2019 | 2020 | 2021 | 2022 | 2023 * | 2018-2023 |
|---|---|---|---|---|---|---|
| **RO** | 0.1843 | 0.1444 | 0.1566 | 0.2328 | 0.1460 | 0.2039 |
| **BE** | 0.1354 | 0.1263 | 0.1758 | 0.2781 | 0.1734 | 0.2371 |
| **CZ** | 0.1413 | 0.1087 | 0.1822 | 0.2833 | 0.0980 | 0.2251 |
| **DK1** | 0.1159 | 0.1175 | 0.1397 | 0.2692 | 0.1478 | 0.2222 |
| **ES** | 0.2298 | 0.1726 | 0.2807 | 0.2568 | 0.2423 | 0.2591 |
| **HU** | 0.1900 | 0.1494 | 0.1659 | 0.2811 | 0.0460 | 0.2294 |
| **NNL** | 0.1509 | 0.1289 | 0.1715 | 0.2795 | 0.1059 | 0.2418 |
| **IT** | 0.1589 | 0.1396 | 0.1901 | 0.3286 | 0.1707 | 0.2431 |
| **GR** | 0.1896 | 0.1539 | 0.1640 | 0.2111 | 0.1988 | 0.2046 |
| **BG** | 0.1438 | 0.1212 | 0.1284 | 0.2021 | 0.1456 | 0.1862 |
| **TTF** | **0.5125** | 0.6244 | 0.5367 | 0.5768 | 0.3212 | 0.5893 |
| **NGNMX** | 0.5708 | 0.5297 | **0.5068** | 0.5212 | 0.3554 | 0.5702 |
| **USD/RUB** | 0.6009 | 0.5473 | 0.5485 | **0.4960** | 0.3627 | **0.5179** |
| ***Note** : H values estimated with reservations, due to a small size of only 90 days (3 months) in 2023 | | | | | | |

**In figure 5** we show the rolling Hurst (window=75 days) of Greek electricity and TTF gas price vs. the reference random Gaussian process for which H=0.5 (red horizontal line). By selecting a window of 75 days (i.e. a 3-month period) we tried to approximate the target short-term maturities that professional traders will choose to trade short-term market shocks in the Futures markets. We observe an anti-persistent (mean reverting) dynamics, with an increasing trend, from H=0.127 in Dec.21 to H=0.470 (closer to EMH limit) in 26 March 2022. The date of minimum deviation from the EMH limit is therefore 26 March 2022, while 28 February, 5 March, 21 May and 10 November of 2022, are the dates corresponding to the closer approach of rolling H curve of TTF with those of GR electricity market.

**Table 2b**: Critical dates in the Rolling Hurst curves of electricity prices and TTF gas price

| Electricity Market | Mode of approach of Rolling Hurst curve to EMH limit (H-0.5) | Dates of min Deviation from H=0.5 and final H value | Dates of closer approach of TTF's rolling Hurst curve with those of electricity markets. |
|---|---|---|---|
| RO | Anti-persistence, from H=0.079 in 25 Dec.2021 to H=0.385 in 26 Feb. 2022 | 26 Feb.2022<br>H=0.385 | 26 February 2022 |
| BE | Anti-persistence, from H=0.18 in 11 Nov. 2021 to H=0.40 after Dec. 2021 | 15 January 2022<br>H=0.470 | 28 February 2022 |
| CZ | Anti-persistence, from H=0.105 in 20 Dec. 2021 to H=0.349 after 24 Feb. 2022 | 24 February 2022<br>H=0.349 | 24 February 2022 |
| DK1 | Anti-persistence, from H=0.08 in 26 Dec. 2021 to H=0.364 in 15 Mar. 2022 | 15 April 2022<br>H=0.417 | 11 June 2022 |
| ES | Anti-persistence, from H=0.246 in 12 Dec. 2021 to H=0.40 in 26 Feb. 2022 | 23 February 2022<br>H=0.489 | 25 February 2022<br>21 May 2022 |
| HU | Anti-persistence, from H=0.003 in 23 Oct. 2021 to H=0.417 in 26 Feb. 2022 | 26 February 2022<br>H=0.417<br>8 Feb. 2023<br>H=0.501 | 26 February 2022<br>10 November 2022<br>8 February 2023 |
| NNL | Anti-persistence, from H=0.06 in 20 Dec. 2021 to H=0.40 in 23 Feb. 2022 | 21 January 2021<br>H=0.455<br>23 February 2022<br>H=0.402 | 9 November 2022<br>23 February 2022<br>19 May 2022 |
| IT | Anti-persistence, from H=0.048 in 22 Oct. 2021 to H=0.499 in 15 Feb. 2022 | 15 February H=0.499 | 23 February 2022<br>19 May 2022 |
| GR | Anti-persistence, from H=0.127 in 24 Dec. 2021 to H=0.470 in 26 Mar. 2022 | 26 March 2022 H=0.470 | 28 February 2022<br>5 March 2022<br>21 May 2022<br>10 November 2022 |
| BG | Anti-persistence, from H=0.094 in Nov. 2021 to H=0.39 in 30 Mar. 2022 | 30 March 2022 H=0.39 | 21 May 2022<br>9 November 2022<br>4 March 2022 |

**Table 2b** contains the above information as well. As a conclusion, we observe that the rolling Hurst of both TTF and Greek prices converge towards the EMH limit of H=0.5, during the period of escalation of Russo-Ukrainian war, i.e. from Nov.2021 -Mar.2022, with the smaller deviation to occur on 26 March 2022, about one month later from the Russian invasion. Instead, TTF market exhibits extensive periods with rolling H>0.5 (persistent) and only short periods with H below 0.5 (anti-persistent) (start of May 2022 to mid-June 2022, and mid Oct- 2022 to end November). Thus, also TTF market is not efficient. **Figure 6** shows the results of the **Spain's electricity market**. We observe an anti-persistent (mean reverting) dynamics, with an increasing trend, from H=0.246 on 12 Dec.21 to H=0.40 (closer to EMH limit) on 26 Feb. 2022, i.e. 2 days after the Russian invasion. Then, Spain's Hurst meets TTF's Hurst for a short period, and finally remains very close to H=0.5 for the entire March 2022. ***As a conclusion, we observe that within the period of Russo-Ukrainian war escalation, the rolling Hurst of electricity prices tend to the Hurst exponent of an efficient market and meets the TTF's Hurst value. This finding is observed for most electricity markets.***

**Figures 7** and **8** show the results of rolling Hurst for the USD/RUB exchange rate and NGNMX vs. TTF price, shown also in tables 2c and 2d containing the dates of critical points and corresponding H values.

**Table 2c:** Dates of critical points (CP) and H values in the dynamics of rolling Hurst curves of USD/RUB rates and TTF gas price, shown in figure 7.

| CP | Date | H |
|---|---|---|
| CP1 | 21 November 2021 | 0.526 |
| CP2 | 01 January 2022 | 0.510 |
| CP3 | 25 February 2022 | 0.499 |
| CP4 | 15 May 2022 | 0.499 |
| | 27 May 2022 | 0.255 |
| | 30 May 2022 | 0.443 |
| | 04 Jun 2022 | 0.482 |
| | 08 Jun 2022 | 0.488 |
| CP5 | 16 June 2022 | 0.499 |
| | 08 August 2022 | 0.234 |
| CP6 | 26 October 2022 | 0.499 |
| | 01 Sept. 2022 | 0.400 |
| CP7 | 7 December 2022 | 0.503 |
| CP8 | 24 February 2023 | 0.490 |

**Table 2d:** Dates of critical points (CP) and H values in the dynamics of rolling Hurst curves of NGNMX and TTF gas prices, shown in figure 8.

| CP | Date | H |
|---|---|---|
| CP1 | 02 November 2021 | 0.461 |
| CP2 | 13 March 2022 | 0.465 |
| CP3 | 16 May 2022 | 0.456 |
| CP4 | 19 Jun 2022 | 0.486 |
| CP5 | 24 October 2022 | 0.447 |
| CP6 | 19 December 2022 | 0.486 |
| CP7 | 27 February 2022 | 0.408 |

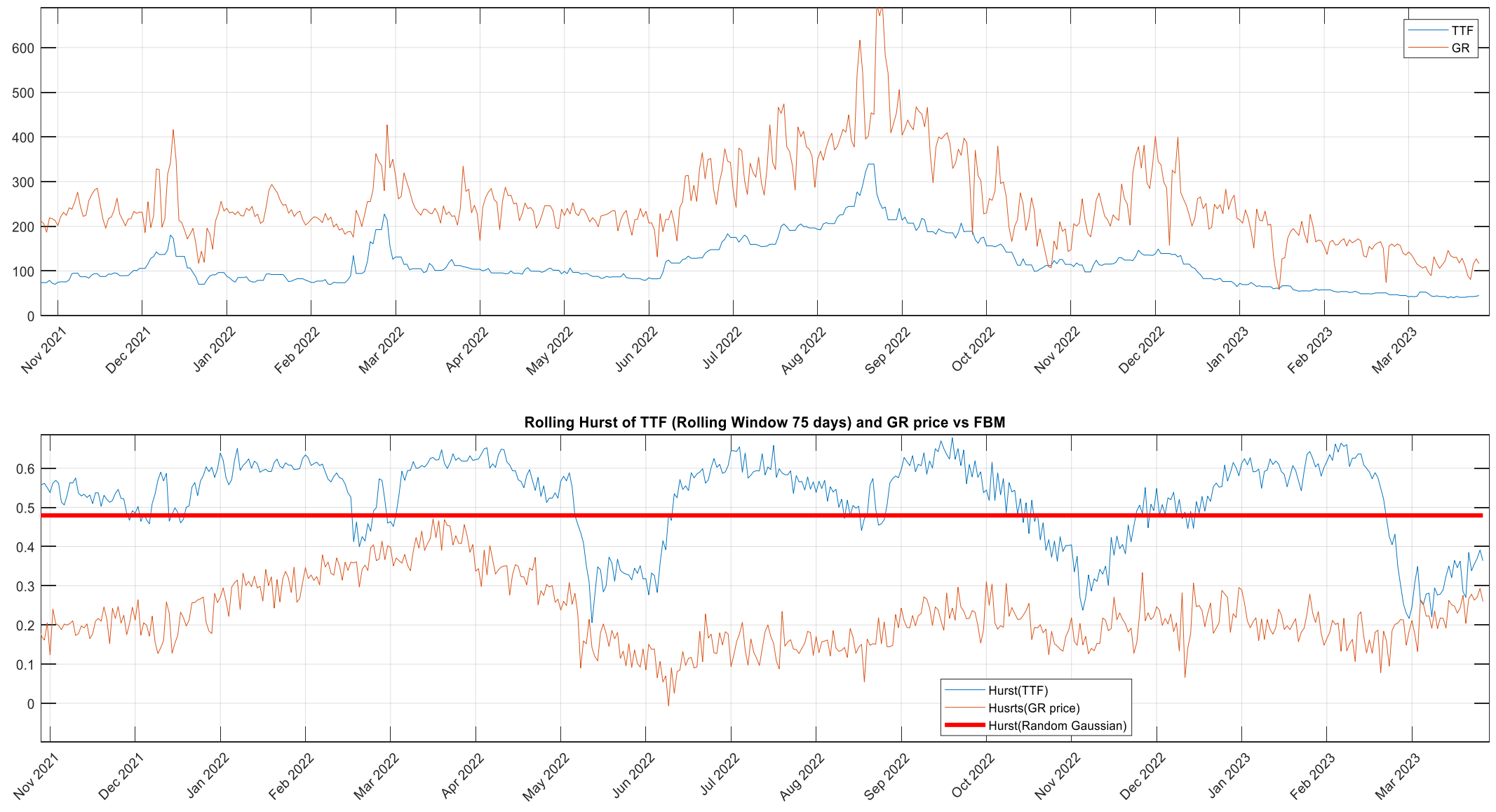

**Figure 5:** Rolling Hurst (window=75 days) of **Greek electricity** market and TTF gas price vs. the reference random Gaussian process for which H=0.5. The period of escalation of Russo-Ukrainian war is from Nov.2021 -Mar.2022.

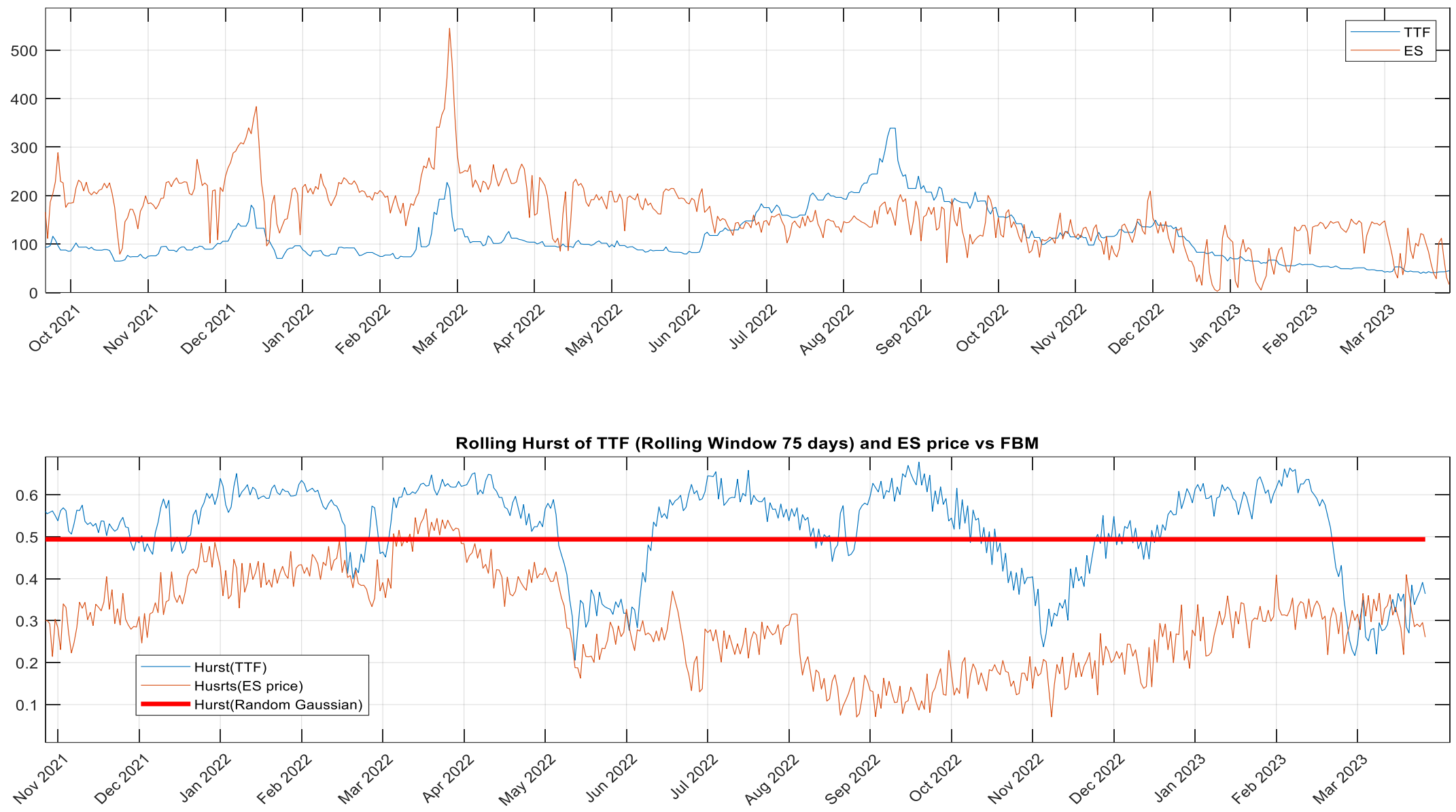

**Figure 6:** a) TTF and Spanish (ES) price time series, b) rolling Hurst (window=75 days) of Spanish electricity and TTF gas price vs. the reference random Gaussian process for which H=0.5.

**Figure 7** shows the dynamics of the ***rolling Hurst for USD/RUB rate and TTF price,*** as it develops in time windows between the critical points of table 2c. In the CP1-CP2 time window both markets are most of the time persistent and inefficient, and only a little before and after Dec.2021 TTF is closer to EMH limit and mean-reverting. Things are drastically different in the CP2-CP3 period where TTF is persistent, and very inefficient up to mid-February 2022, *and after this* ***moves quickly towards the EMH limit and meets the USD/RUB's Hurst on 25 Feb.22 (one day after Russian invasion). Just after invasion, the USD/RUB market becomes more inefficient and persistent, while on the opposite, the TTF market remains closer to EMH limit, during the entire period until E4 (15 May 22).***

Our finding is compared with the finding of the papers by (Sari E.L., et al., 2023), (see section 3.2.1 for a short description of their main results). Our work support the results of the above paper, in respect to the response of the financial market IDX in their paper, with the response of our financial market (USD/RUB exchange rate), since the rolling Hurst of USD/RUB price (see fig.7 and **table 2c**) increases also significantly from H=0.443 on May 30, one day before the announcement of imposing an oil import embargo on Russia, to H=0.499 (almost at the EMH limit), on 16 June 2022. An increase is also observed from H=0.234 (08 August 2022), i.e. one week after 1$^{st}$ August 2022 of the announcement of the first wheat export ship's departure from Odesa port. Thus, the two financial markets (IDX and USD/RUB) reacted similarly to the same type of new information. We also observe that in the same periods *the TTF gas market tends to be more efficient than the USD/RUB financial market.* During the CP4-CP5 period (from 15 May to 16 June 2022), both markets become mean-reverting. In CP5-CP6 time window, TTF market is persistent most of the time (except in mid-August 2022 where alternates above and down and closer to EMH limit), while USD/RUB market is mean-reverting and more efficient than the TTF market, with a peak closer to EMH limit in mid-August 2022, at the same time where a drought occurred by TTF. Both markets are mean-reverting and inefficient in CP6-CP7 period (with TTF more inefficient, on average), and finally in CP7 (7 Dec.22)-CP8(24 Feb.23) period, TTF is persistent and more inefficient than USD/RUB, both markets become more efficient (they come closer) at mid-Dec.22, and , USD/RUB is almost efficient (H=0.49) during Feb.23, while TTF market becomes drastically mean-reverting and inefficient just after CP8 (24 Feb.23) and USD/RUB is alternating above and down EMH limit, and on average more efficient.

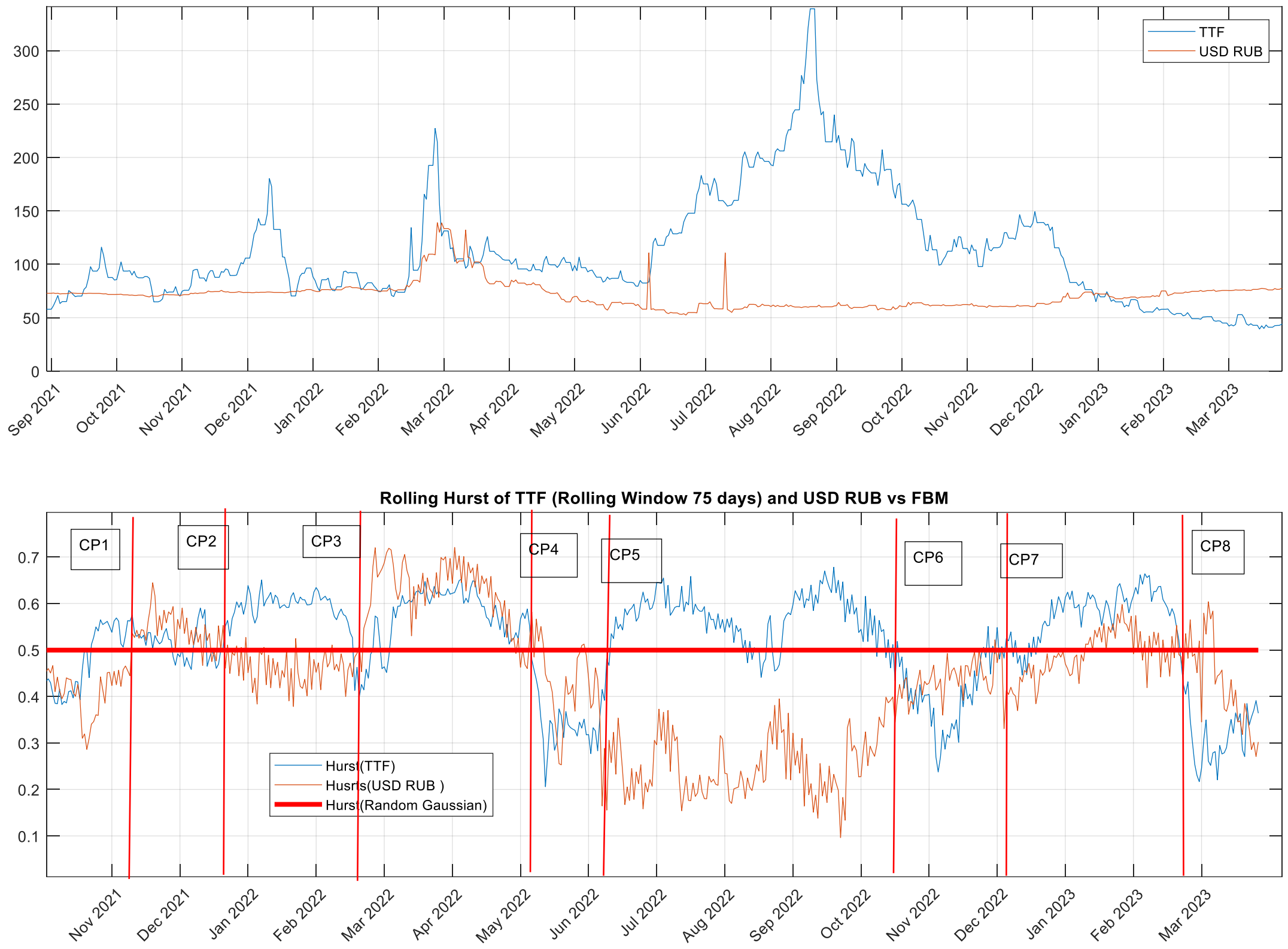

**Figure 7**: a) TTF and USD/RUB price time series, b) rolling Hurst (window=75 days) of USD/RUB and TTF gas price vs. the reference random Gaussian process for which H=0.5.

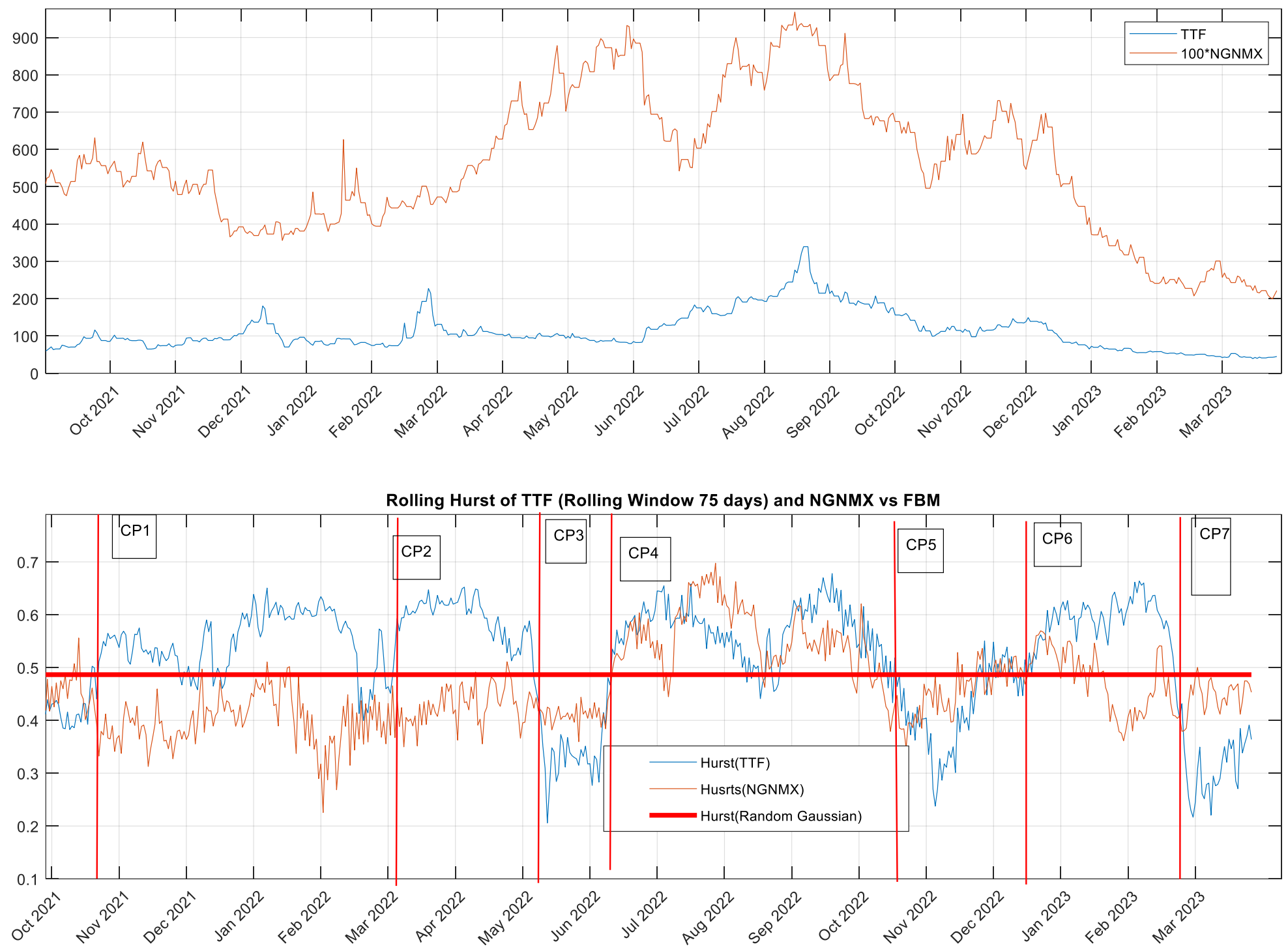

**Figure 8**: a) TTF and NGNMX price time series, b) rolling Hurst (window=75 days) of GNMX and TTF gas price vs. the reference random Gaussian process for which H=0.5.

For the case of NGNMX, in **figure 8** we observe that in the time window CP1 (2 Nov.22, H=0.461) to CP2(13 Mar.22, H=0.465), on the average, the curve of the rolling Hurst of TTF is above 0.5, indicating a persistent market, while that of NGNMX is anti-persistent, i.e. both markets are inefficient. *However, in mid-January 2022, NGNMX's Hurst approaches 0.5, i.e. it becomes more efficient, a finding that supports the results of (Aslam F., et al., 2022), that in the onset of war the gas market tends to be more efficient than before. On 24 Feb.22 (Russian invasion), the two energy markets come closer (co-evolve) and both tend towards the EMH limit*. A similar behavior is exhibited by the two markets in window CP2-CP3, while in the CP3-CP4 period exhibit a mean-reverting dynamics, deviated further from the EMH limit. Both markets become persistent in CP4-CP5 time window, mean-reverting in CP5-CP6 but still inefficient until end of Nov.22, after which their dynamics alternates about the EMH limit, i.e. they become more efficient.  Finally, in CP6-CP7 period, TTF's Hurst value is persistent, inefficient, while NGNMX alternates from persistent to mean-reverting.

***As a conclusion, during the crisis (from the onset, during and after) the energy markets (TTF and NGNMX) become systematically more efficient.*** *This result also suggests* ***that investors in these markets are likely to show herding, especially on the onset and just after the invasion***.

As opposed to traditional financial markets where the dynamics of the Hurst exponent usually oscillates at equal paces between above-and-below 0.5 levels (example being the USD/RUB Hurst levels displayed in **Figure 7** ), the corresponding dynamics displayed in the TTF gas market in Figure 8 and similar Figures of the **Supplementary material A,** for the analyzed ***electricity markets***, expressed by Hurst exponent, is staying below the 0.5 threshold for long-lasting periods, indicates strong mean-reverting movements during most of the time examined in our analysis. However, this strong mean-reversion is not necessarily associated with "predictable" market patterns under the context of real trading applications and should be more cautiously examined and commented on. More specifically, trading the specific asset class, as already stated, is highly affected by market frictions like high transaction costs, slippages, regulatory and business barriers-to-entry, as well as liquidity shocks, which might alter to a great extend the trading performance results obtained in the real world, as compared to a 'paper/demo' trading environment based on purely data-based simulations. Even so, in all markets, even under a data-only based analysis *we can observe an interesting "dragging-upwards" move towards the 0.5 threshold territory during specific time periods. Since the 0.5 threshold is associated with the efficient market assumption (EMH), we will refer to this **"pulling" mechanism as "pull-to-efficiency"** for the several markets*. Such periods include the start of 2020, which is associated with the COVID-19 pandemic, and also start of 2021 and 2022 years, as well, both of which are associated with periods of intense talks and actions around post-pandemic inflationary pressures around the commodities markets (energy prices soared to historic levels) at first, and subsequently on the possible *diffusion of the inflation effects* to the rest of the financial markets, as well. Diving into a bit more detailed comparative analysis between the several markets Hurst dynamics, we can observe that in some markets this "pull-to-efficiency" mechanism as previously discussed is higher for some markets relative to others, meaning the relative move from the low/average of the rolling Hurst to the 0.5 threshold is greater. An example list includes e.g. Spain versus Greece, or Denmark versus Italy, or even Spain and Denmark versus Hungary etc. As a further comment to this, we speculate that the effects of any macro shocks arrived at 'different paces' to each country and to the respective market, potentially identifiable by the causality measuring methodologies outlined in the rest of the sections below.

The above results, that all electricity markets are inefficient during the entire period of analysis and specifically the finding that the *trend of the rolling Hurst curve* of most energy markets analyzed here, when *approaches a breakpoint*, tends towards the efficiency limit H=0.5, *agree with the results of the work of (Kaharan, C.C., et al., 2024) (the breakpoints in their paper correspond to the dates of the coupling-decoupling events of the analyzed markets with other markets).* They found that none of the market prices are generated by entirely efficient processes, and instead, the processes oscillate between mean-reverting and persistent behaviors in most markets, and so this inefficiency may provide opportunities for profitable trading strategies for market agents. They also indicated that the day-ahead prices are more volatile (with strong long-term memory features) than the futures prices. In fact, we have also found that the TTF's *futures* prices are more volatile than the *spot* prices of electricity markets.

A very useful and interesting subsequent result, emerging from the Hurst analysis of the individual gas and electricity markets considered in this work is ***the detection of similarities of the dynamic evolution of the efficiency of the markets***. **Figure 9** depicts the *correlation matrix of the rolling Hurst exponent time series of all pairs of the given markets.* We observe that some markets show very similar i.e. highly correlated (correlations > 70%) efficiency evolutions: RO-HU 92%, BE-NNL 89%, BE-CZ 84%, HU-BG 80%. An interesting also result is that no similarities exist in the development of their market efficiency of IT, GR and ES with all other markets and finally no significant correlation exist between the evolution of TTF's market efficiency and all other market efficiencies.

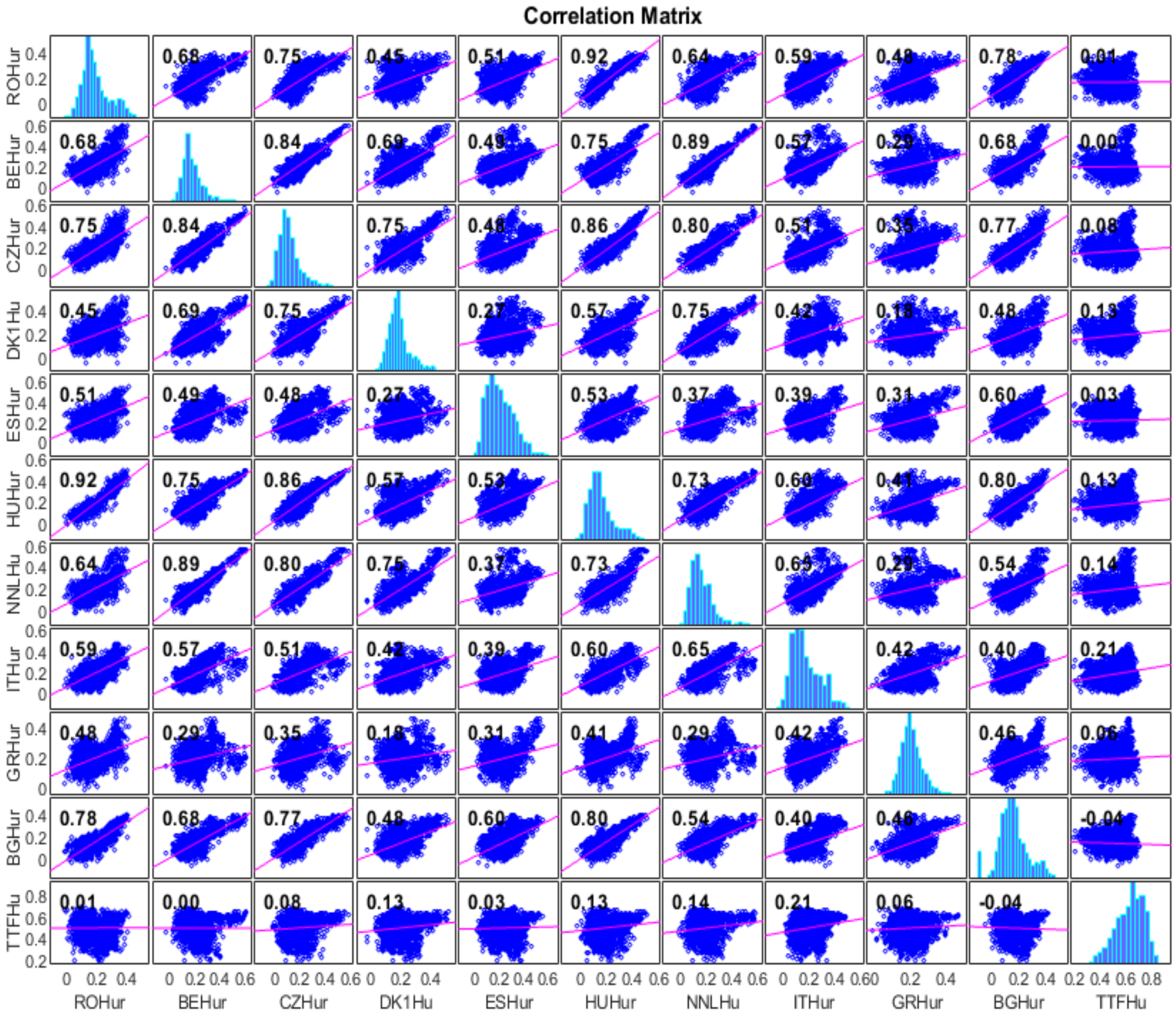

Figure 9: Correlation matrix of rolling Hurst exponents (window of 75 days) of all energy markets. A large correlation between two markets means that their efficiency follows a similar dynamic pattern for the period examined.

### *6.2. Mutual Information analysis in rolling window for European electricity prices against TTF, NGNMX prices and USD/RUB exchange rate.*

In **Figure 10** we display the 60 days rolling window calculated Mutual Information (MI) values of European electricity prices against the USD/RUB exchange rate, spanning the period Nov 2021 to May 2022, while in figures 20 and 21, we plot the rolling MI of Greek and TTF prices against USD/RUB. Based on the graphs, we comment that the MI(GR, USDRUB) and MI(TTF, USDRUB) values are quite greater in level than the rest of the pairs, and also that they have been experiencing a relatively similar dynamics over time (even matched exactly the same level of MI as well). This converged behavior seems to be" broken" on February 04, 2022, where a deviation between the two paths is observed and which seems to be maintained until the end of the dataset in focus (May 2022). This behavior change could be connected to the Russian-Ukrainian War and its effects on global trading of Natural Gas and the subsequent effects on rubbles. There seems to be a time gap of almost three or four weeks prior to the official announcement of the invasion and the MI behavior change which could potentially be explained as a market's discounting mechanism that the invasion probability was getting higher and higher as time went by starting from the start of the same month. The date of the decoupling event is consistent with the date found by the BEAST. More specifically the decoupling took place within the period 2 January-14 February 2022. Critical event E5 also happened on 8 January 2022, i.e. within this period TTF price exhibited an earlier reaction, shown by the existence of changepoint on 4$^{th}$ January 2022, thus, the gas market reacted as an almost 'efficient market', since during this period the rolling Hurst exponent of TTF approached very close the value of H=0.5 value (see table 2, and figure 17). In other words, TTF market showed a strong discounting mechanism.

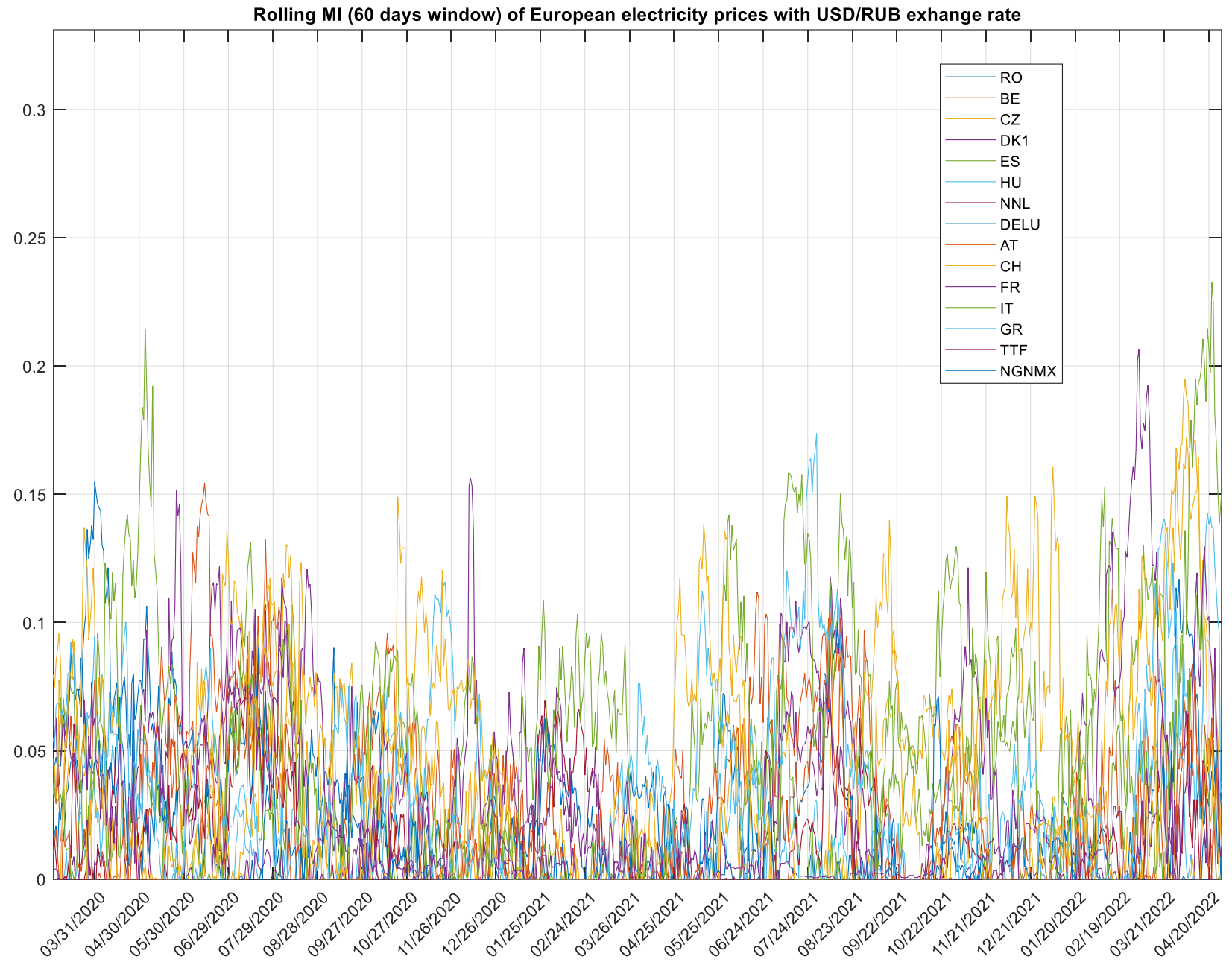

Figure 10: 60 days rolling window calculated Mutual Information (MI) values of all European electricity prices against the USD/RUB exchange rate, spanning the period Nov 2021 to May 2022.

In figure 12 we show a zoom of figure 11, focusing more closely on the period just before and after the Russian invasion of 22 February 2022. While the two curves of MI (GR against YSD/RUB and TTF against USD/RUB) are shown to have evolved almost together (a dynamic coevolution) over a large time span up to the point where the arrow is located on the graph, a ***clear decoupling starts to take place at about 10 February 2022***, becoming larger and larger over this month, peaking at 22 February and finally decreasing afterwards, emphasizing the fact that the decoupling have started to happened days before the Russian invasion, indicating that a considerable amount of information had reached in the markets causing them to respond as early as possible. The detected date of decoupling is inside the period of the onset the Russo-Ukraine conflict, a finding that agrees with the results of the work of (Lyocsa and Phihal, 2022), mentioned in 3.2.1, therefore our work ***contributes*** *to the emerging* *literature in detecting the 'exact' period of the onset of the conflict and therefore the onset of its impacts on the European energy markets.*

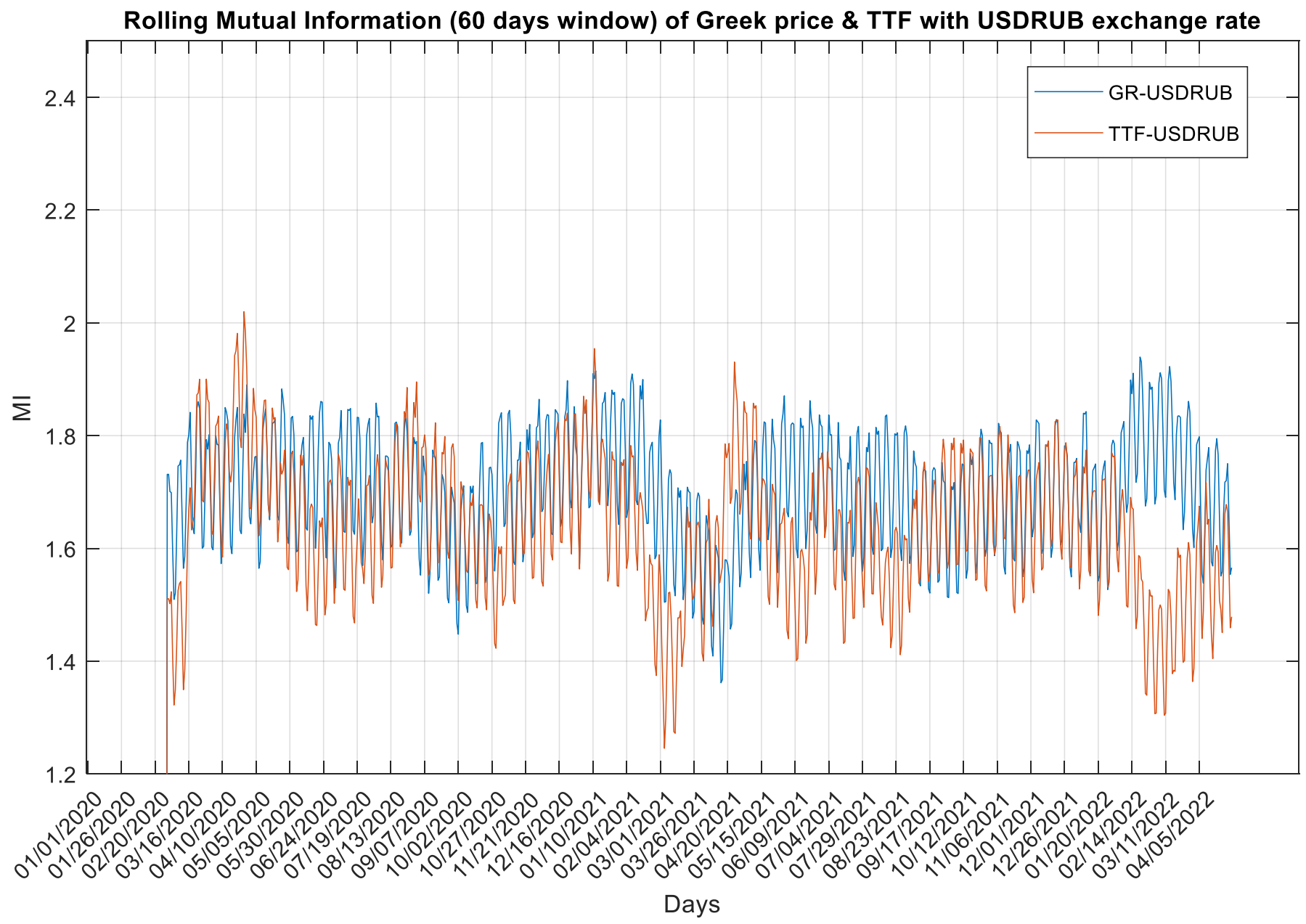

Figure 11 Rolling Mutual Information (window 60 days), of Greek price-USD/RUB and TTF- USD/RUB

The decoupling of mutual information (MI) of the Greek price and the USD/RUB exchange rate, GR-USD/RUB, is also supported by the network graph of fig. 23. The exchange rate is strongly correlated with TTF price, which in turn affects the GR price.

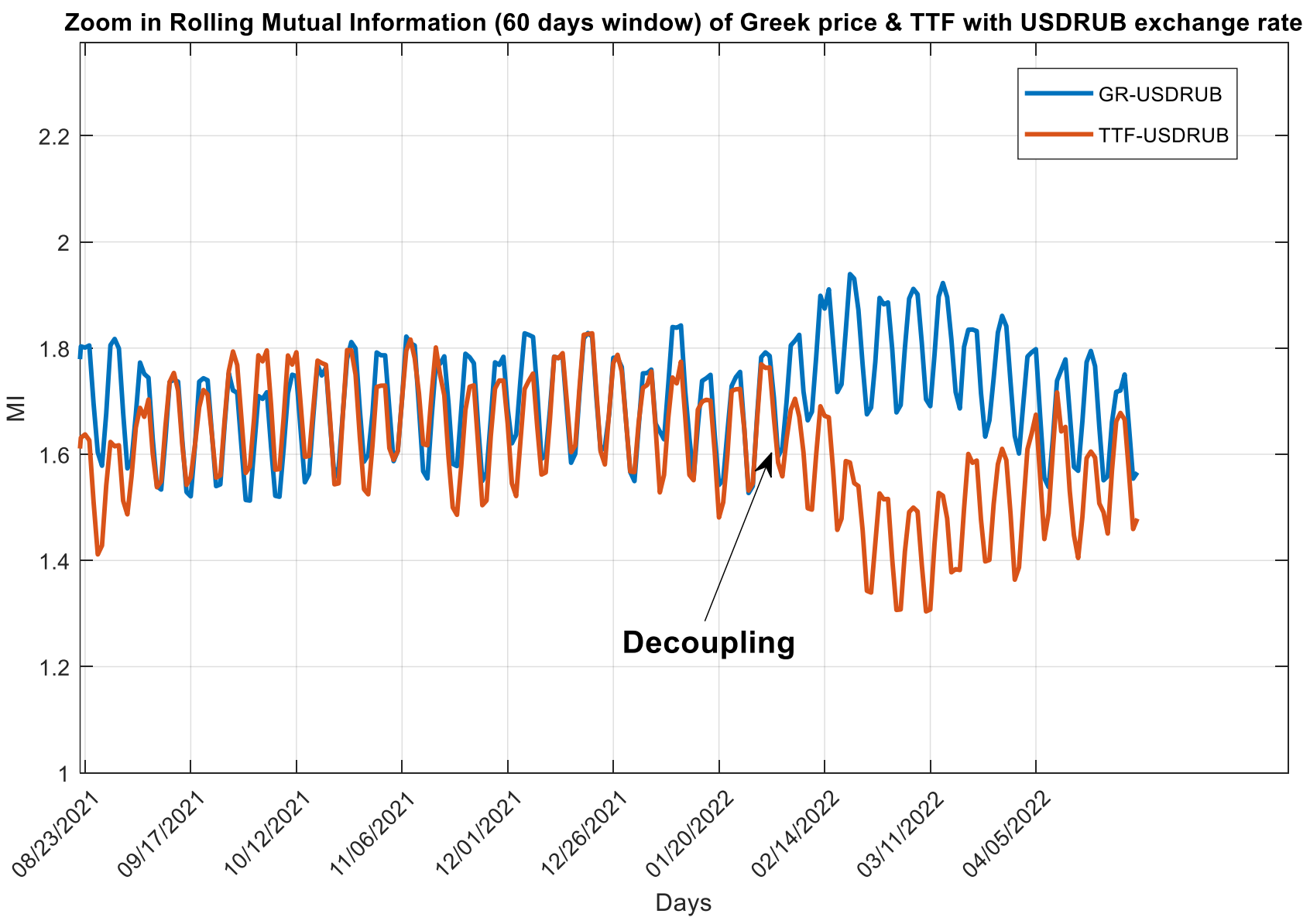

Figure 12: A Zoom in figure 11 Rolling Mutual Information (window 60 days), of Greek price-USD/RUB and TTF- USD/RUB. The date of the decoupling, around 10th February 2022, is now clear.

### *6.3. Results of PMIME modelling*

#### 6.3.1 PMIME results of Greek, TTF NGNMX price and USD/RUB exchange rate time series.

**Figure 13a** shows the *heat map* of the information flow, as derived from the PMIME 'tool', among the three variables that are assumed to affect the dynamic evolution of the Greek market price. We observe that very strong information flow exists between GR price and TTF and NGNMX gas prices (with 0.944 and 0.766 partial mutual information, respectively), with a direction as expected (shown in **figure 13b**) from TTF and NGNMX towards GR, with the first direction to be strongest (heavier arrow line) than the second one, i.e. the dynamics of the Greek price is more affected by dynamic changes od TTF. However, and this very interesting, Greek market is affected by NGNMX because there is a dynamic mutual interaction between TTF and NGNMX, as shown by the two arrows connecting them, in the network graph. Heat maps and Network graphs of partial mutual information are provided for all markets analyzed in this work, in Supplementary material. For comparison, we present also the network graph of PMIME results for the Czech market, in **figure 22c**. In this case the impact of TTF and NGMNX on CZ price is not strong (as in the case of Greece), shown by the two slim arrow lines. *Therefore, we observe that the two gas prices have different effects on the GR and CZ electricity prices, as the network graph and the heat maps (not shown for the CZ market) evidently indicate.* Similar network graphs and heat maps are provided for all markets in the Supplementary material.

From the network graph, an interesting finding is evident, in connection with the rolling MI and mainly with the decoupling of USD/RUB with TTF and GR price. The TTF-USD/RUB interaction shown in the network, and the impact of TTF on GR price, supports further the decoupling shown in figures 11-12. In conclusion, rolling MI, PMIME and rolling Hurst exponent results, and the results from BEAST analysis, described below, are all consistent. *We also observe that the gas prices TTF and NGNMX are strongly partially correlated with USD/RUB exchange rate, with information flowing from gas markets (fat arrow lines) to USD/RUB. After Russia's invasion of Ukraine, the USD to RUB value lost significant ground, reaching a low of 135 rubles in March 2022, as we saw previously in figure 12.*

In section 3.2.1 we mentioned the work of (Lyocsa S. and Plihal T, 2022), which is related to our work, in respect to the effects of Russian invasion on USD/RUB variations and more specifically, to the connection of one of their main findings to our finding given here: the *ruble-related attention* was found to be positive and statistically significant with a considerable effect for the USD/RUB exchange rate, the general market attention was found to be relevant, positive, and significant for the USD/RUB exchange rate, ***and finally and more important, the price fluctuations appear to be higher for the EUR/RUB exchange rate after the observation of greater attention levels toward the variables related to: a) the economic sanctions imposed to Russia, b) the removal of Russian Banks from the SWIFT interbank, c) Russia's asset freeze, d) the disruption Nord Stream2 gas pipeline, and e) export controls (gas included), of the Russian economy during the crisis.*** The *removal of selected Russian banks from the SWIFT interbank system, and the prohibition of the Central Bank of Russia from access to foreign exchange reserves, is an effect of the invasion.* Therefore, we may logically assume that the drastic increase in TTF and NGNMX prices due to invasion, might be the driving factors of the USD/RUB dynamics, and this is the case as it is shown in figures 11 and 12. In conclusion, our result, that the TTF and NGNMX gas prices are mutually interacting with USD/RUB, is consistent with / and further enhance the results of (Lyocsa S. and Plihal T, 2022), explaining adequately the ‘causality’, i.e. the directed information flow from the two gas markets to USD/RUB financial market.

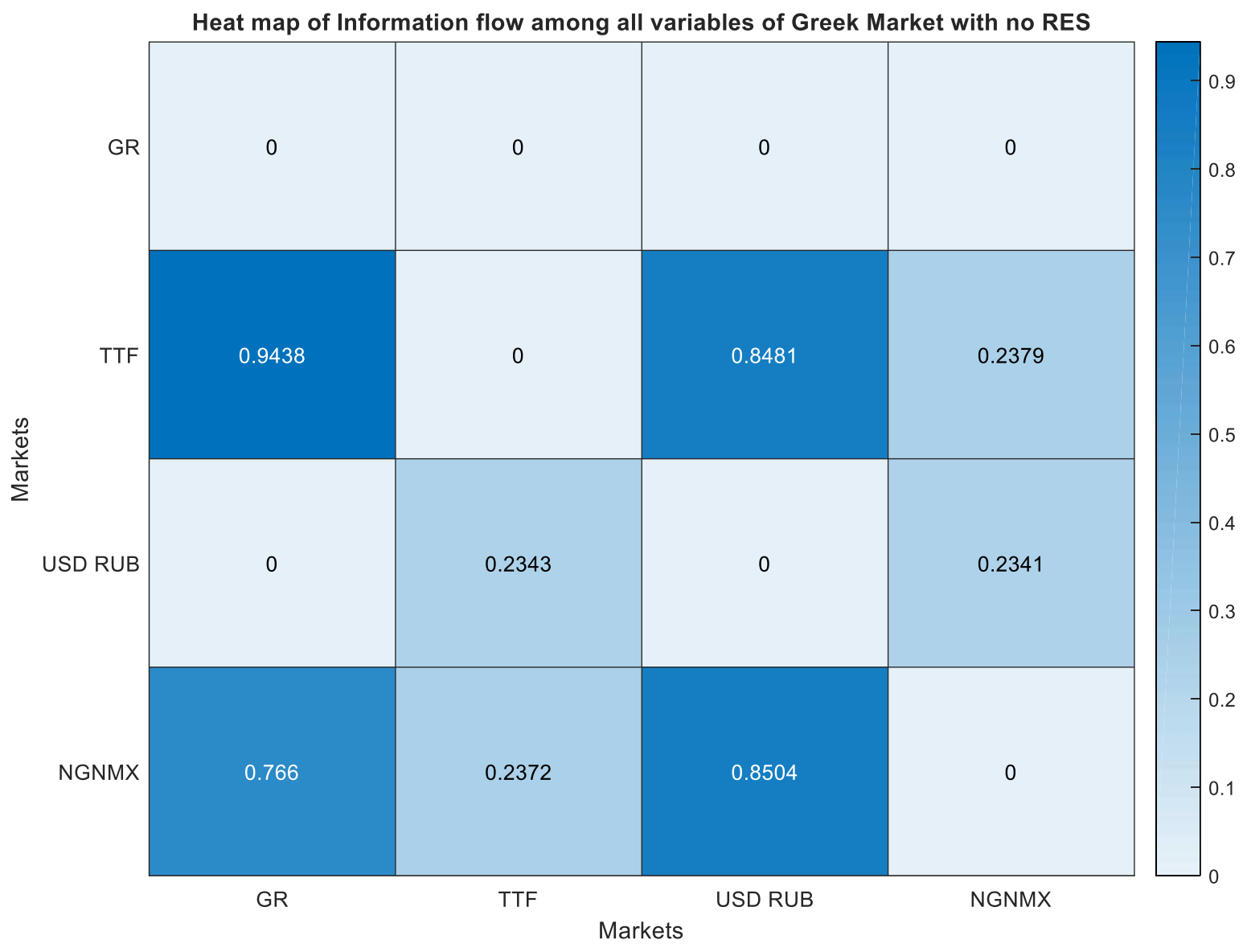

Fig. 13a: Heat map of information flow, derived by PMIME approach, among the Greek price, with TTF and NGNMX prices and USD/RUB rate (log price returns). Cells (areas) of the map with more intense blue color, indicate strongest partial mutual information flow.

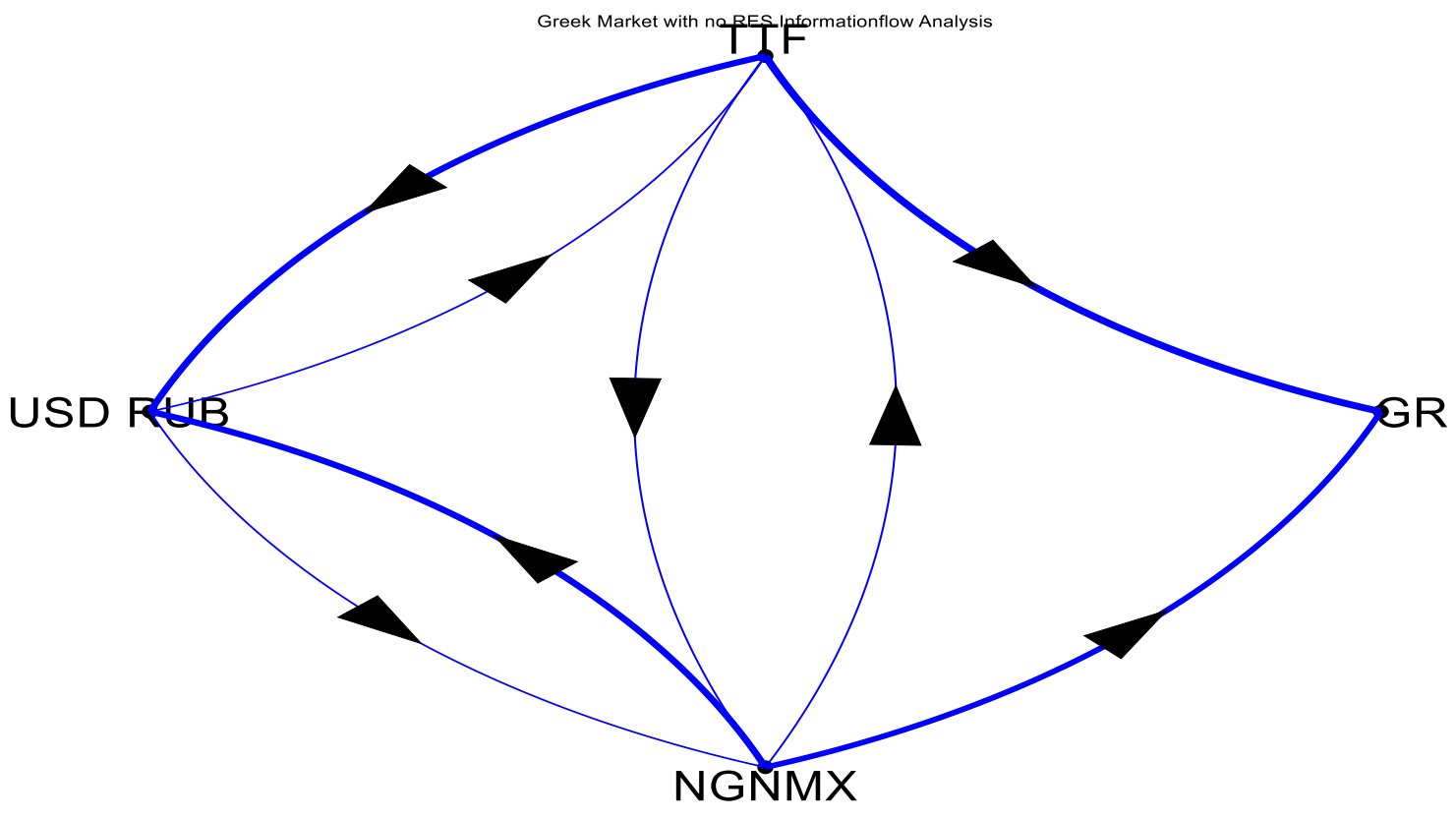

**Fig. 13b**: Network graph of information flow, derived by PMIME approach, of the Greek price, with TTF and NGNMX prices and USD/RUB rate (signed log price returns). Heavy arrow lines, connecting two nodes, indicate the strongest partial mutual information flow. Greek price is strongly affected by TTF and NGNMX gas prices (fat arrow lines), which also mutually interact to each other.

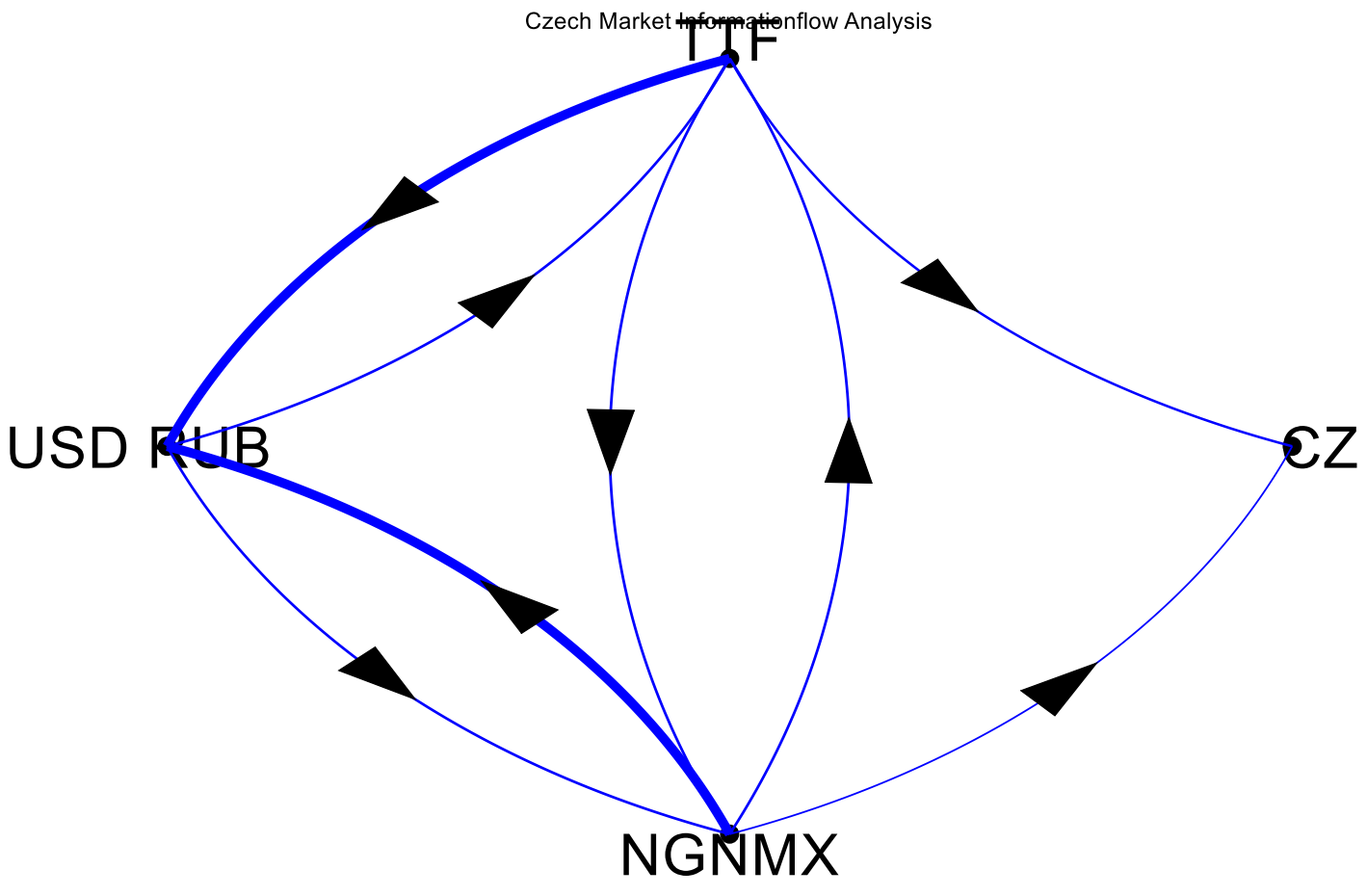

**Fig. 13c**: Network graph of information flow, derived by PMIME approach, of the Czech price, with TTF and NGNMX prices and USD/RUB rate (signed log price returns). Heavy arrow lines, connecting two nodes, indicate the strongest partial mutual information flow. Czech price is *moderately* affected by TTF and NGNMX gas prices (slim arrow lines), which also mutually interact to each other.

Table 3 summarizes the results of PMIME analysis. *The Greek and Italian electricity markets are shown to be the most strongly affected markets from variations in TTF and NGNMX prices.* Only Spain is shown to be weakly affected by the two gas markets.

Table 3: PMIME values between TTF-NGNMX gas and all electricity prices

| **Market** | **TTF** | **NGNMX** |
|---|---|---|
| **RO** | 0.220 | 0.143 |
| **BE** | 0.294 | 0.296 |
| **CZ** | 0.221 | 0.156 |
| **DK1** | - | 0.284 |
| **ES** | 0.321 | 0.367 |
| **HU** | 0.123 | 0.095 |
| **NNL** | 0.235 | 0.286 |
| **IT** | **0.751** | **0.632** |
| **GR** | **0.944** | **0.766** |
| **BG** | 0.180 | 0.147 |

The above results, comply with and supports the work of *(*Manelli A., *et al., 2024),* as described in section 3.2.1, in respect of the mutual or bidirected interaction between a financial and a commodity market. They found that the during the Russia-Ukraine conflict, the *Eurostoxx50 market affected the TTF market and vice versa*, a result very similar to ours, i.e. the mutual interaction between USD/Rub and TTF. Therefore, our paper contributes to the recent literature focusing on the flow of 'causality' between energy-commodity and financial markets during the war.

## 6.4 BEAST approach results

### 6.4.1 BEAST decomposition of energy and financial time series and change points (breakpoints) detection.

TTF data despite being familiar to large audiences for its rapid price increases due to the February 2022 Russian invasion of Ukraine, its true underlying seasonal and trend dynamics are unknown, except that we know that several critical events, as described in section 3, are expected to have shaped crucially its dynamic evolution. However, questions like *how many breakpoints* have occurred and *when* (not only in TTF time series but also in the electricity markets that are affected by TTF), what are the dynamical characteristics of the underlying trend and seasonal components of the analyzed time series, are some of the questions that will be answered in this section. Another, relevant question is whether the existence of the detected by BEAST breakpoints is justified only by the 'drivers' incorporated in the critical events E1-E13 listed in table 1, or also by other *hidden event (factors) occurred in different dates than those linked to the critical events,* that are revealed by the BEAST model. BEAST has unveiled both the large-magnitude and subtle changepoints in the trend and seasonal dynamics of TTF and electricity markets and their probability of occurrence, as shown in **Table 4** and the figures that follow.

**Table 4** provides the 'big picture' of the changepoints detection by BEAST, for all energy (electricity and gas) and financial (USD/RUB) time series ('markets') analyzed here. The first row contains all the critical events and their dates of occurrence, given also in table 1 and section 3. The rest of the rows contain the dates of changepoints detected for each market (day, month, year), their probability distribution, and finally whether this date leads, lags or is concurrent with the date of the critical event. As an ***example***, *for* ***Romania***, 89% of the individual models for the trend component, sampled in the Monte Carlo Markov Chains (MCMC) of the BEAST, have indicated the existence of a structural changepoint in the trend component of the electricity price on 20 October 2021, i.e. the Romanian market reacted with a one-week lag to the EC's tools of measures, announced on 13 October 2021 (event E1), and almost 'concurrently' to event E2 (Gazprom ceased selling volumes at EU gas hubs). Also, for the same country, five changepoints detected in advance of E4-E7 critical events, no market reaction (no changepoints detected) to E8-E9 events, while Romania's market reacted with lag to events E12, E13. In total, ten out of 13 (~77%) structural changepoints were detected for the price of this country, a result shown also in the figure of the trend component produced by BEAST which is provided in the *Supplementary material C*. Looking carefully at table 4.1, we observe that BEAST provides a different number of structural changepoints, as 'reactions' or 'responses' of the markets under analysis to the critical events E1-E13. We see that eight out of ten (i.e. 80%) of the *electricity* markets, one of the largest proportions, have reacted to ***E5 critical event***, i.e. that Russian troops had begun arriving in Russia's ally Belarus for military exercises, on 17 January 2022. This critical event proved to be much more critical than the Russian invasion in Ukraine itself, occurred days later, on 24 February 2022. This indicates that most of the electricity markets *responded earlier, in advance, of the day of invasion*, following an increase of the TTF gas price, occurred even earlier about four days, on 4$^{th}$ January 2022, a finding which is further supported by the *strong correlations* between *TTF price and electricity prices*, depicted in figure 26. This is also shown by their strong increases of the trend component curves and the associated observed changepoints (see figures 14, 15 and 20, 21, of trend curves of TTF and Greek prices respectively, and similar ones in supplementary material, for all other markets).

Figures 14-19 show the results of the BEAST decomposition of the TTF price time series, figures 20-25 show the results of the Greek price data, while figure 26 shows the correlation matrix between the TTF's trend component and the trend components of all electricity markets. Figures of trend components of the rest of the electricity markets are provided in the **Supplementary material C (SMC).** Finally figures 27 and 28 show the trend components of NGNMX and USD/RUB data.

*Table 4.1: Comparison of crucial events E1-E13's dates with the structural changepoints' dates detected by BEAST.*

| | E1<br>13 Oct.<br>2021 | E2<br>Mid-Oct.<br>2021 | E3<br>10 Nov.<br>2021 | E4<br>17 Dec.<br>2021 | E5<br>17 Jan.<br>2022 | **E6<br>24 Feb.<br>2022** | E7<br>27 Apr.<br>2022 | E8<br>May.<br>2022 | E9<br>23 June<br>2022 | E10<br>21 July<br>2022 | E11<br>30<br>Aug.<br>2022 | E12<br>14 Sep.<br>2022 | E13<br>1 Nov.<br>2022 |
|---|---|---|---|---|---|---|---|---|---|---|---|---|---|
| RO | 20 Oct.<br>2021<br>89%<br>Lag | 20 Oct.<br>2021<br>89%<br>concurrent | 29 Nov.<br>2021<br>87%<br>lead | 6 Dec.<br>2021<br>87%<br>lead | 8 Jan.<br>2022<br>100%<br>lead | 13 Feb.<br>2022<br>100%<br>lead | 20 Apr.<br>2022<br>99%<br>lead | | | 2 Jul.<br>2022<br>100%<br>lead | | 20 Sep.<br>2022<br>100%<br>lag | 18 Nov.<br>2022<br>100%<br>lag |
| BE | 30 Oct.<br>2021<br>99%<br>lag | 30 Oct.<br>2021<br>99%<br>lag | | | 8 Jan.<br>2022<br>100%<br>lead | 13 Feb.<br>2022<br>99%<br>concurrent | | 12 May.<br>2022<br>100%<br>lead | 28 Jun.<br>2022<br>100%<br>lag | | | 20 Sep.<br>2022<br>100%<br>lag | 29 Nov.<br>2022<br>100%<br>lag |
| CZ | 27 Oct.<br>2021<br>100%<br>lag | 27 Oct.<br>2021<br>100%<br>lag | | | 1 Jan<br>2022<br>100%<br>lead | 10 Feb<br>2022<br>99%<br>lead | 12 Apr.<br>2022<br>100%<br>lead | | 6 June<br>2022<br>99%<br>lead | 20 Jul.<br>2022<br>99%<br>concurrent | | 13 Sept.<br>2022<br>100%<br>'concurrent' | 21 Nov.<br>2022<br>100%<br>lag |
| DK1 | 27 Oct.<br>2021<br>100%<br>lag | 27 Oct.<br>2021<br>100%<br>lag | 21 Nov.<br>2022<br>100%<br>lagged | | 8 Jan.<br>2022<br>100%<br>lead | | 20 Apr.<br>2022<br>100%<br>lead | 11 May<br>2022<br>100%<br>lead | 3 June<br>2022<br>100%<br>lead | | | 13 Sept.<br>2022<br>100%<br>'concurrent' | 21 Nov.<br>2022<br>100%<br>lag |
| ES | | | 10 Nov.<br>2021<br>100%<br>concurrent | | 8 Jan.<br>2022<br>100%<br>lead | | 20 Apr.<br>2022<br>100%<br>lead | | | 6 Jul.<br>2022<br>100%<br>lead | | 9 Sept.<br>2022<br>97%<br>lag | 7 Nov.<br>2022<br>91%<br>lag |
| HU | | | | | 8 Jan.<br>2022<br>100%<br>lead | | 27 Apr.<br>2022<br>100%<br>concurrent | | | 2 July<br>2022<br>100%<br>concurrent | | 9 Sept.<br>2022<br>100%<br>lead | |
| NNL | | | | 6 Dec.<br>2021<br>96%<br>lead | 8 Jan.<br>2022<br>100%<br>lead | | | 8 May<br>2022<br>99%<br>lead | 28 June<br>2022<br>100%<br>lag | | | 20 Sept.<br>2022<br>100%<br>lag | |
| IT | | | 10 Nov.<br>2021<br>100%<br>concurrent | | | | 16 Apr.<br>2022<br>100%<br>lead | 12 May<br>2022<br>22%<br>lead | 8 June<br>2022<br>100%<br>lead | 6 Jully<br>2022<br>100%<br>lead | | 20 Sep.<br>2022<br>100%<br>lag | 14 Nov.<br>2022<br>100%<br>lag |
| GR | 16 Oct. | 16 Oct. | 14 Nov. | 10 Dec. | | | 20 Apr. | | | 9 July | | 9 Sept. | 14 Nov. |
| | 2021 82% Lag | 2021 82% Lag | 2021 100% lag | 2021 18% lead | | | 2022 99% lead | | | 2022 99% lead | 22 Aug. 2022 40% lead | 2022 100% lead | 2022 100% lag |
| BG | 20 Oct. 2021 82% lag | 20 Oct. 2021 82% lag | | 6 Dec. 2021 47% lead | 8 Jan. 2022 100% lead | | 20 Apr. 2022 98% lead | | | 9 July 2022 100% lead | | 9 Sept. 2022 100% lead | 18 Nov. 2022 100% lag |
| TTF | | | 10 Nov. 2021 97% concurrent | | 4 Jan. 2022 99% lead | | | | | 2 July 2022 97% lead | | 13 Sep. 2022 100% concurrent | |
| NGNM X | | | 7 Nov. 2021 74% lead | 13 Dec. 2021 66^ lead | | 13 Feb. 2022 99% lead | | 19 May 2022 97% 'concurrent' | | 2 July 2022 100% lead | 29 Aug. 2022 99% lead | | |
| USD/R UB | 27 Oct. 2021 72% lag | 27 Oct. 2021 72% lag | | | | | 16 Apr. 2022 100% lead | | | | 4 Aug. 2022 100% lead | | |

As already mentioned, most of the electricity markets have shown breakpoints that happened on dates that are lagged, leading or are concurrent to the date of Russian invasion. *No 'direct' changepoints (i.e. 'immediate' reactions) to E6 critical event (24 Feb.22, Russian invasion) were observed for the cases of DK1, ES, HU, NNL, Greek and Italian markets.* We present at this point some explanations for this finding, i.e. why there is a deviation of the dates of breakpoints from that of the date of invasion, just for the Greek and Italian markets only, due to space limitations as well as that similar explanations can be given for the rest of the markets, based on specific for each market information for the crisis period, a work which is beyond the scope of the present paper.

Greek market seems to have reacted in advance to E6 event, and especially after Putin's proposition for a prohibition on Ukraine joining NATO, on 17 December 2021 (E4), reflecting the Country's concern that this geopolitical conflict could escalate badly and might have an impact on the security of gas supply from Russia, taking also into account the country's tightly dependence on Gazprom supplies. For the case of Italy, we try to explain this finding based on the recent information in the report of the Friedrich-Ebert-Stiftung Politics for Europe, the case of Italy (Andreolli F., et al., 2023). The main reason that no direct changepoint in the trend component of the Italian price is observed, that during the end of 2021 and beginning of 2022 period, the share of the Russian gas in the electricity consumption was drastically reduced. Prior to the war in Ukraine, in 2021, Italy was strongly dependent on Russian natural gas imports with around 40% of total gas imports (72.6 billion standard cubic meters) coming from Russia. Italy's gas imports from Russia were halved in 2022 (to 19% of the total) and, at the same time, tripled its exports. In addition, even though at least $1/5^{th}$ of the electricity consumed in this country in 2021 was generated using Russian gas, *this share was further drastically reduced to around one tenth in 2022.* During 2022, 72.4 bn cubic meters of natural gas were imported, 19% of which came from Russia. Thus, Italy's dependence on Russian supplies were halved, due to supplies from other countries via pre-existing infrastructure, despite that the total natural gas imports virtually were unchanged. For the Italian electricity generation, especially, in 2021, electrical energy demand was 320 TWh, 51% of which came from non-renewable sources, 36% from renewable sources and the rest from imports. Natural gas share was the largest in the approximately 180 TWh produced from fossil fuels, all of which was approximately imported. According to the above report, from the Italian gas imports in 2021, approximately 40% was Russian. As a conclusion, if we assume thatthe Russian share of gas imports is divided up equally between the industrial sectors consuming it, at least 1/5$^{th}$ of the electricity consumed in Italy during 2021 was produced with Russian gas. This share was further dropped to around 1/10$^{th}$ in 2022, given the reduction in Russian gas imports as a proportion of total imports. In addition to this, the substitute of coal in Italy's electricity generation (which was dependent on Russian imports to the extent shown previously) proved to have a marginal importance during 2021 but increased during the crisis. As shown also in table 3, TTF gas price reacted almost half a week in advance of E5, while NGNMX gas market, and the USD/RUB rate showed to be insensitive. Therefore, due to the very early drastic reductions of Italy's gas imports from Russia, as well as 'the signals' received by the strong TTF's price increase, as a reaction to E5 event.

A large proportion of the electricity markets' reactions to critical events is also shown for events E7(8 out of 10, 80%), E10 (70%), E12 (100%) and E13 (80%). Table 4.2 connects the critical events E1-E13 to the result of BEAST for each market. The table shows the number of markets in which a breakpoint due to a specific critical event is detected, as well as the 'profiles' of trend curves as decomposed by BEAST (shown in the figures that follow) that can be attributed to the critical events E1-E13.

We observe that 9/13 critical points are associated with smooth or abrupt trend increases, 2/13 with decreasing trends, and a sudden drop (when Nord stream went out of operation). We observe also that E5, E7, E10 and E12 events have caused breakpoints detected in many markets, and in only 3 markets a breakpoint directly connected to E6 has been detected, i.e. on the date of Russian invasion.

**In Fig.14**, the first top subplot depicts the raw data, while the second one shows the curve of seasonal changepoints (scp), with the probability of occurrence P(scp) shown on the 3$^{rd}$ subplot. The 4$^{th}$ subplot depicts the number of seasonal orders of the optimally selected model. The trend curve with its changepoints (tcp) is in 5$^{th}$ subplot with their probabilities in the 6$^{th}$ sbplot. The evolution of the order of the model for the trend curve is on the 7$^{th}$ subplot. The 9$^{th}$ subplot shows the probability of the slope (of the trend) being *positive (red part)* (i.e., increasing trend) for the trend component. *Zero slope* is the *green* part and *negative slope the blue part*. As an example, if the probability is 0.80, at a given point in time, it means that 80% of the individual trend models sampled in the Monte Carlo Markov Chains (MCMC) of BEAST model have a positive slope at that point.

Table 4.2: Profiles of trend curves from BEAST decomposition of the ten electricity markets attributed to the critical events E1-E13.

| Sort Description | symbol | date | year | Number of markets detected with a breakpoint by BEAST | trend |
|---|---|---|---|---|---|
| EU toolbox | E1 | 13 Oct. | 2021 | 7 | increasing |
| Gazprom ceased selling at EU gas hubs | E2 | Mid-Oct. | 2021 | 7 | increasing |
| Unusual movement of Russian troops | E3 | 10 Nov. | 2021 | 7 | Increasing (small slope) |
| prohibition on Ukraine joining NATO | E4 | 17 Dec. | 2021 | 5 | increasing |
| Russian military exercises | E5 | 17 Jan. | 2022 | **9** | increasing |
| Russia invades Ukraine | E6 | 24 Feb. | 2022 | 3 | flat |
| Gazprom cuts off gas supplies | E7 | 27 Apr. | 2022 | **9** | Steep increasing |
| EU eliminate Russian energy imports | E8 | May. | 2022 | 5 | steep increasing |
| Germany raising alert level | E9 | 23 June | 2022 | 5 | Steep increasing |
| EU measures to Russia | E10 | 21 July | 2022 | **9** | Steep increasing |
| Nordstream out of operation | E11 | 30 Aug. | 2022 | 3 | sudden drop |
| EU announces tax energy companies | E12 | 14 Sep. | 2022 | **11** | decreasing |
| EU minimum filling target | E13 | 1 Nov. | 2022 | 8 | decreasing |

As it is shown from the graph, the algorithm has detected, on average, nine *seasonal changepoints* (scp) as well as nine *trend* ones (tcp), represented by the vertical lines (black). These breakpoints divide the whole period in ten-time phases and the question now is how the dates of these detected changepoints are compared with the dates of the critical events found in the ACER-CEER and other similar reports. The comparison thus assesses the effectiveness of the BEAST tool used here and subsequently the usefulness of the present work in detecting not only the obvious but also any ‘hidden’ breakpoints not reported in the available literature.

Examining carefully figure 14 we observe that the seasonal dynamics in the first phase (beginning of 2020 to the end of $3^{rd}$ quarter 2020, is almost ‘flat’, with no periodicity, and the amplitude of change is insignificant, indicating a ‘tranquility’ in TTF prices. Things are changed drastically then, with a turmoil in prices occurring at the spikes-breakpoints, shown to occur with varying probabilities. The breakpoints with highest probabilities are those corresponding to table 4.1 for TTF.

Figure 15 is an isolation of trend component in figure 14, showing the 9 trend changepoints (tcp) their probability of occurrence, P(tcp) as well as the probability of the trend slope being positive (red part) (i.e., increasing trend), for being zero (green) and finally for being negative (blue), at each time point, of the trend component of TTF data.

In figure 16 we show the probability distribution of having a changepoint in the trend component of TTF, at each point of time. A higher peak indicates a higher chance of being a changepoint, however, only at that point in time and does not necessarily mean a higher chance of observing a changepoint around that time.

Figure 17 depicts the probability distribution of having a certain number of trends changepoint over the range of min=0 to max=20, set in this work, for TTF data. We observe that the largest probability is for having 10 trend changepoints is approximately 42%, while is 15% for having 11 changepoints and zero for having 15 or 20 changepoints. This result is consistent with the maximum number of 12 critical events identified from literature, given in table 1.

The Probability distribution of total number of trend curve changepoints in TTF price, decomposed by BEAST, is given in Figure 18. Thus, the probability of having one total changepoint is 100 %, of 2 or 3 99.9 %, and only 10% for having 10 total changepoints.

Finally, in Figure 19 we show the sudden changes (jumps) in trend component of TTF price at changepoints. So, for example, at changepoint number 3, the fitted trend curve contains approximately 34 positive jumps, while the 4$^{th}$ changepoint, is associated with 13 negative jumps.

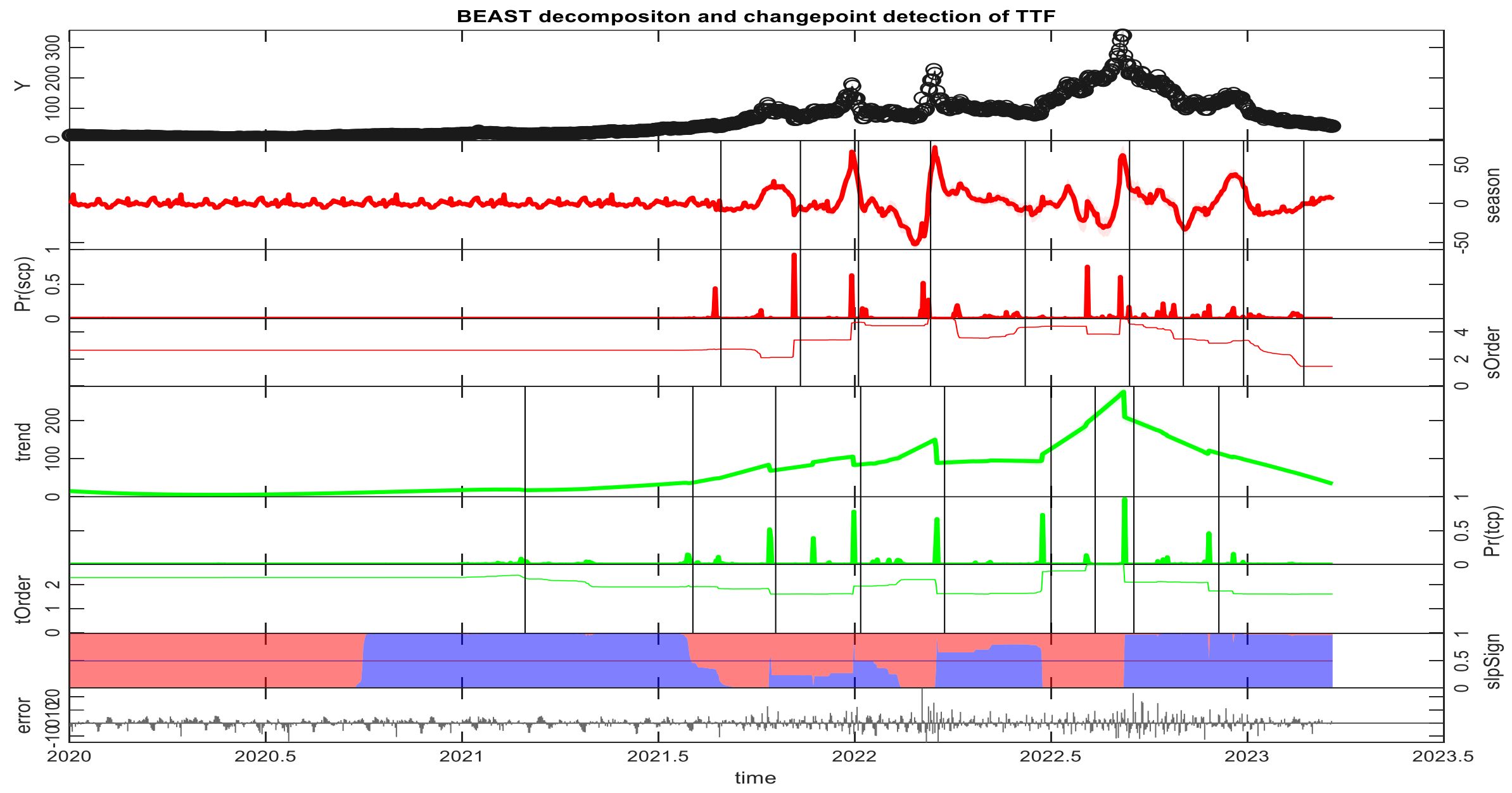

Figure 14: Result of BEAST decomposition for the TTF data. The nine subplots contained useful information as described in the main text. In the 8$^{th}$ subplot (slpSign): The sign of the trend slope is positive (red), negative (blue), zero (green). Black vertical lines correspond to the nine dates of changepoints, both in the trend and seasonal components, detected by BEAST.

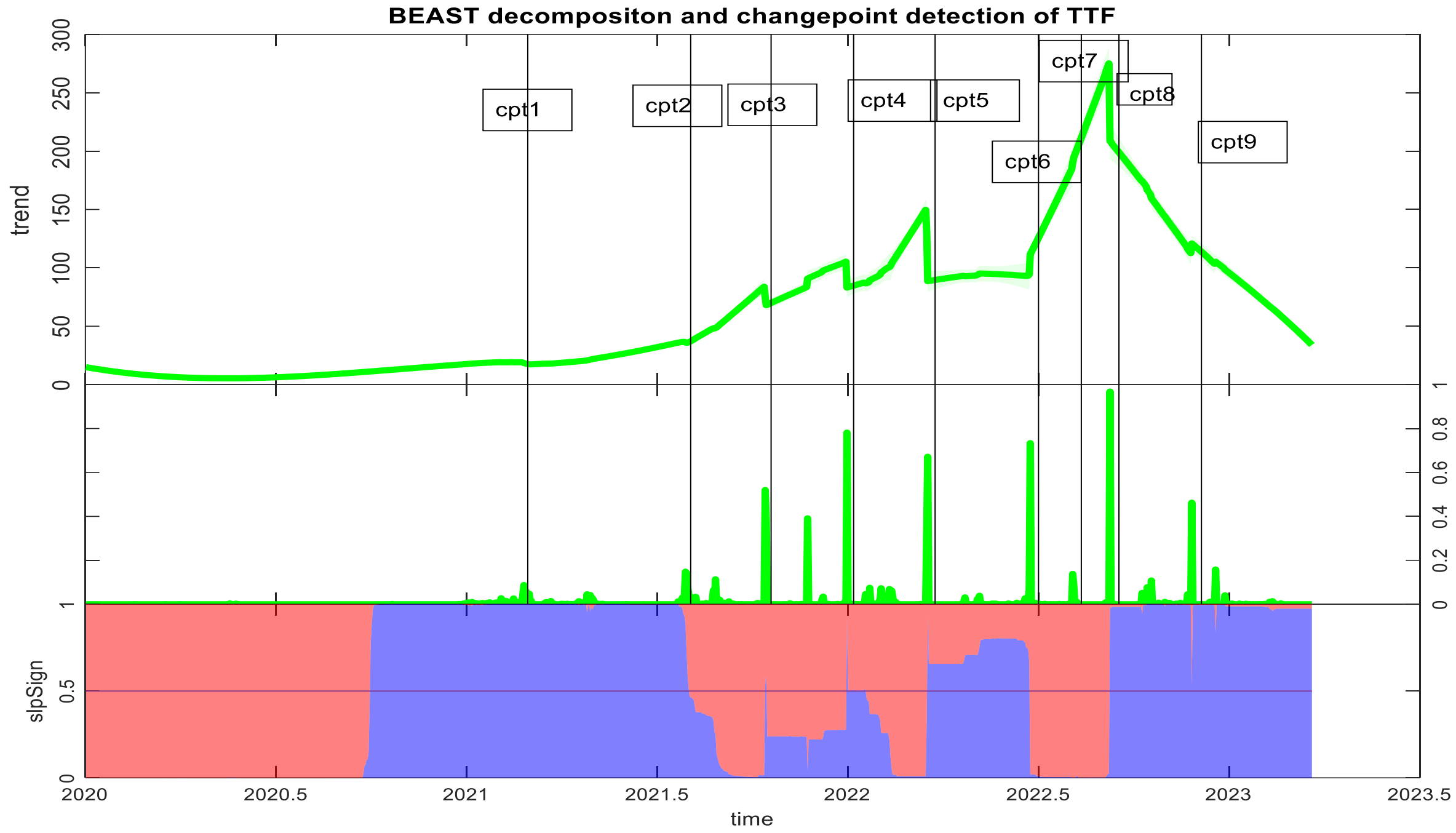

Figure 15: Isolation of trend component in figure 44, showing the 9 trend changepents (tcp) their probability of occurrence, P(tcp) as well as the probability of the trend slope being positive (red part) (i.e., increasing trend), for being zero (green) and finaly for being negative (blue), at each time point, of the trend component of TTF data. Black vertical lines correspond to the nine dates of changepoints, both in the trend component, detected by BEAST.

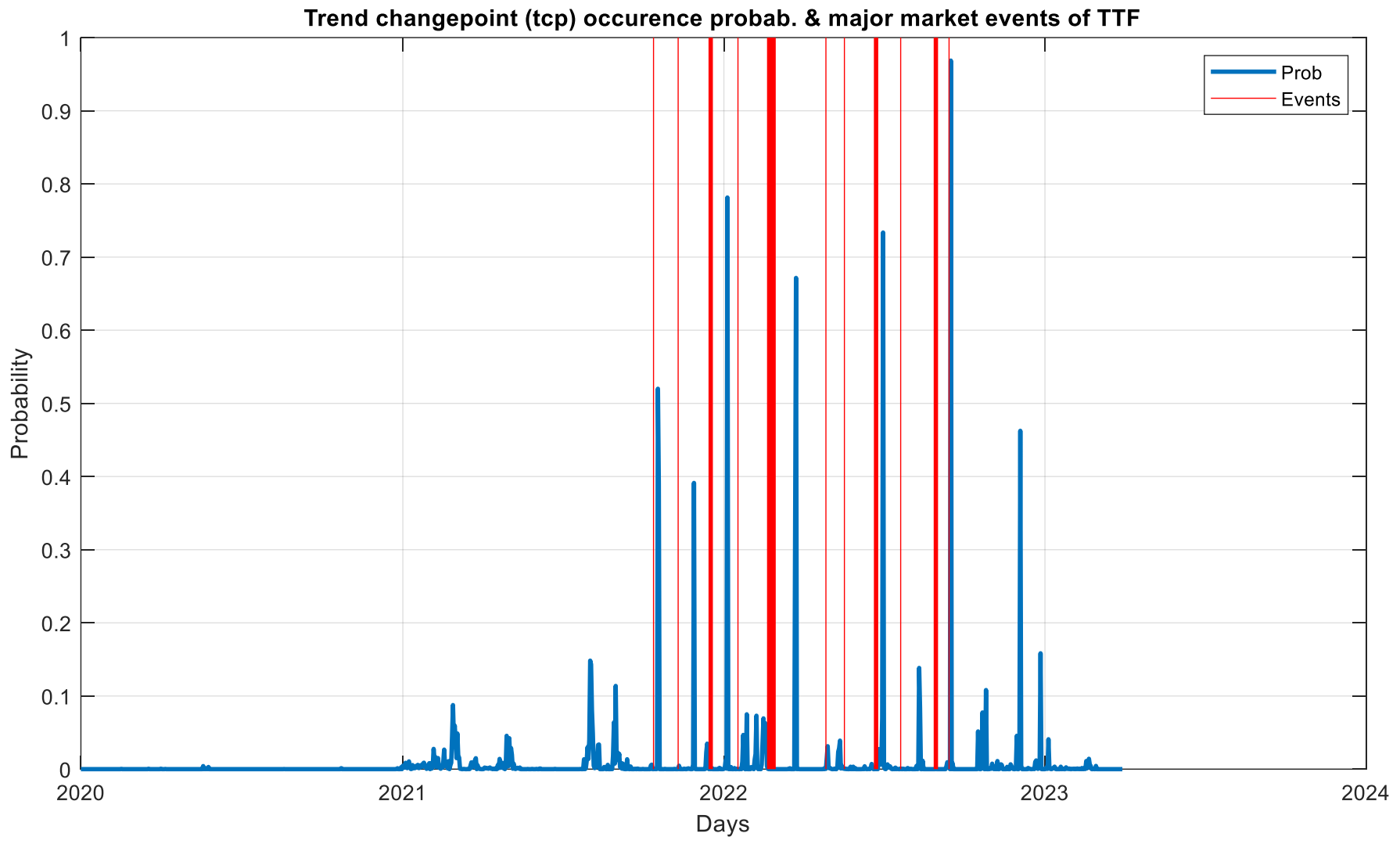

Figure 16: Probability distribution of having a changepoint in the trend component of TTF data, at each point of time. Red vertical lines correspond (with the same order) to the critical events described in table 1.

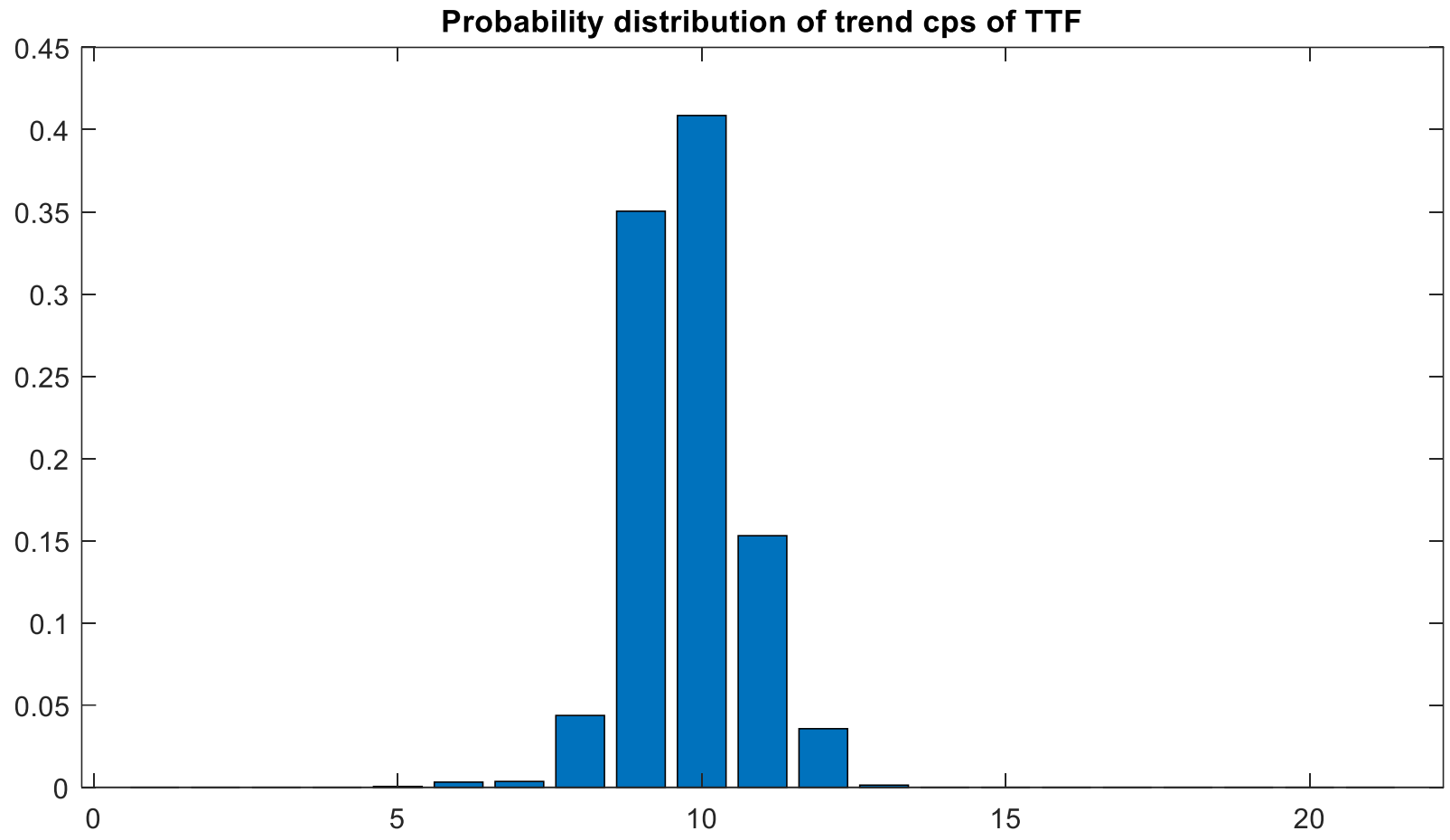

Figure 17 Probability distribution for the number of changepoints, in the trend component of the TTF price, decomposed by BEAST tool.

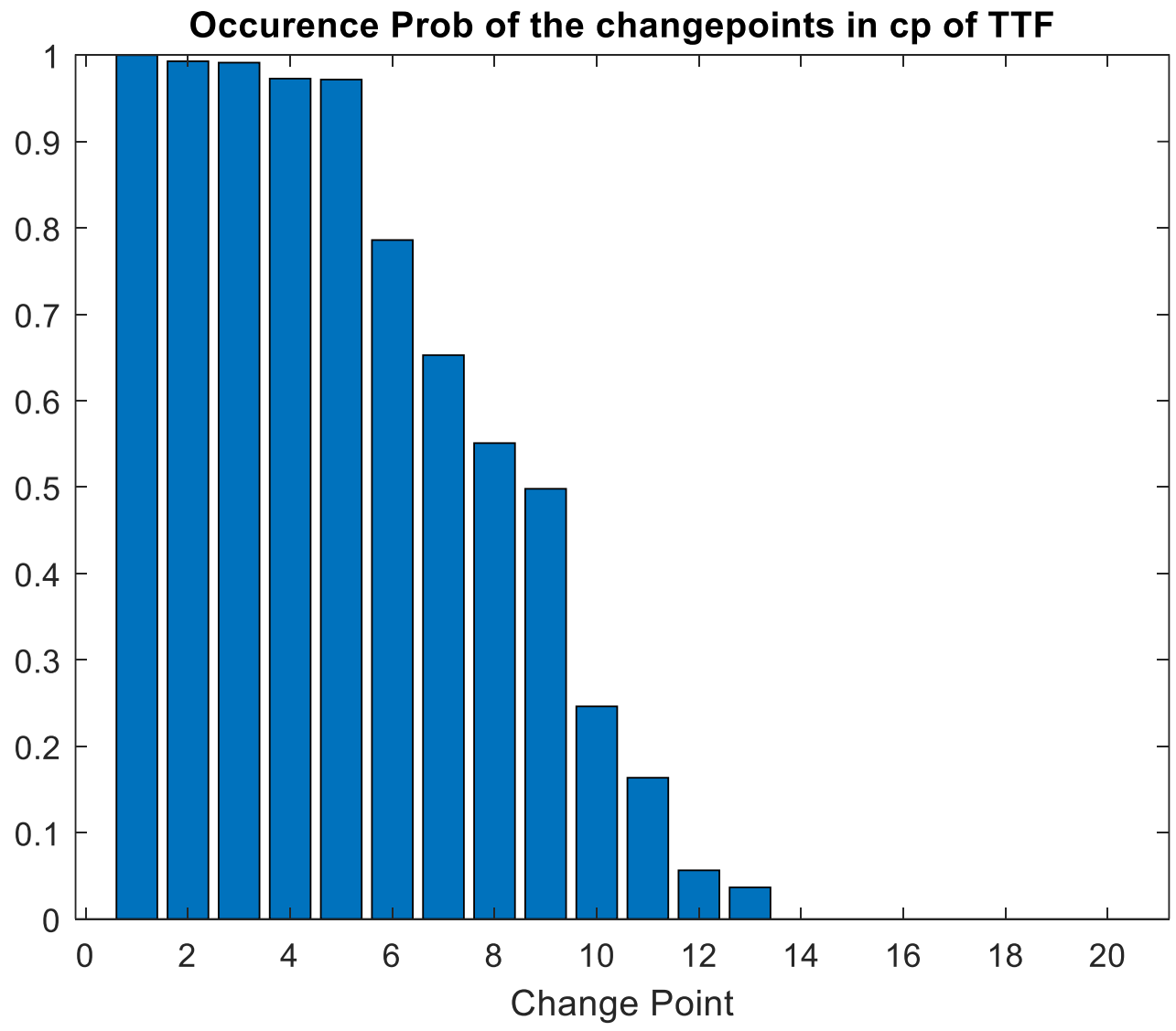

Figure 18. Probability distribution of total number of trend changepoints in TTF price, decomposed by BEAST.

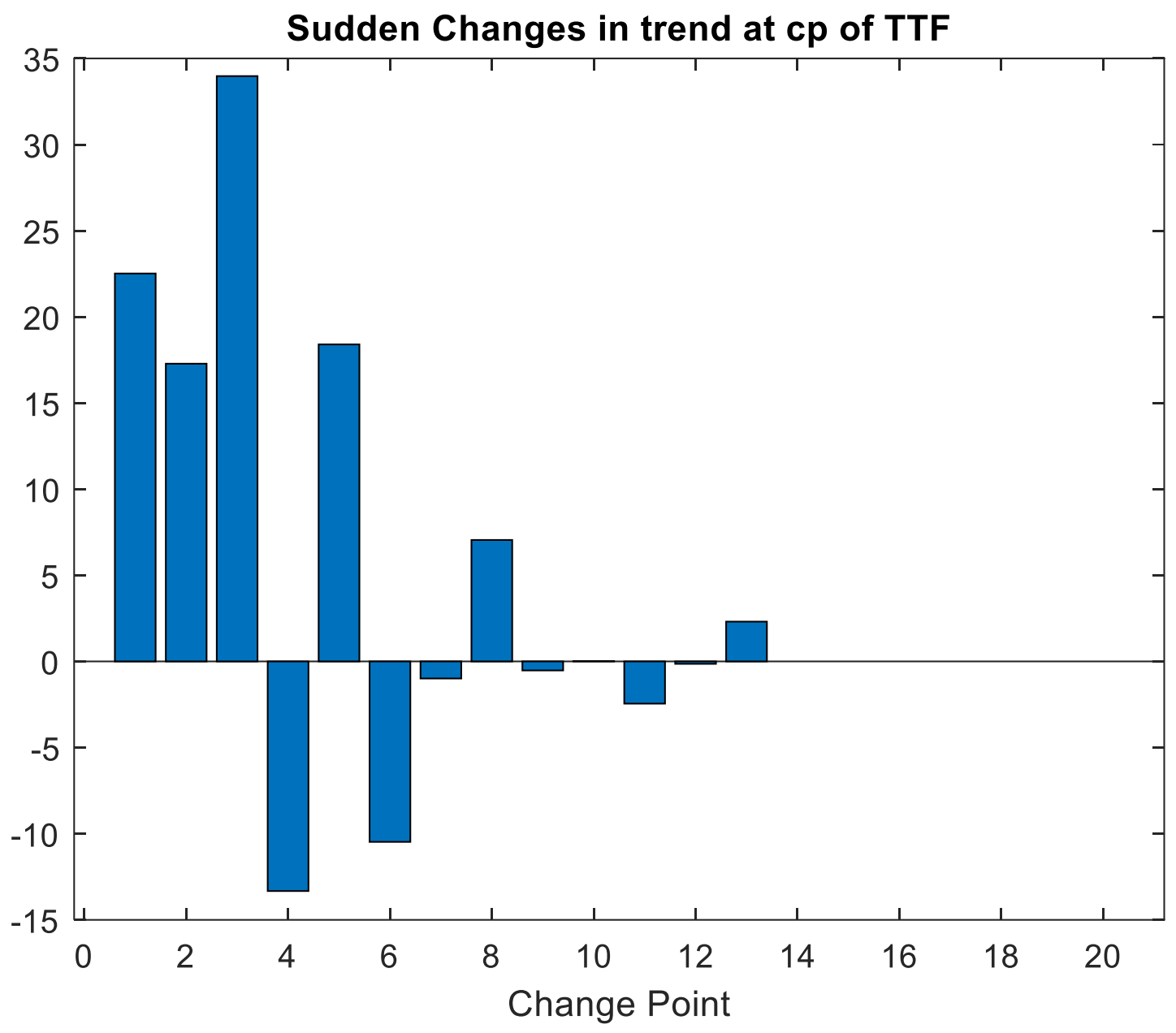

**Figure 19:** The sudden changes (jumps) in trend component of TTF price at changepoints.

### 6.4.2 BEAST decomposition of Greek price time series and changepoints (breakpoints) detection

In figures 20 to 26, we present the results of the BEAST decomposition of the Greek price time series. All figures contain information like the figures 24-30 above, for the TTF price.

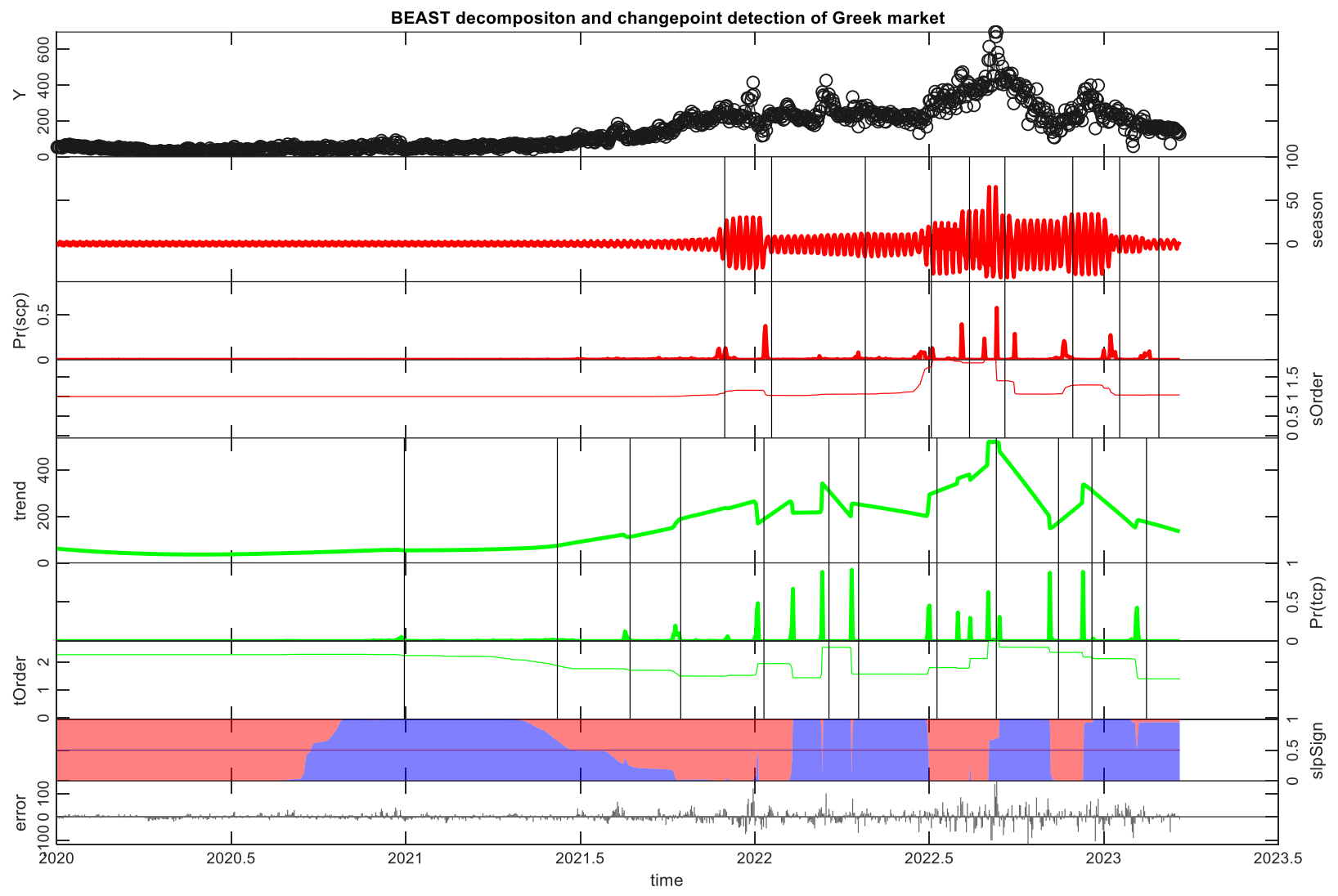

**Figure 20:** Result of BEAST decomposition for the Greek electricity price data. The nine subplots contained useful information as described in the main text. In the 8th subplot (slpSign): The sign of the trend slope is positive (red), negative (blue), zero (green). Black vertical lines correspond to the nine dates of changepoints, both in the trend and seasonal components, detected by BEAST.

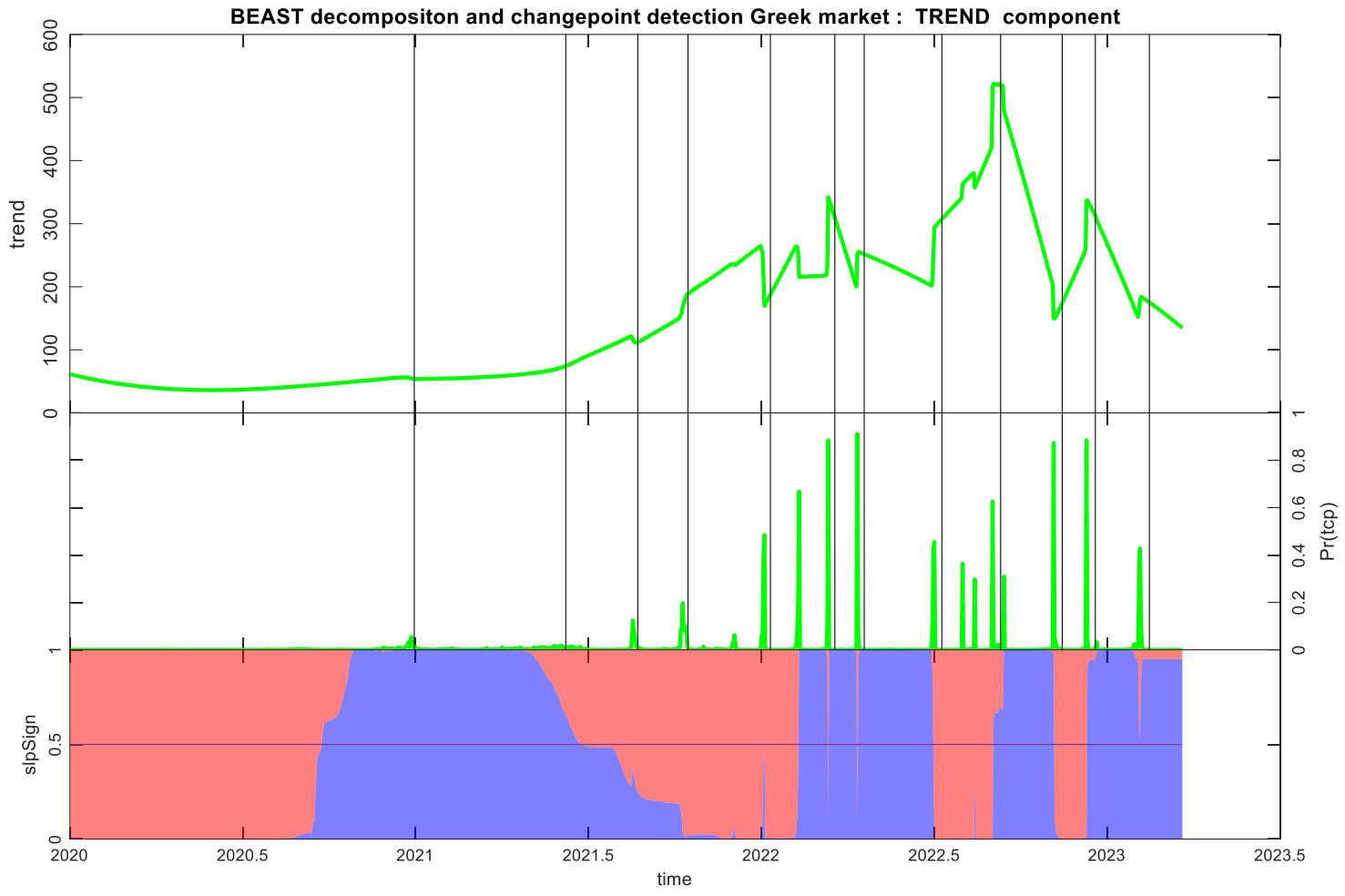

**Figure 21:** Isolation of trend component in figure 20, showing the 12 trend changeponts (tcp) their probability of occurrence, P(tcp) as well as the probability of the trend slope being positive (red part) (i.e., increasing trend), for being zero (green) and finaly for being negative (blue), at each time point, of the trend component of Greek price data. Black vertical lines correspond to the nine dates of changepoints, both in the trend component, detected by BEAST.

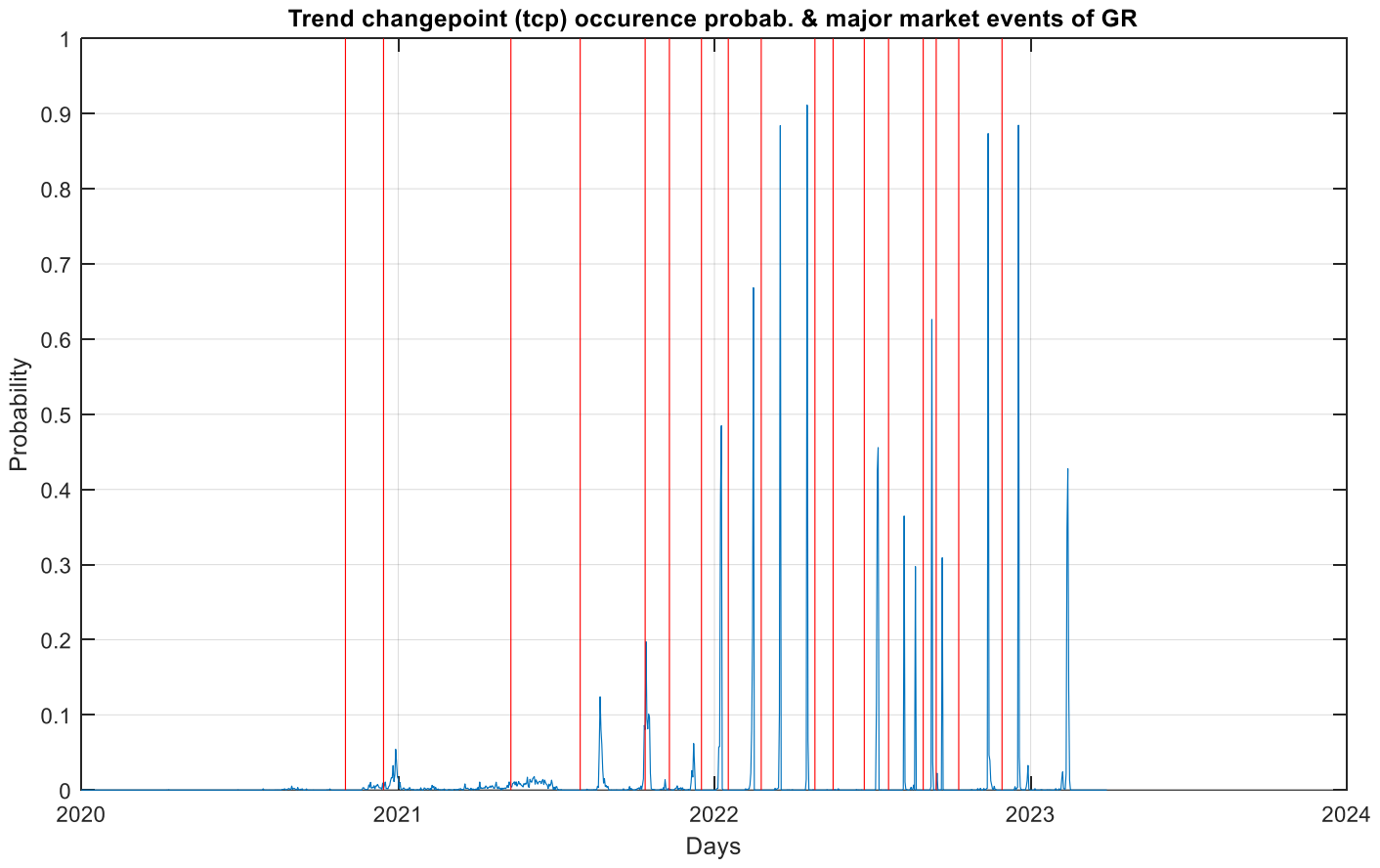

**Figure 22:** Probability distribution of having a changepoint in the trend component of Greek price data, at each point of time. Red vertical lines correspond (with the same order) to the critical events described in table 1.

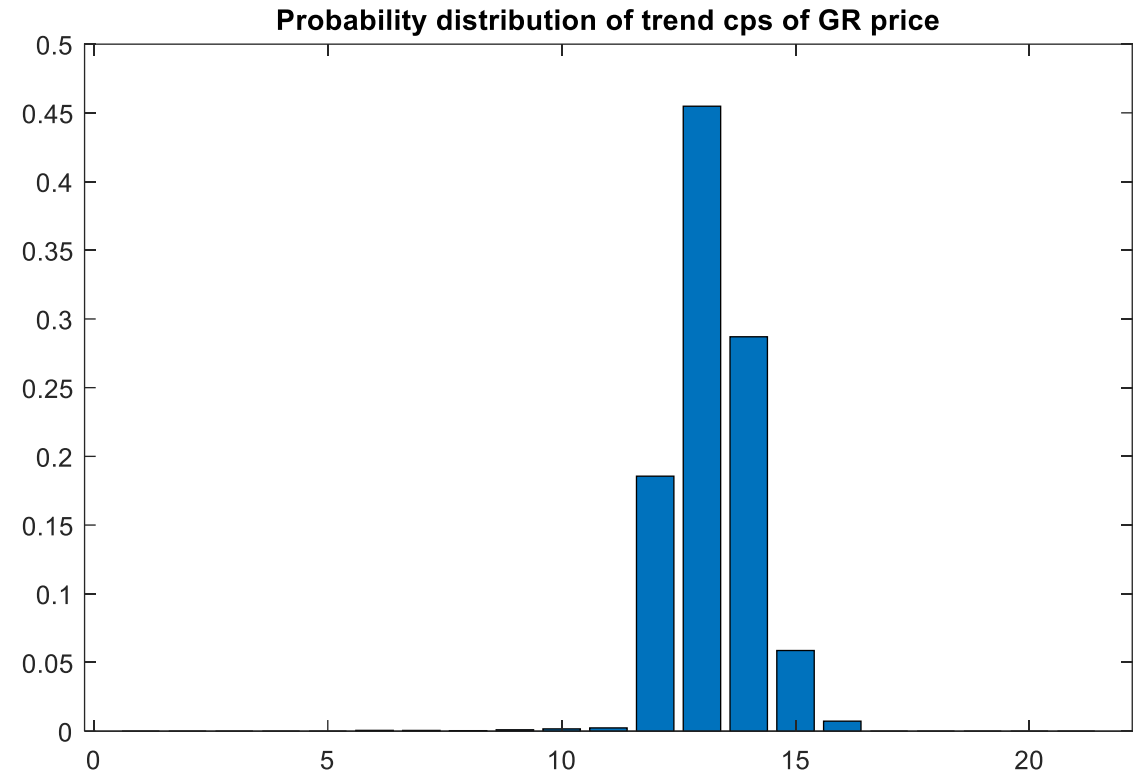

**Figure 23:** Probability distribution for the number of changepoints, in the trend component of the Greek price, decomposed by BEAST tool.

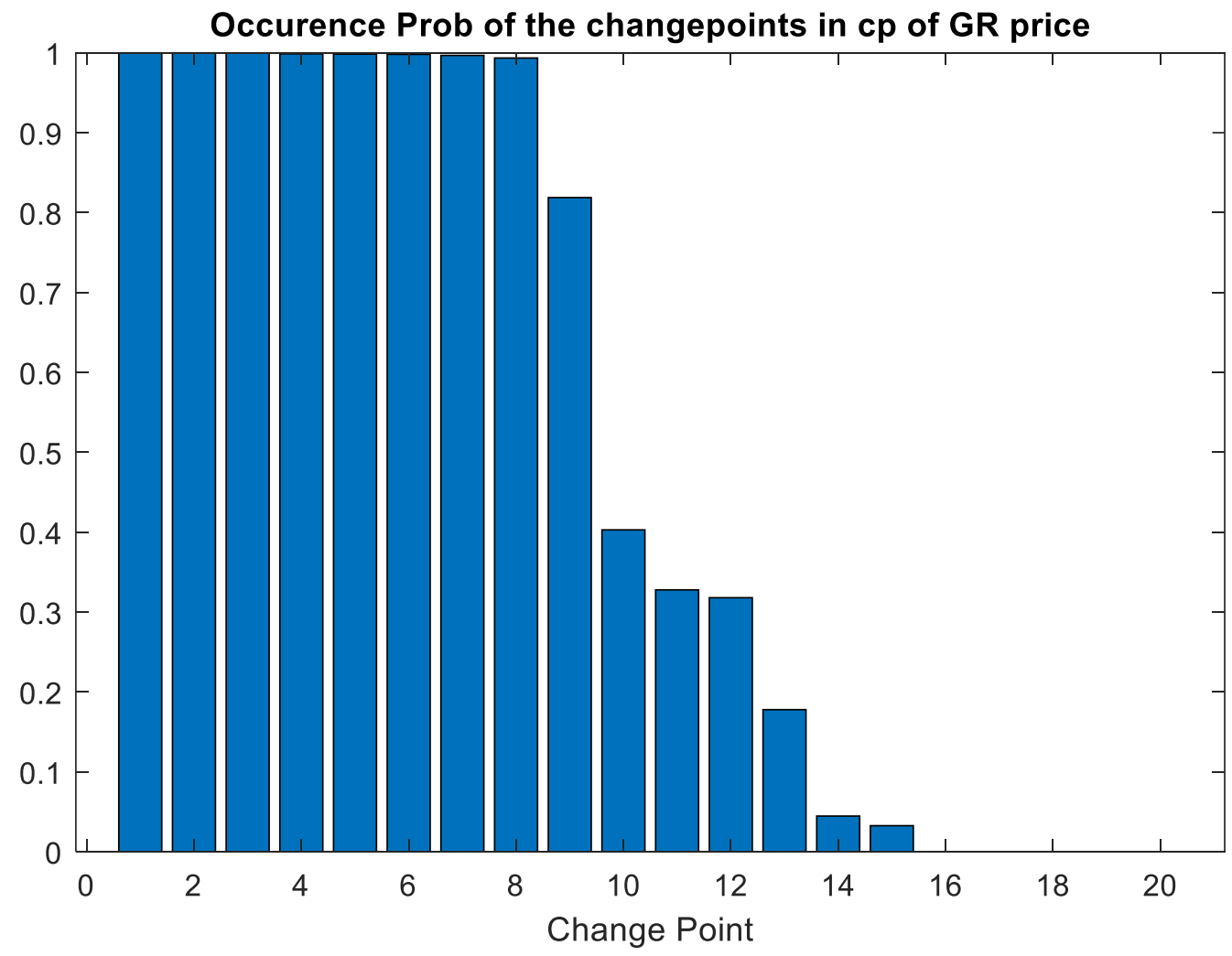

**Figure 24:** Probability distribution of total number of trend curve changepoints in Greek price, decomposed by BEAST.

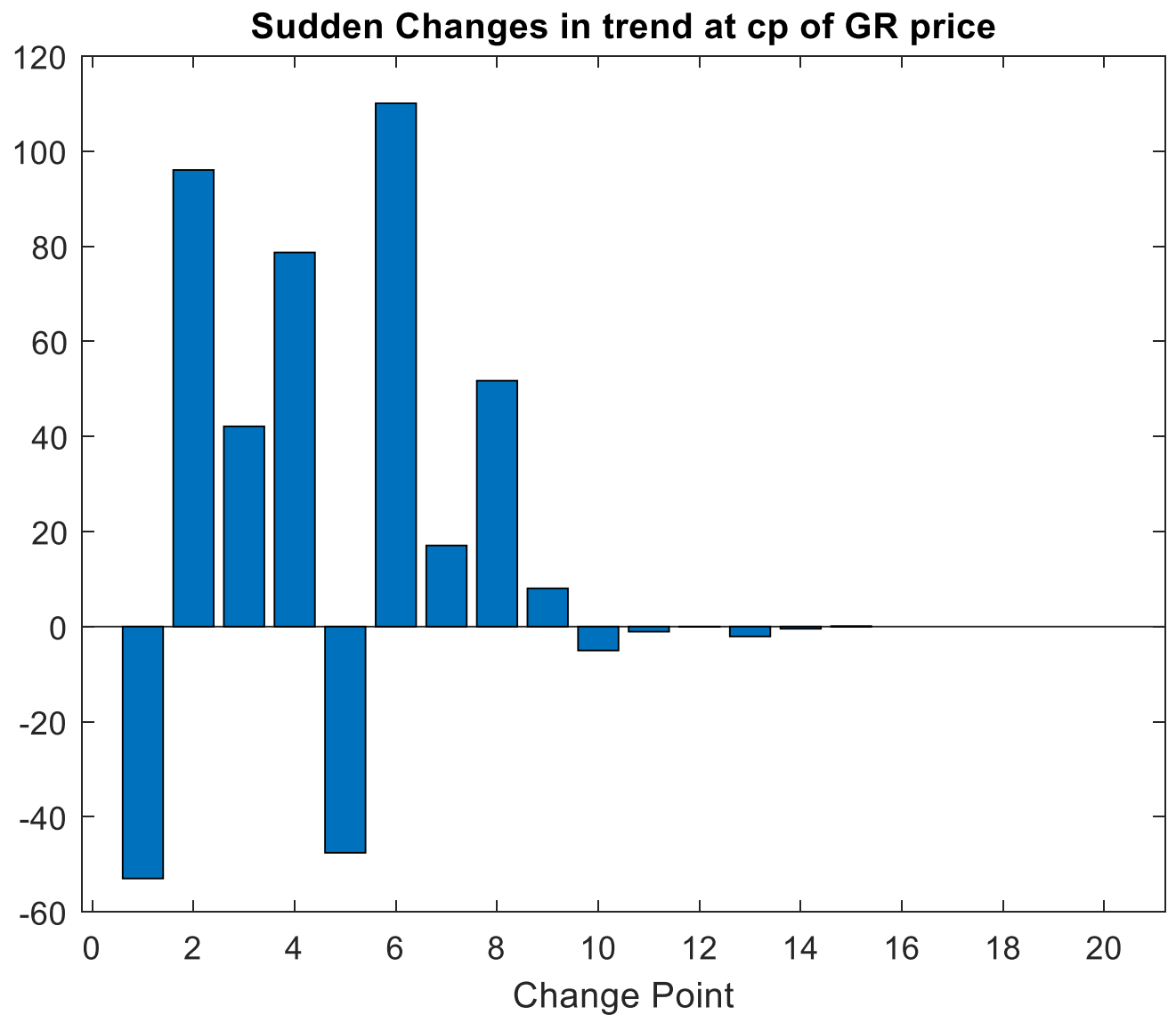

**Figure 25:** The sudden changes (jumps) in trend component of Greek price at changepoints.

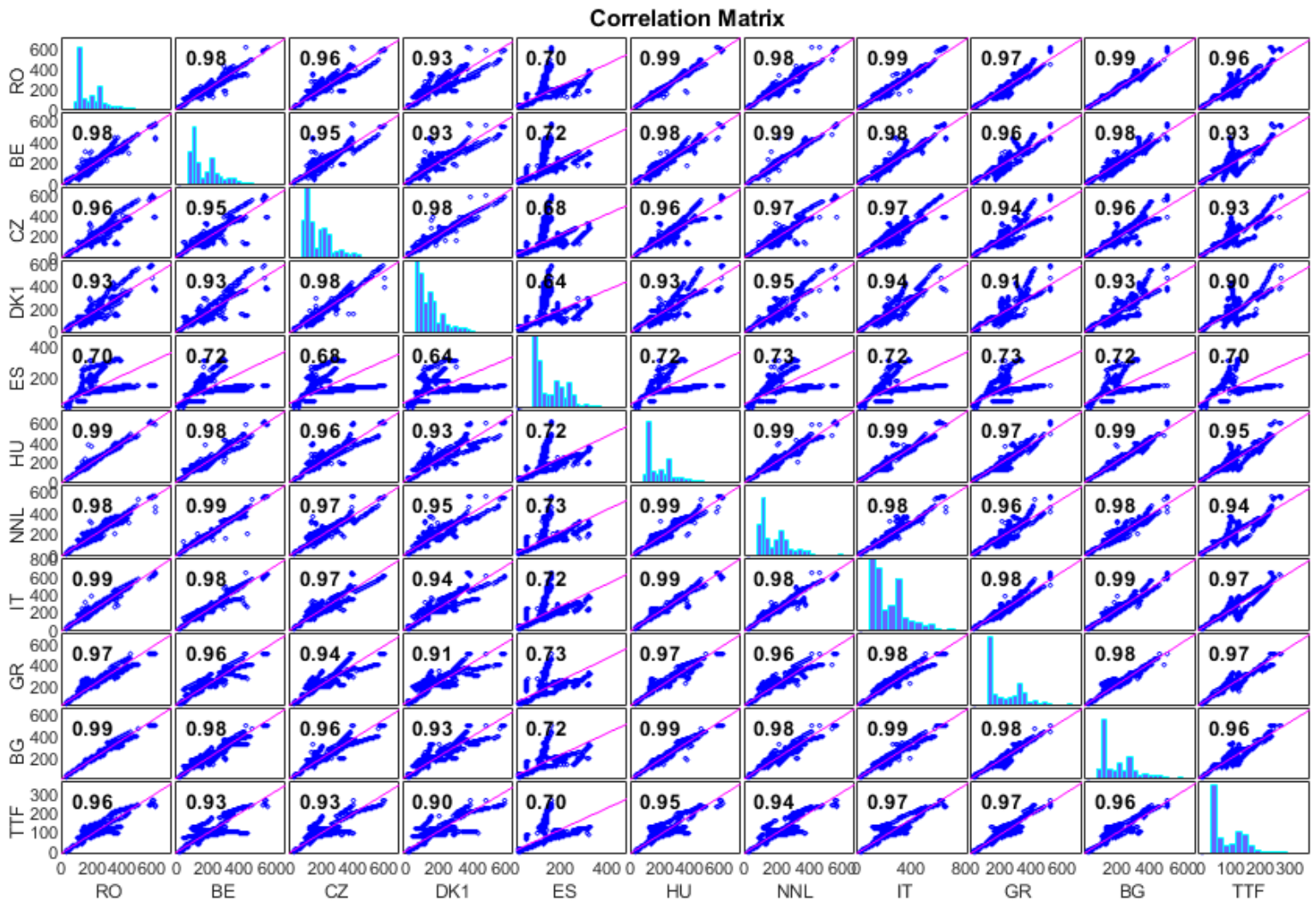

Figure 26: Correlation matrix of trend curves of all electricity markets and TTF. High correlation values indicate a 'similarity' of the dynamics of the trend curves of a pair of markets.

Figure 26 shows the *correlation matrix* among *trend components curves* produced by BEAST between all electricity prices and TTF price. A strong value in the matrix reflects a strong dynamic similarity of trend curves. From the figure, we observe that Italian and Greek prices trend curves present the highest correlation with TTF prices, while the smallest value is between Spanish prices and TTF price trend curves. Therefore, we can assume that because of the dynamic similarity of two markets, an absence of a structural changepoint in one market, as its response to a specific critical event, is rationally expected to occur also in the other market, at a similar date. In fact, in the case of Greek and Italian prices, the two markets reacted similarly, exhibiting no changepoint at the E5 critical value (see comments for Table 3).

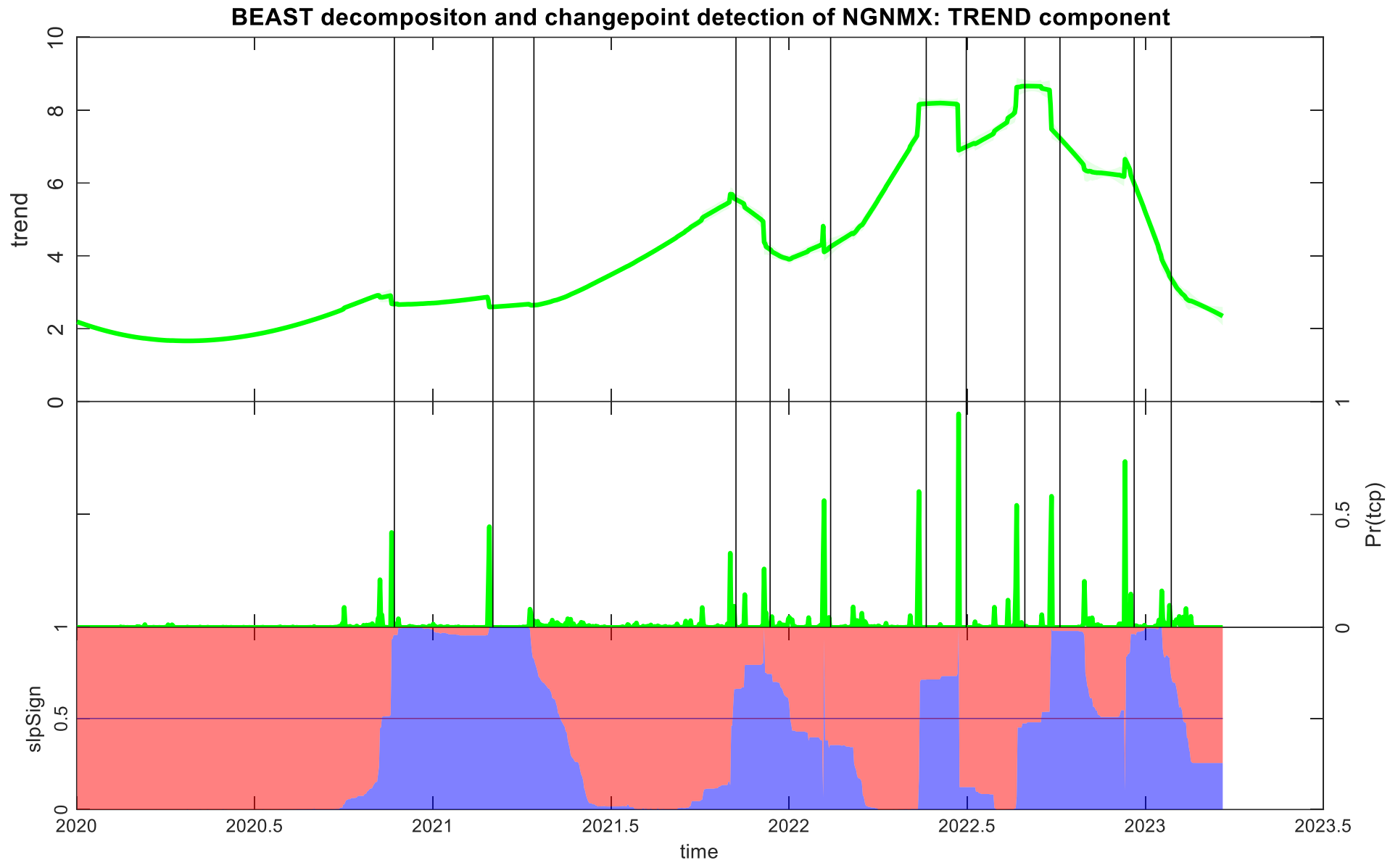

Figure 27: Trend component of BEAST analysis for NGNMX, showing the 12 trend changepents (tcp) their probability of occurrence, P(tcp) as well as the probability of the trend slope being positive (red part) (i.e., increasing trend), for being zero (green) and finaly for being negative (blue), at each time point, of the trend component of NGNMX price data. Black vertical lines correspond to the nine dates of changepoints, both in the trend component, detected by BEAST.

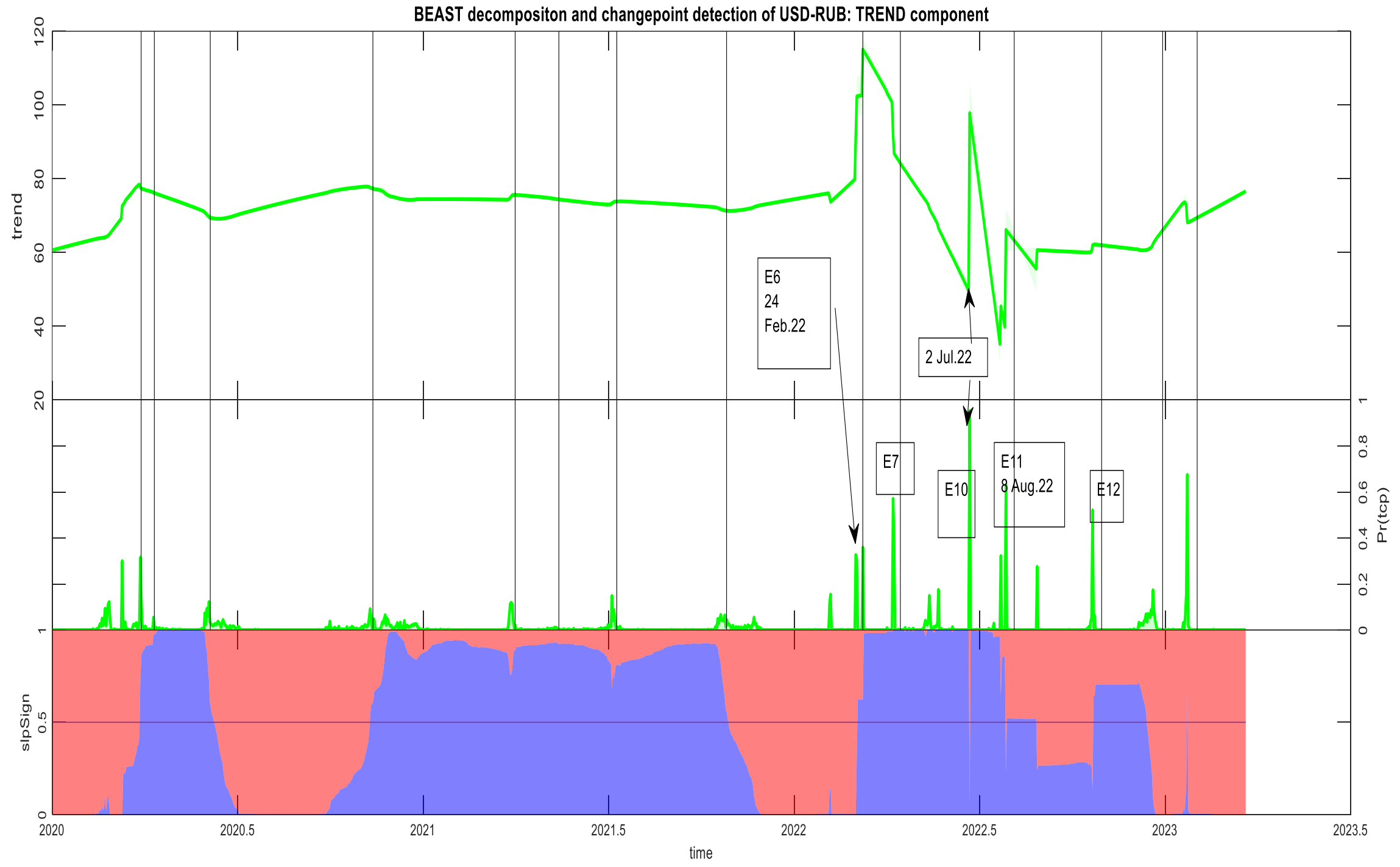

Figure 28: Trend component of BEAST analysis for USD/RUB data, showing the 12 trend changepents (tcp) their probability of occurrence, P(tcp) as well as the probability of the trend slope being positive (red part) (i.e., increasing trend), for being zero (green) and finaly for being negative (blue), at each time point, of the trend component of USD/RUB price data. Black vertical lines correspond to the nine dates of changepoints, both in the trend component, detected by BEAST.

Table 4.3. Breakpoints in the BEAST ‘s trend curve of USD/RUB with probability of occurrence and their connection with critical events E1-E13 of table 1.

| Breakpoint | Date from BEAST | Probability | Connection with Critical Event |
|---|---|---|---|
| 1 | 24 Feb.22 | 0.326 | E6 |
| 2 | 20 Apr.22 | 0.570 | E7 (27 Apr.2022), leading |
| 3 | 2 Jul.22 | 0.960 | E10 (21 Jul.2022), leading |
| 4 | 8 Aug.22 | 0.630 | E11 (30 Aug. 2022), leading |
| 5 | 20 Oct.22 | 0.520 | E12 (14 Sep.2022), lagged |

Table 4.3 shows the breakpoints detected in the BEAST trend curve of the ‘financial market’ USD/RUB with probability of occurrence and their connection with critical events E1-E13 of table 1, that are also depicted in figure 28. We observe that point 1, with date 24 Feb.2022, the date of the Russian invasion, is not so probable to happen, in this financial market, so remotely located from the European scene of the crisis. Point 2, 20 April 2022 shows a larger probability of occurrence (>50%), reflecting the lagged reaction of this market on the decision of Gazprom to cut off gas supplies to Bulgaria and Poland (Event E7, 27 April 2022). The highest probability is that of breakpoint (occurrence) 3, in 2 Jully 2022, which reflects the leading reaction of the market to critical event E10, regarding the Western countries’ response to the Russian invasion, in the form of new measures of sanctions. The USD/RUB market reacted also with a leading behavior, with significant occurrence probabilities, to E11 and with a lagged response to E12 critical event.

**7. Discussion - Conclusions and Policy recommendations**

In the current study, we have combined several ‘innovative’ tools to shed light on the European energy crisis of 2022. We showed that by combining *structural breakpoints detection (BEAST approach), rolling Hurst exponent analysis, rolling Mutual (MI) and Partial Mutual Information with Mixed Embedding (PMIME),* we could gain a comprehensive framework for understanding and shed light on the factors that have led to the ‘recent’ European energy crisis of 2022. We have shown how these techniques can be integrated to reveal different aspects of the 2022 crisis:

1. **Structural Breakpoints Detection**: we identified periods of significant structural changes in the price time series data of ten European electricity prices and two gas time series (European TTF and New York’s NGNMX) energy prices, associated with their idiosyncratic market behaviors, using an innovative approach, the Bayesian ensemble structural breakpoints detection technique (BEAST). The dates of the breakpoints extracted by BEAST tool, revealed that *markets have reacted with leading, lagging or concurrent responses to the ‘known critical events’ E1-E13 connected to the Russo-Ukrainian crisis elicited from literature.* The tool has extracted also ***new ‘hidden critical events’***, manifested as truly new dates or leading/lagged deviations of the newfound dates from the known dates of E1-E13. ***In short, this analysis revealed critical junctures or anomalies in the price time series data analyzed, that coincide with the key events or developments during the entire period of energy crisis escalation of 2022, associated with the extensive consequences of the geopolitical tension of the Russian invasion in Ukraine, expressed in the form of reactions as, for example, the associated European policy interventions, the Russian supply disruptions, as a ‘retaliation’ to these interventions.***

   The news that Russian troops had started advancing and bombings, initiated the escalation in the analyzed markets volatility, as we have shown previously in related figures. More importantly, we have shown that some markets responded more swiftly than others, a result supported by the works of (Cataline Gheorghe, et al., 2023), and (Yousafet, I., et al., 2022). In the latter paper, their analysis showed that the stock market of *Hungary, reacted* almost ‘concurrently’ to the military events, exhibiting *negative returns the period before the critical event,* while the stock markets of *Italy, Spain, Romania*, were adversely affected in the after invasion period. All above countries, are included also in our preset study. Thus, the results in the papers above, even though not related to energy markets, supports our work in respect of the similarity in the time and strength of their responses to the Russian invasion event.

2. ***Rolling Hurst Exponent Analysis***: we calculated the rolling Hurst exponent for all electricity and gas price time series to assess the long-term memory or persistence of the associated market dynamics. We showed that changes in the Hurst exponent over the period of our analysis indicated ***shifts in the electricity and gas markets efficiency (as expressed by EMH)***, their ***predictability***, or ***volatility*** during the crisis period, and especially how these markets have responded to the critical events linked to the 2022 crisis. For example, we emphasized that for most of the markets their efficiencies increased when the time approached the dates of extracted breakpoints, and their rolling Hurst curves approached that of the TTF's curve at these breakpoints. **In general, we emphasized how at the found breakpoints an observed shifting of *the Hurst exponent towards the EMH limit of H=0.5, could suggest heightened market efficient or stability as opposed to market inefficiency when H at a breakpoint moves away from that limit.*** This finding, ***the connection of market efficiency with structural breakpoints***, is very important and contributes to the current literature, since, to the best of our knowledge, this result is the second one (Kaharan, C.C., et al., 2024) mentioned to the existed literature we reviewed.

The Hurst exponent has provided to us a valuable perspective on how the Russian invasion of Ukraine caused turmoil in energy prices by analyzing *the impact on market stability and the predictability of price movements*. Here's how the Hurst exponent has helped us explain this situation:

a) By understanding Market Turbulence:

- **Pre-Invasion Period (H > 0.5):** As we have seen, before the invasion, in the case the energy markets exhibit a Hurst exponent greater than 0.5, it indicates a ***stable market*** with predictable, persistent price trends. This stability might have been due to a well-functioning supply chain, consistent demand, and geopolitical stability.
- **Post-Invasion Period (H < 0.5):** The invasion disrupted global energy markets, particularly in Europe, where many countries rely on Russian gas and oil. This disruption likely caused a shift in the Hurst exponent towards values less than 0.5, indicating *increased volatility* and *unpredictability* in energy prices. *The market's previous patterns of stability were likely broken, leading to price swings and increased uncertainty.*

b) By considering the impact of Geopolitical Shocks, related to Russo-Ukrainian war, on the evolution of Hurst Exponent:

- **Increased Volatility:** as already mentioned, the invasion led to supply chain disruptions, sanctions on Russian energy exports, and the search for alternative energy sources. These factors created a highly volatile environment, where prices fluctuated widely in response to news, policy changes, and shifts in supply and demand. Therefore, a *Hurst exponent moving towards or below 0.5 during this period has reflected these sudden and erratic price changes we have observed in this study.*
- **Market Reversals:** The Hurst exponent being less than 0.5 indicates anti-persistence, where *price increases are followed by decreases and vice versa*. This behavior in the energy markets we analyzed, is reflected as *prices spiked with each new development* or critical events (e.g., pipeline disruptions, embargoes) and then partially corrected as markets adjusted (e.g., new supply agreements, government interventions).

c) By examining post-invasion market dynamics:

- **Adaptation and Stabilization:** Our results have shown that as markets began to adapt to the new geopolitical realities (e.g., Europe reducing reliance on Russian energy, increased LNG imports, renewable energy investments), the Hurst exponent gradually returned towards 0.5 or even above it, indicating a new form of market stability, though at a higher and more volatile price level than before.
- **Continued Uncertainty:** we observed also that in market cases where the Hurst exponent remained below 0.5 for an extended period, we could conclude that the market is still in a state of flux, with continued uncertainty and instability, which can be explained by ongoing geopolitical tensions, unpredictable policy responses, or new disruptions in energy supply.

All above, we think, may have some implications in the areas of quantitative analysis and risk Management of energy markets, specifically in periods of geopolitical conflicts and other crucial events that affect their prices and volatility. First, regarding the challenges of forecasting a *lower Hurst exponent* in the post-invasion period could indicate that traditional forecasting models based on past trends would be less reliable. The energy market's behavior would be dominated by short-term reactions to events, making it challenging to predict prices. For risk management, energy companies, traders, and policymakers would need to adapt their risk management strategies in response to the *lower Hurst exponent*. This might involve hedging against extreme price fluctuations, diversifying energy sources, or increasing reserves to buffer against supply disruptions. Similarly, for *policy and regulation,* governments and regulatory bodies might use the *Hurst exponent to monitor market stability and intervene*, when necessary, for example, by implementing price caps (as in the case of Greece, in the period of energy crisis), releasing strategic reserves, or supporting alternative energy sources to reduce volatility. Finally, for *investment decisions***,** investors could use the Hurst exponent to assess the risk of energy-related investments during this period. A lower Hurst exponent would signal a higher-risk environment, potentially leading to more conservative investment strategies or a focus on assets less correlated with energy prices. As a conclusion, the turmoil in energy prices due to the Russian invasion of Ukraine can be explained through the lens of the Hurst exponent by showing how the invasion disrupted market stability and increased volatility. A shift in the Hurst exponent from a stable, persistent value (H > 0.5) to a more volatile, anti-persistent value (H < 0.5) reflects the significant impact of geopolitical shocks on energy markets, highlighting the challenges in forecasting and managing risks during such turbulent times.

3. ***Rolling Mutual Information (MI) and rolling PMIME Analysis***: we computed the rolling MI and PMIME between electricity prices and other relevant variables, such as TTF natural gas price, the USD/RUB exchange rate, an economic indicator, in the era of intense geopolitical event. This analysis helped us to ***identify the 'causalities' (strength and direction of relationships) between all the factors (electricity- gas prices and exchange rate), considered in this work, during the crisis.*** We showed ***how changes in MI and PMIME patterns have highlighted significant correlations, dependencies, or causal relationships that influence the dynamics of the analyzed markets, during the crisis.***

By integrating these above techniques, we have gained deeper insights into various aspects of the European energy crisis of 2022, as follows:

- ***by identifying new 'hidden' critical events (breakpoints) occurred at different dates from the ones of the 'known' E1-E13 critical events, elicited from literature,*** *reflecting* the degree of responsiveness of each market, as a manifestation of their structural characteristics.

- ***by assessment of Market Efficiency at the breakpoints***: By analyzing changes in the rolling Hurst exponent alongside structural breakpoints detection, it has provided to us insights into shifts in *market efficiency* (a EMH testing), that is related to *transparency, and liquidity during the crisis*. This assessment has helped us to make assumptions, indirectly, about *the effectiveness of market mechanisms, regulatory interventions, or policy responses in addressing the crisis's underlying causes.*

- ***by understanding Causal Relationships via hot maps and PMIME network graphs***: Rolling MI and PMIME analysis enabled us in exploring the *causal relationships and dependencies* between electricity and gas prices and USD/RUB exchange rate variables, considered significant in this work in contributing -driving the crisis. By identifying significant partial mutual correlations or information flow between these variables, we better understood ***the complex interactions and feedback mechanisms shaping electricity, gas and financial markets' dynamics during the 2022 crisis.***

As a conclusion, we can state that integrating structural breakpoints detection, rolling Hurst exponent analysis, and rolling MI and PMIME analysis has provided us a powerful framework for examining the European energy crisis of 2022 from multiple perspectives, including identifying new ('hidden') or reassuring known (from literature) critical events, assessing market efficiency at these breakpoints, and understanding causal relationships between the different factors, considered in this work, influencing energy market outcomes.

The main contribution of this paper is the combination of three methodologies in shedding light on how the nonlinear interactions of several critical energy and financial factors, assumed to be significant, have influenced the development of the 2022 energy crisis, by focusing on the structural anomalies occurred by critical events related to the Russo-Ukrainian war. We believe that the results of the paper could be useful in the design of future ***European Energy Policies***, to prevent a similar crisis as in 2022. We believe that integrating tools as the above-mentioned approaches into the design of European energy policies can enhance policymakers' ability to prevent or mitigate crises like the one experienced in 2022. We describe how such a tool could be utilized:

1. ***Early Warning System***: The combination of structural breakpoints detection, rolling Hurst exponent analysis, and rolling MI and PMIME can serve as an early warning system for identifying emerging vulnerabilities or instabilities in energy markets. By continuously monitoring key indicators and patterns, policymakers can detect potential crisis triggers and take proactive measures to address underlying issues before they escalate into full-blown crises.

2. ***Identifying Structural Changes***: *Structural breakpoints detection* helps identify periods of significant structural changes in energy markets, such as shifts in supply-demand dynamics, regulatory interventions, or crucial geopolitical events, as the Russian invasion in Ukraine. By understanding the timing and nature of these changes, policymakers can tailor policy responses to address specific challenges or vulnerabilities in the energy system.

3. **Assessing Market Efficiency and Resilience**: Rolling Hurst exponent analysis provides insights into the efficiency, predictability, and resilience of energy markets over time. By assessing changes in market dynamics and persistence, policymakers can gauge the effectiveness of existing policies, regulatory frameworks, and market mechanisms in promoting market stability and mitigating risks.

4. ***Understanding Interdependencies***: Rolling MI and PMIME analysis helps policymakers understand the complex interdependencies and causal relationships between different factors influencing energy market outcomes. By quantifying the strength and direction of relationships between variables (as the PMIME network graph and hot maps) policymakers can identify potential sources of systemic risk, market failures, or unintended consequences of policy interventions.

5. ***Policy Impact Assessment***: Integrating these analyses into policy design enables policymakers to assess the potential impact of *proposed policy measures on energy market dynamics and resilience*. By simulating different policy scenarios and their effects on *structural breakpoints, Hurst exponent values, and Mutual Information patterns*, policymakers can identify *optimal policy strategies that enhance market efficiency, resilience, and sustainability.*

6. ***Dynamic Policy Adjustment***: The tool can facilitate dynamic policy adjustment and fine-tuning in response to changing market conditions, emerging risks, or unexpected events. By continuously monitoring *structural breakpoints, Hurst exponent trends, and Mutual Information dynamics*, policymakers can adapt policy interventions in real-time to address evolving challenges and prevent crises from escalating.

Overall, integrating structural breakpoints detection, rolling Hurst exponent analysis, and rolling MI-PMIME analysis into the design of European energy policies offers a *data-driven approach* to enhancing market efficiency, resilience, and sustainability. *By leveraging these analytical tools, policymakers can proactively identify vulnerabilities, tailor policy responses, and mitigate risks to prevent future energy crises like the one experienced in 2022. Toward this target we believe that this work can be very useful and provide insights to other researchers to apply other innovative tools in the analysis of complex phenomena such as the energy crisis of 2022.* This work contributes toward this aim.

## SUPPLEMENTARY MATERIAL A : Rolling Hurst of Energy Markets

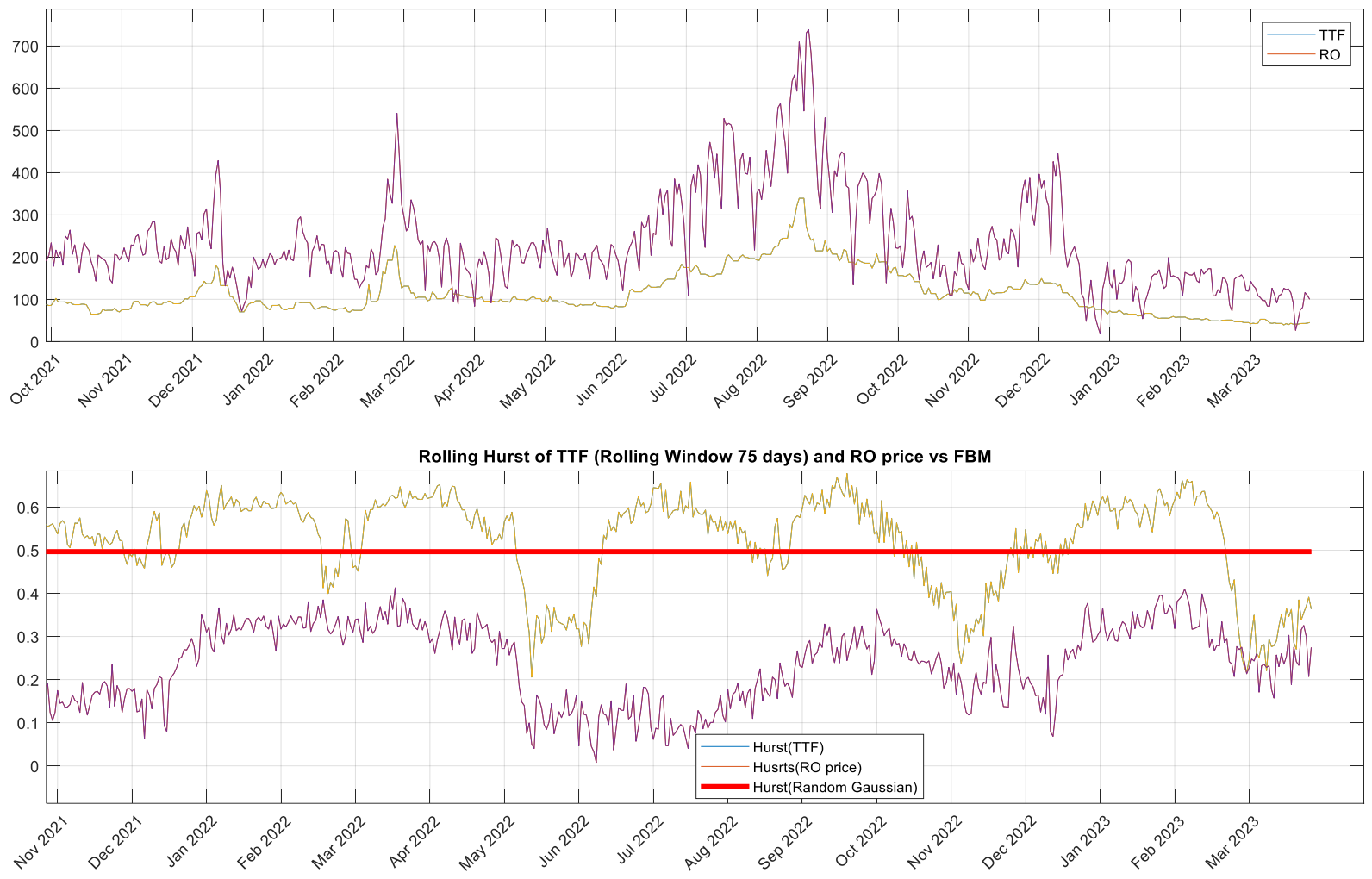

**Figure SM-A.1 Rolling Hurst (75 days window) of Romanian (RO) electricity price.**

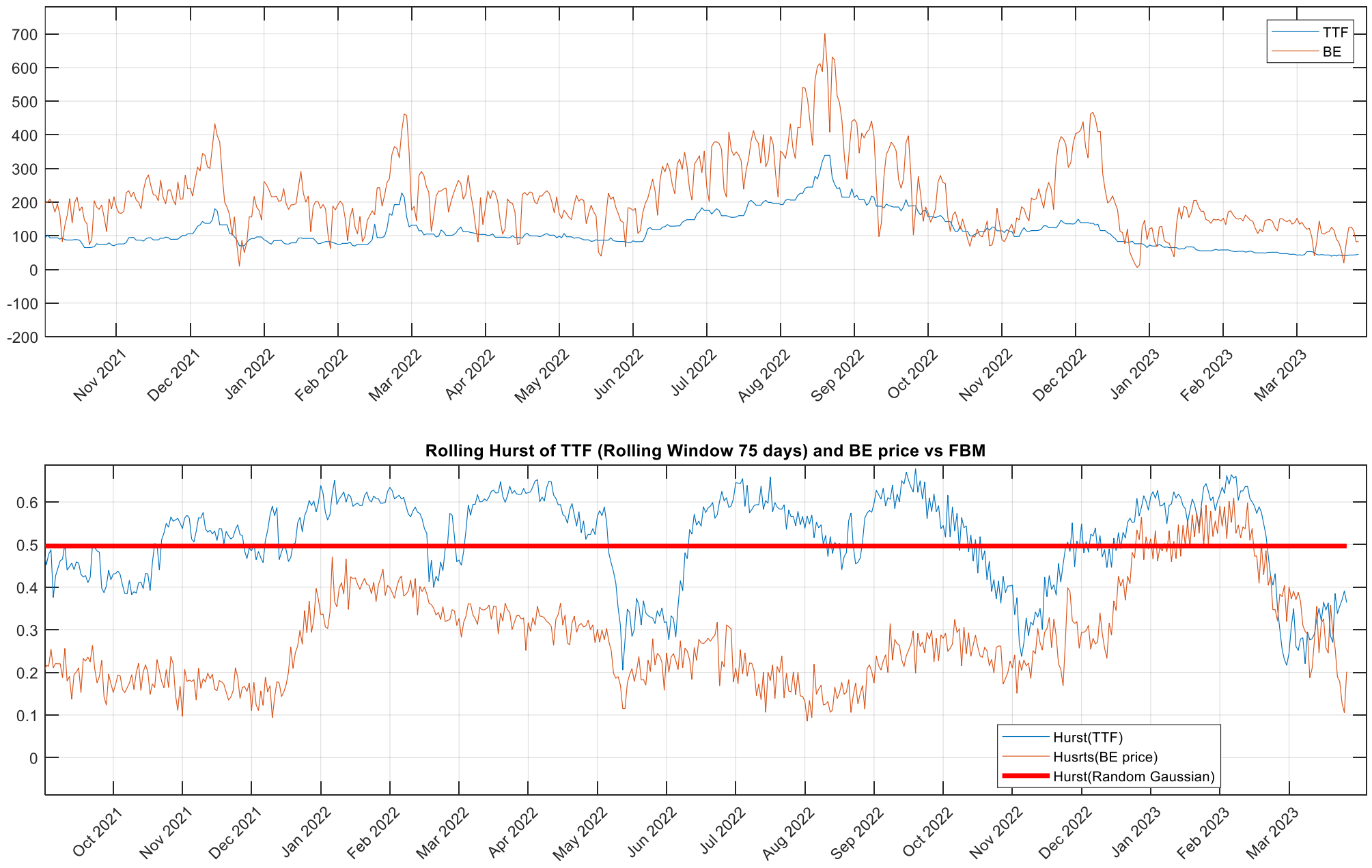

**Figure SM-A.2 Rolling Hurst (75 days window) of Belgian (BE) electricity price.**

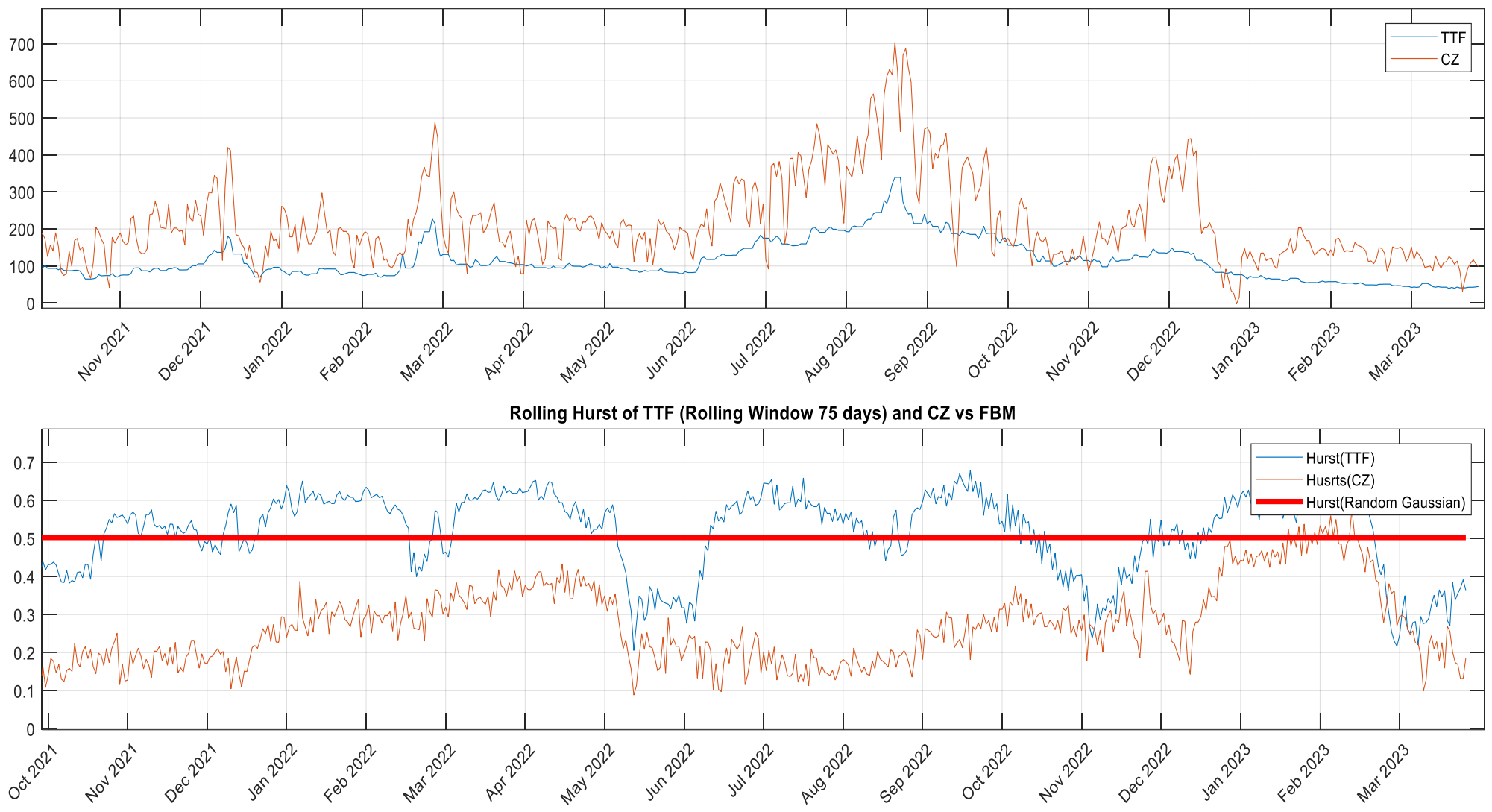

# Figure SM-A.3 Rolling Hurst (75 days window) of CZ electricity price.

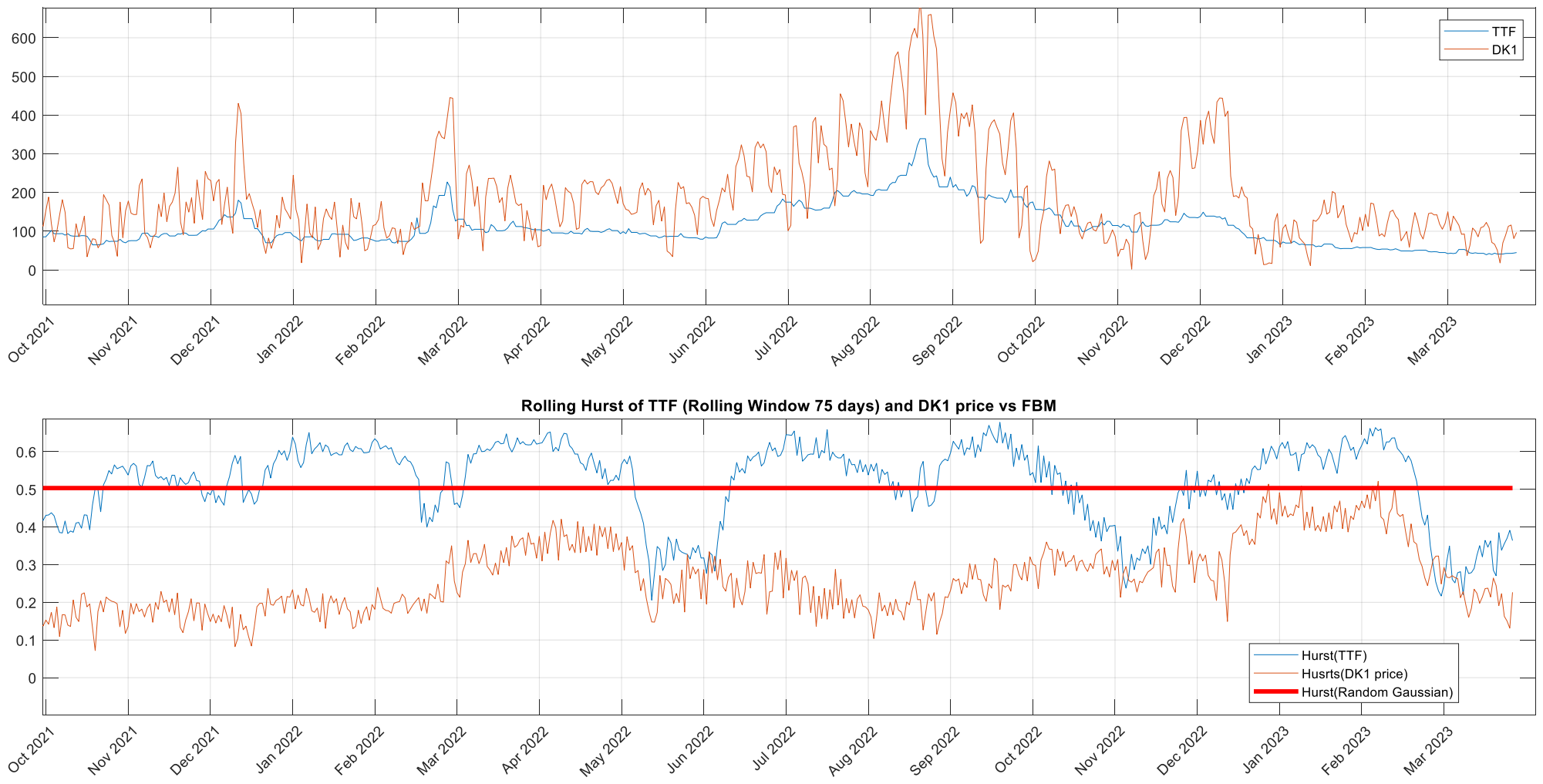

# Figure SM-A.4 Rolling Hurst (75 days window) of DK1 electricity price.

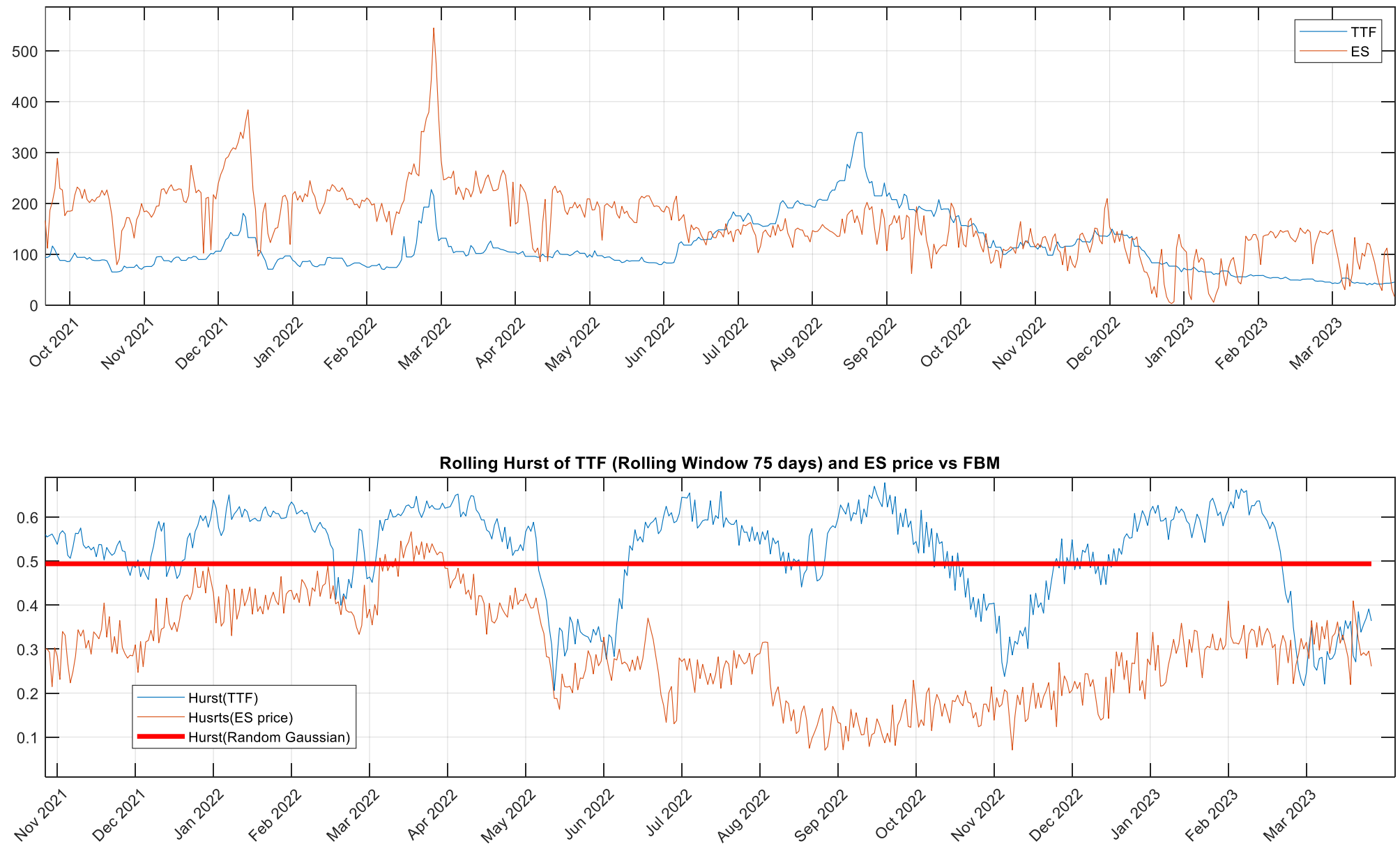

**Figure SM-A.5 Rolling Hurst (75 days window) of Spanish (ES) electricity price.**

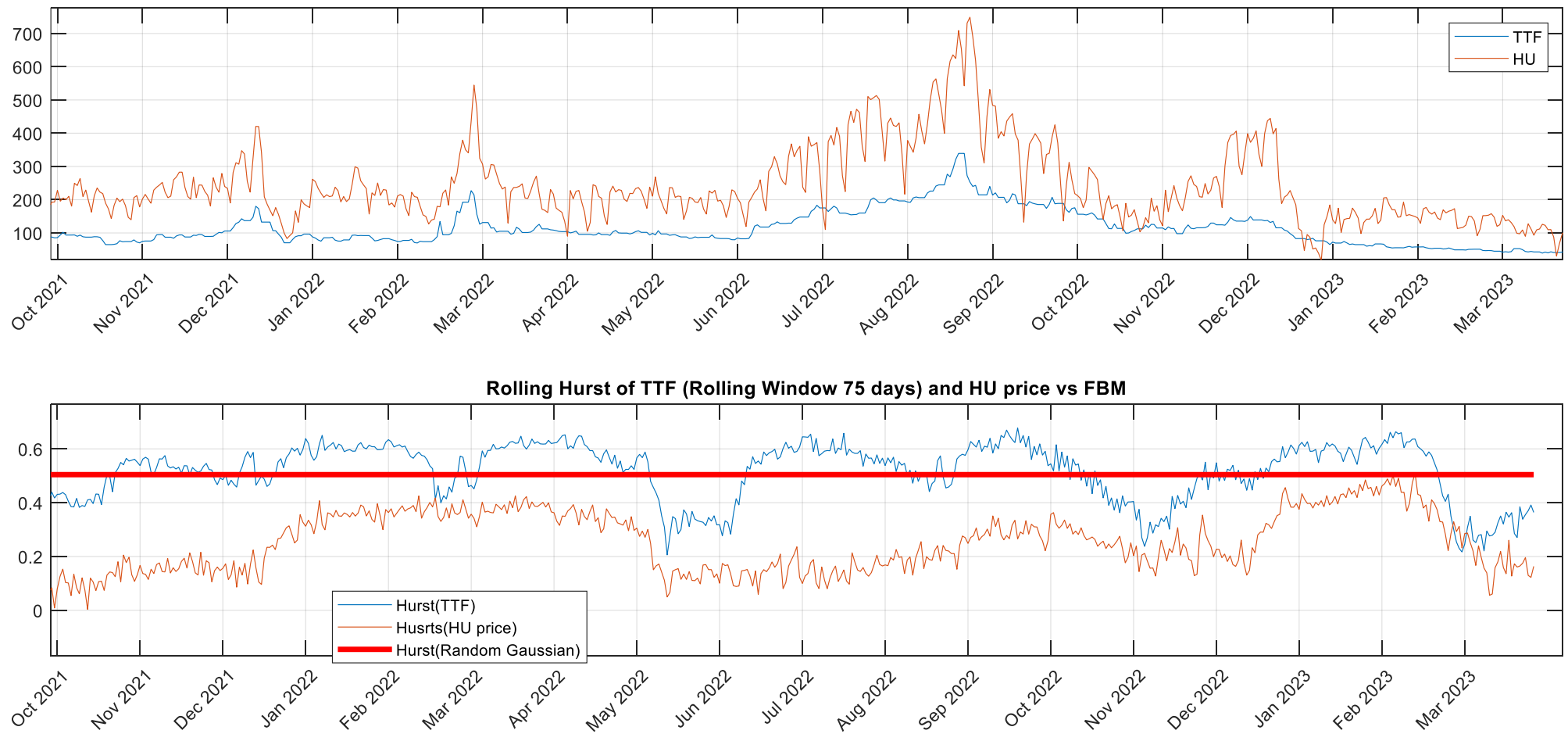

**Figure SM-A.6 Rolling Hurst (75 days window) of Hungarian (HU) electricity price.**

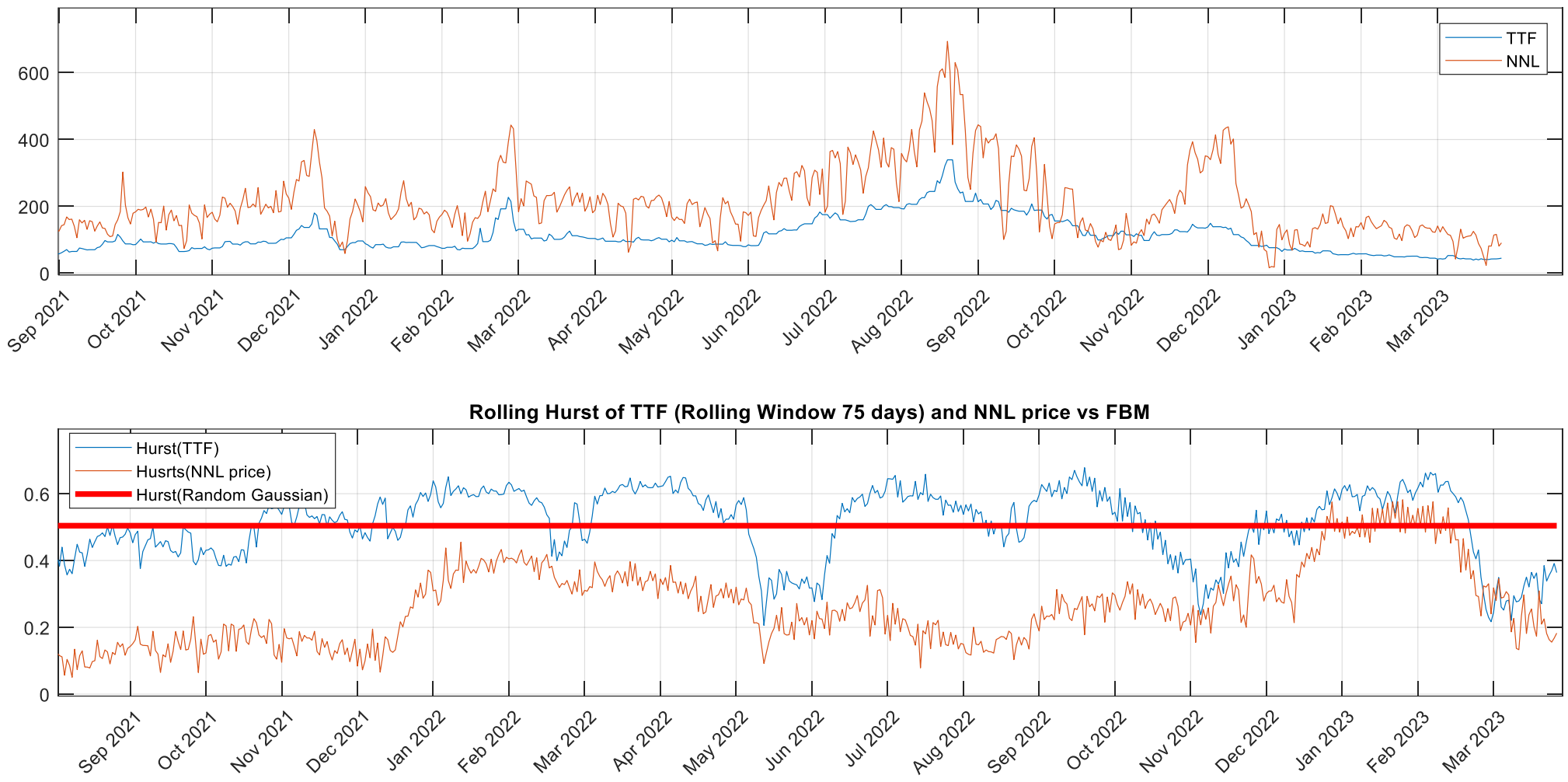

**Figure SM-A.7 Rolling Hurst (75 days window) of NNL electricity price.**

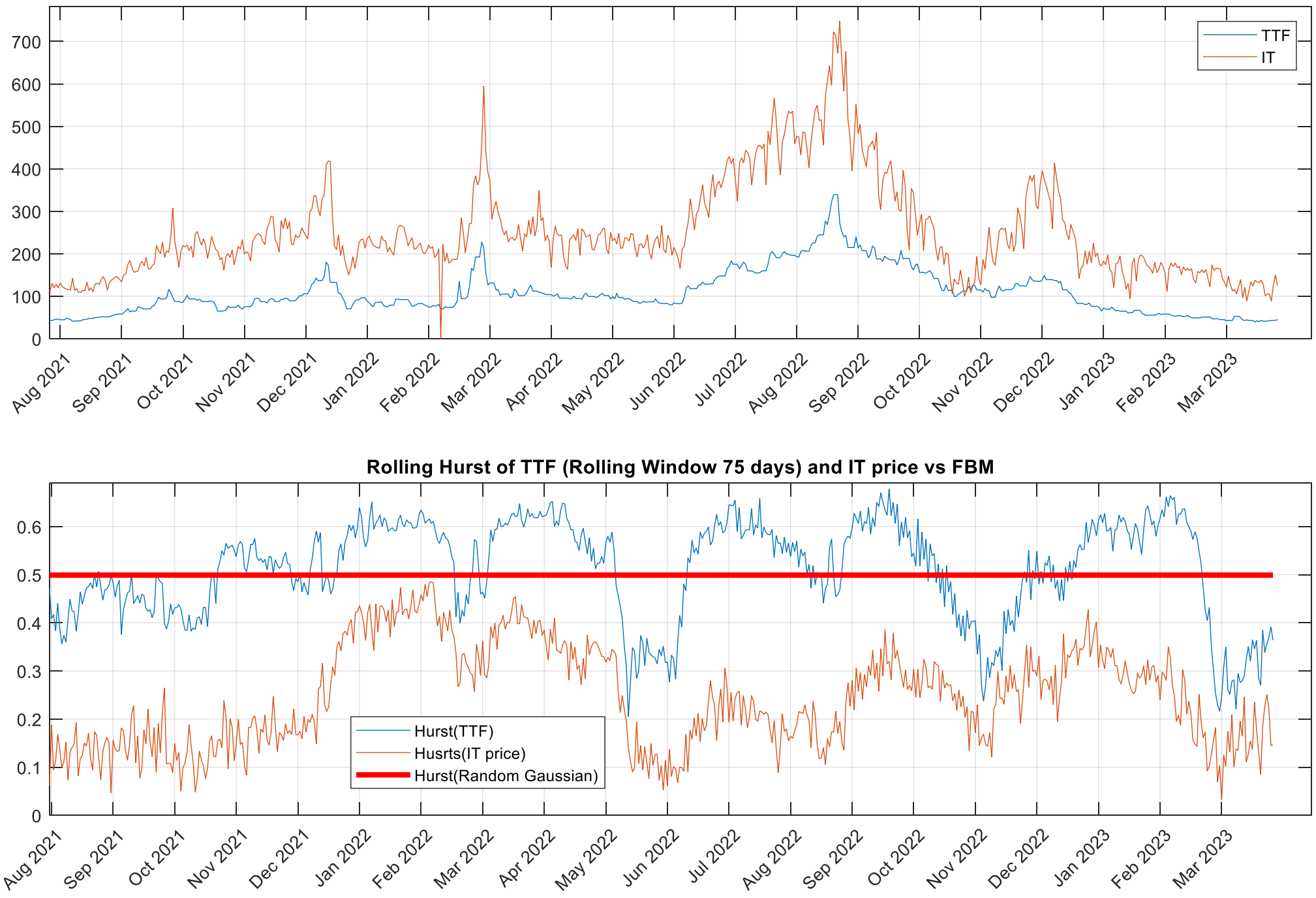

**Figure SM-A.8 Rolling Hurst (75 days window) of Italian (IT) electricity price.**

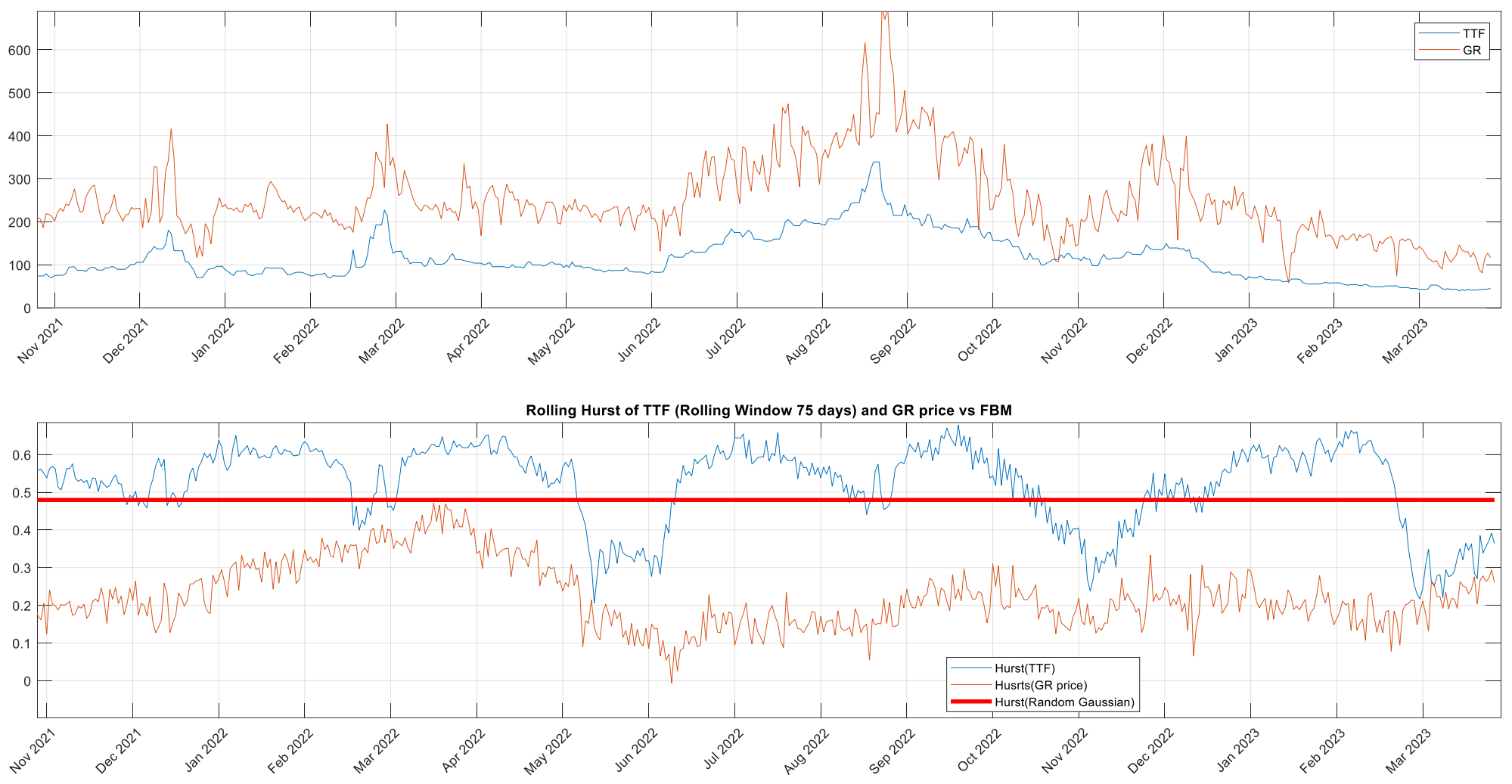

# Figure SM-A.9 Rolling Hurst (75 days window) of Greek (GR) electricity price.

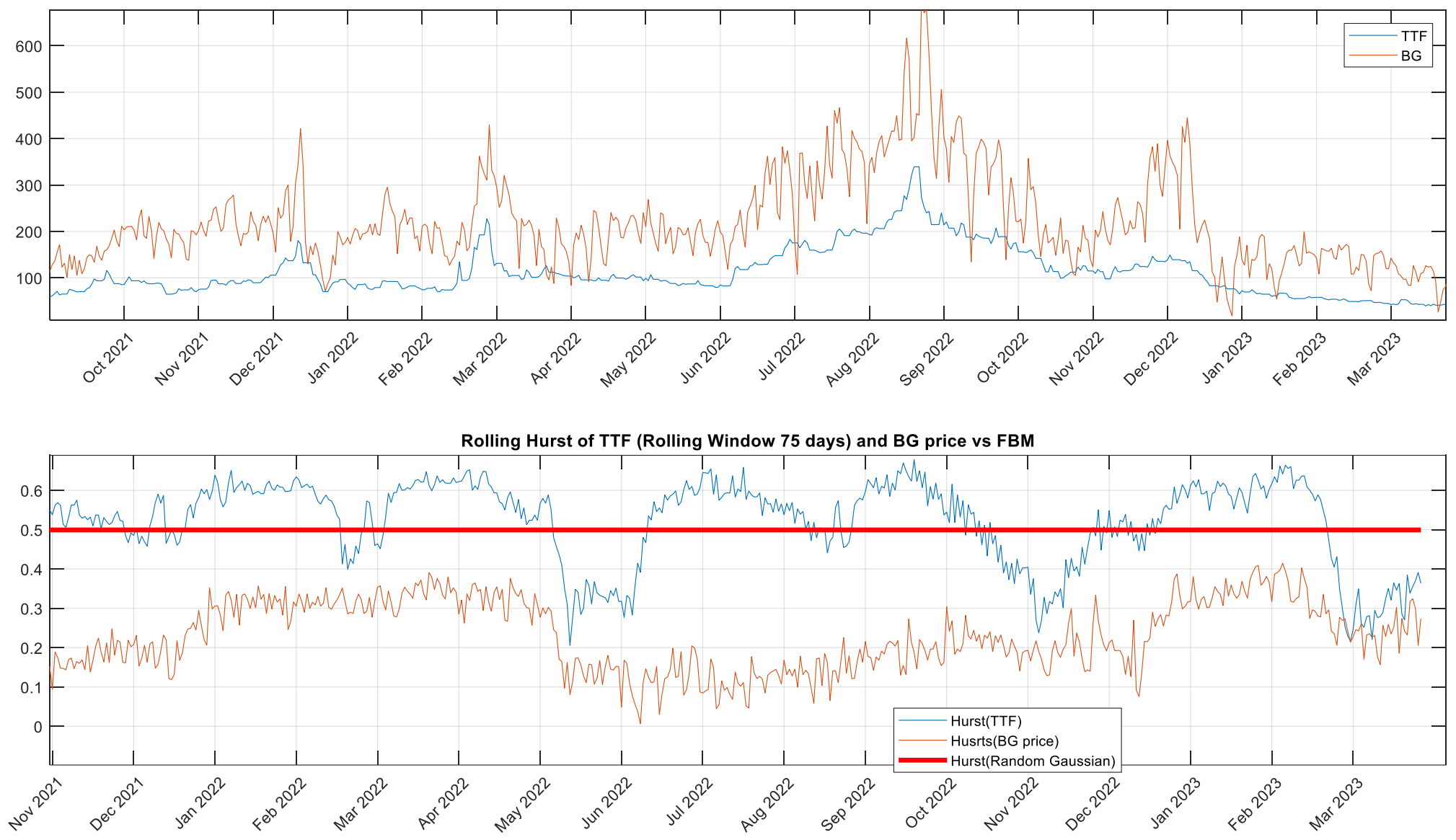

# Figure SM-A.10 Rolling Hurst (75 days window) of Bulgarian (BG) electricity price.

# SUPPLEMENTARY MATERIAL A
## PMIME analysis: Heat maps and Network graphs of all markets

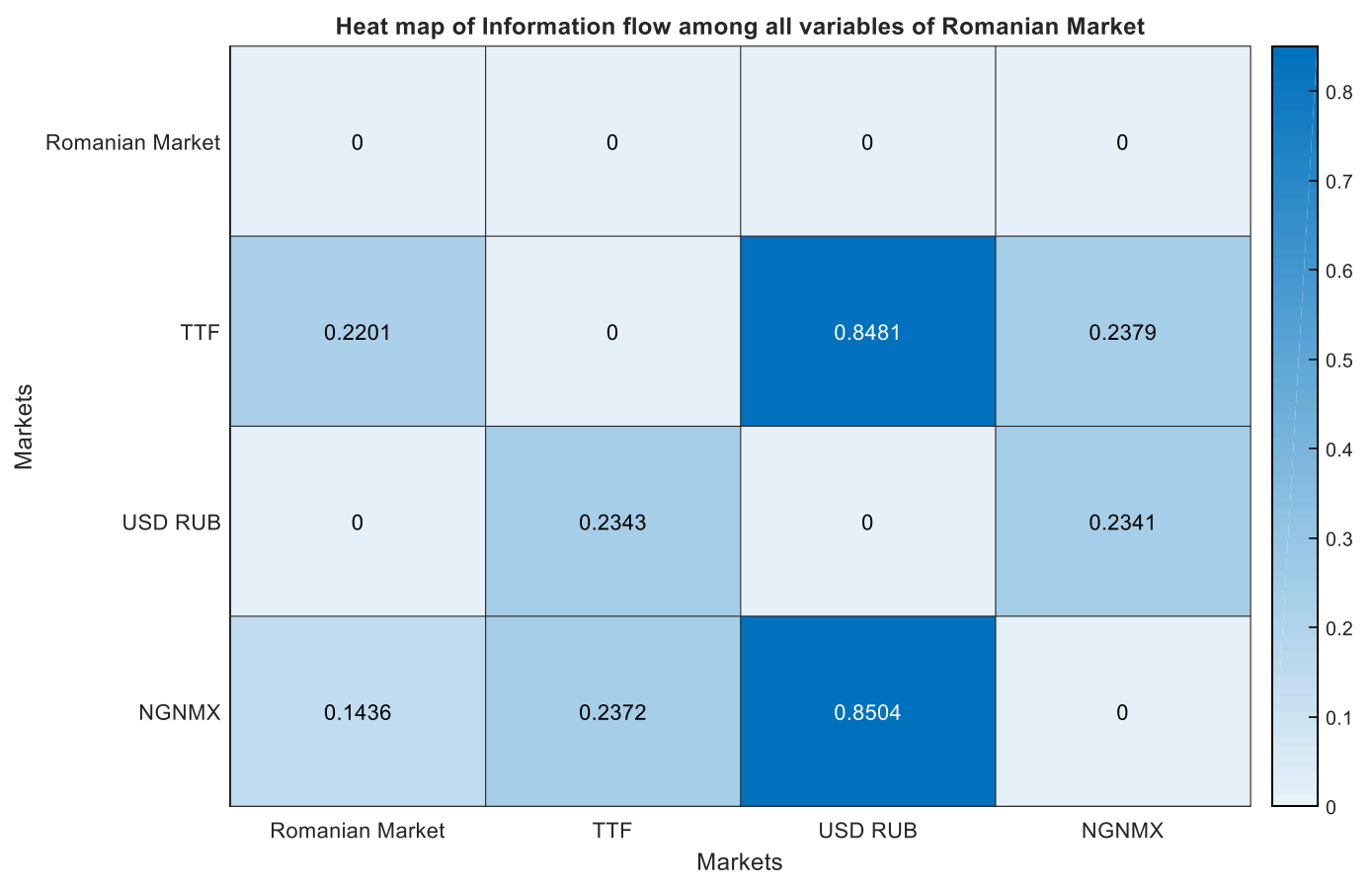

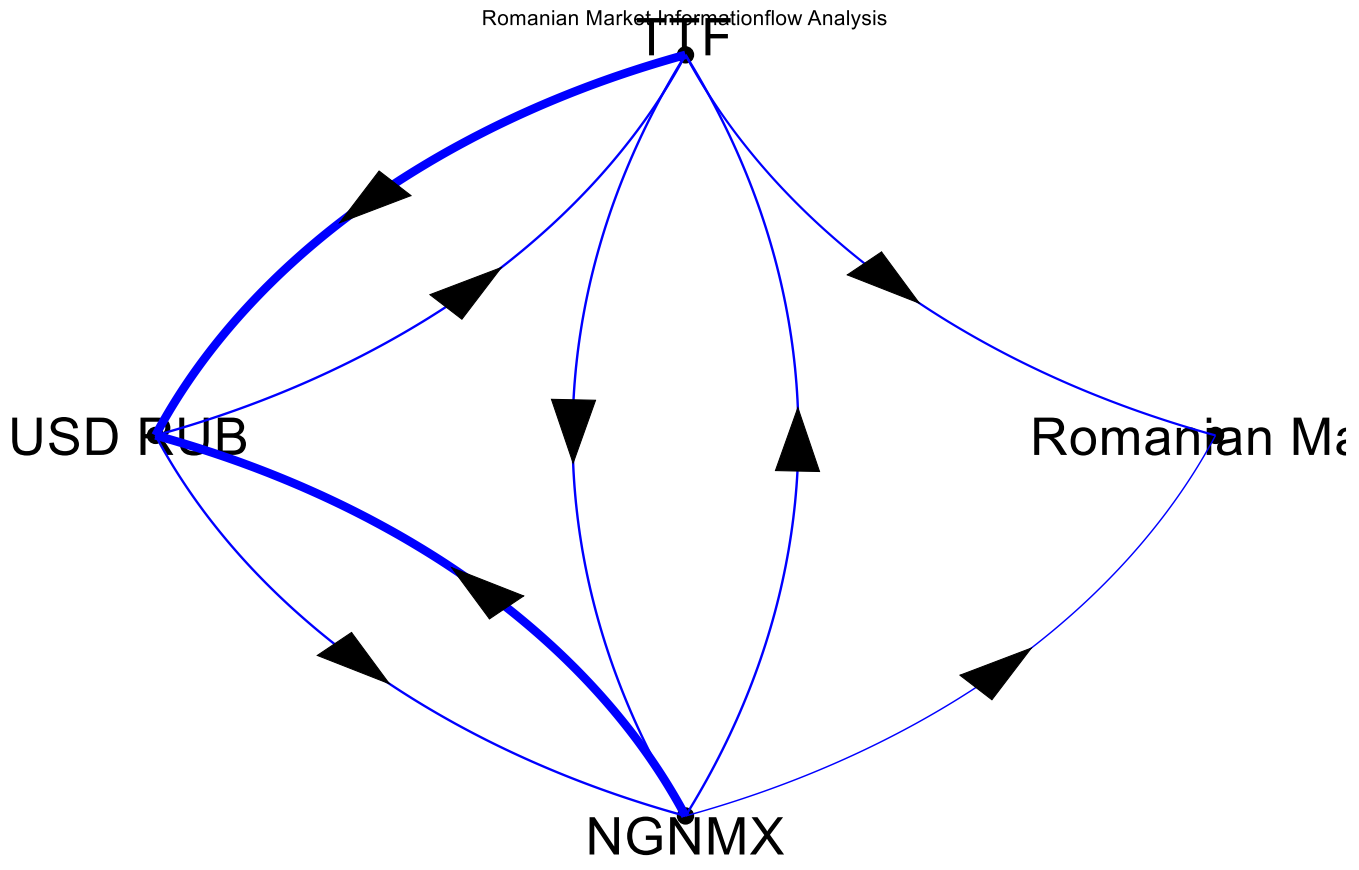

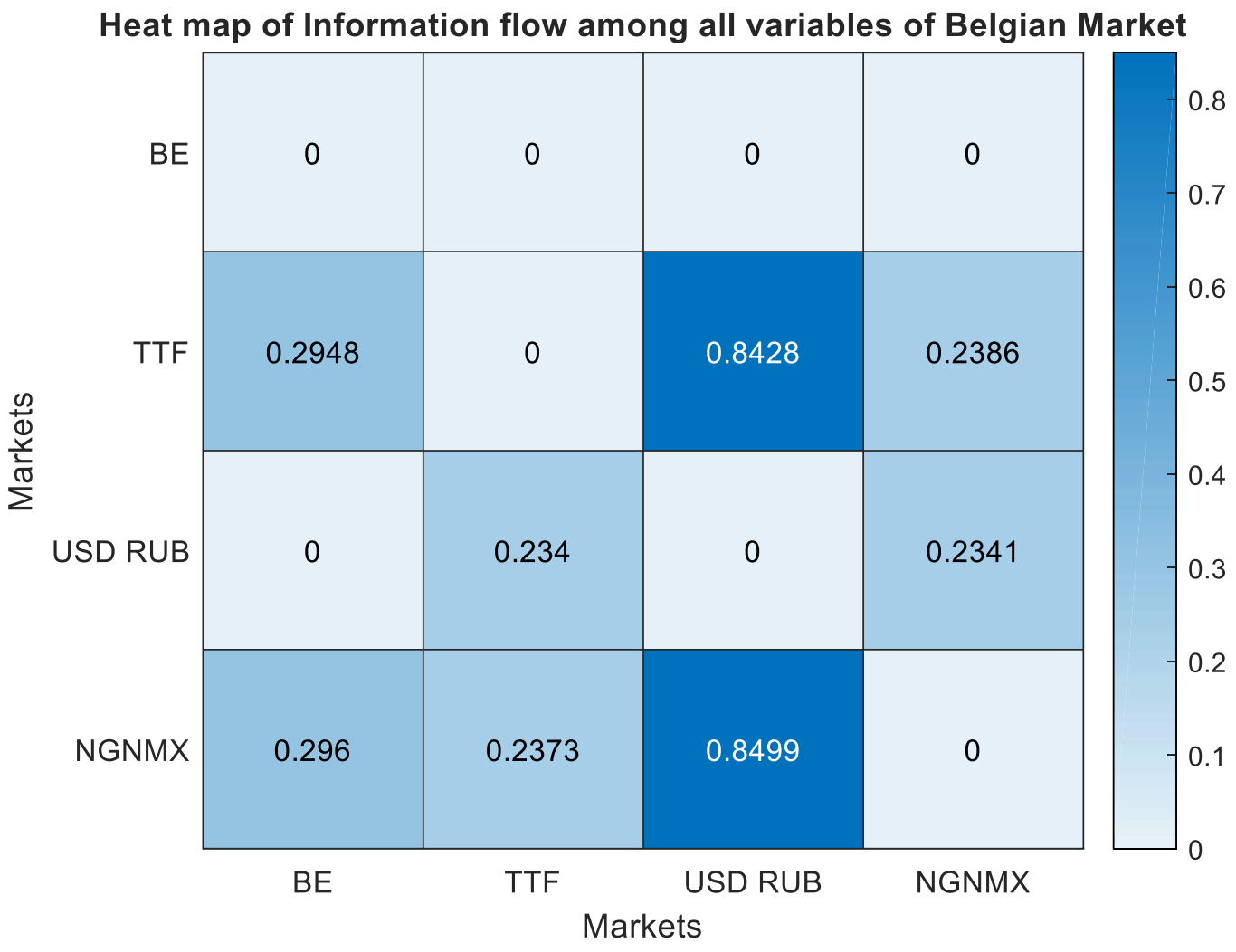
Heat map of Information flow among all variables of Belgian Market
Markets
BE
TTF
USD RUB
NGNMX
0
0
0
0
0.2948
0
0.8428
0.2386
0
0.234
0
0.2341
0.296
0.2373
0.8499
0
BE
TTF
USD RUB
NGNMX
Markets
0.8
0.7
0.6
0.5
0.4
0.3
0.2
0.1
0

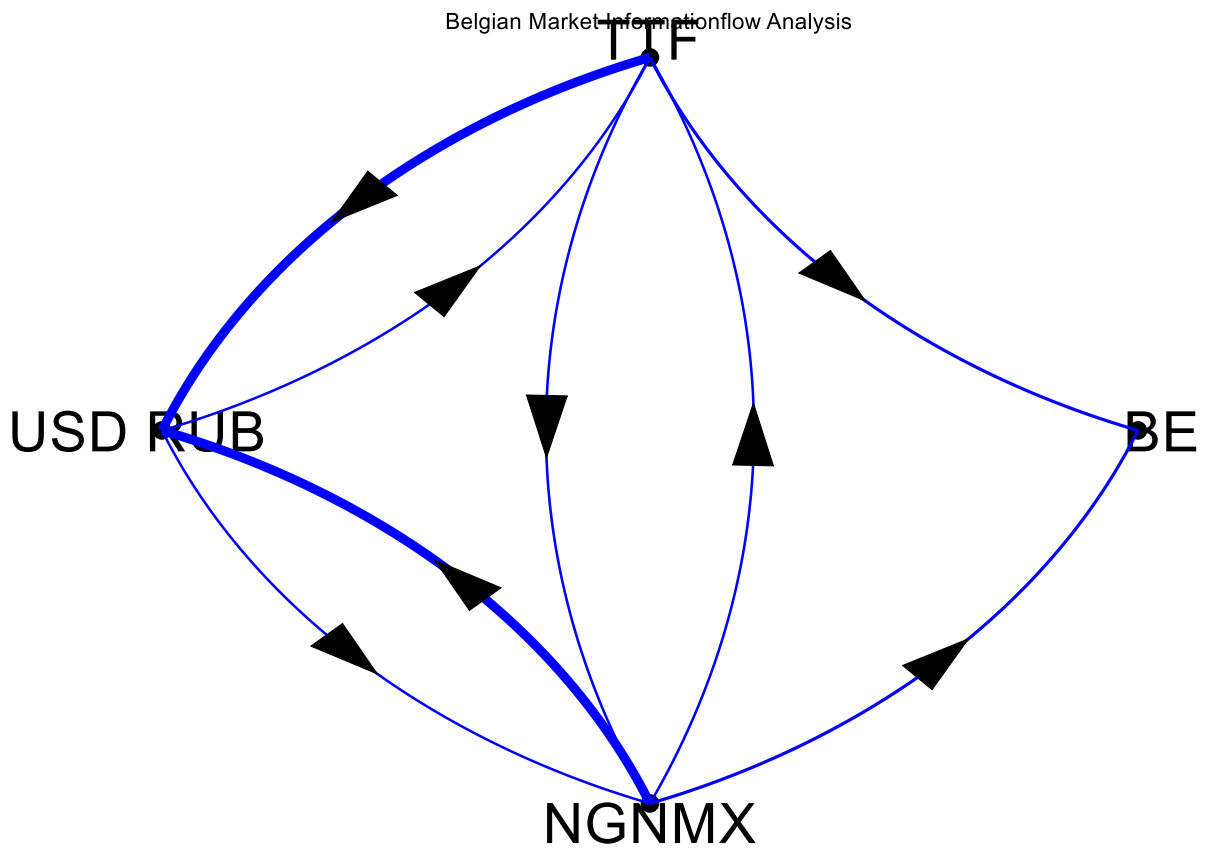
Belgian Market Informationflow Analysis
TTF
USD RUB
BE
NGNMX

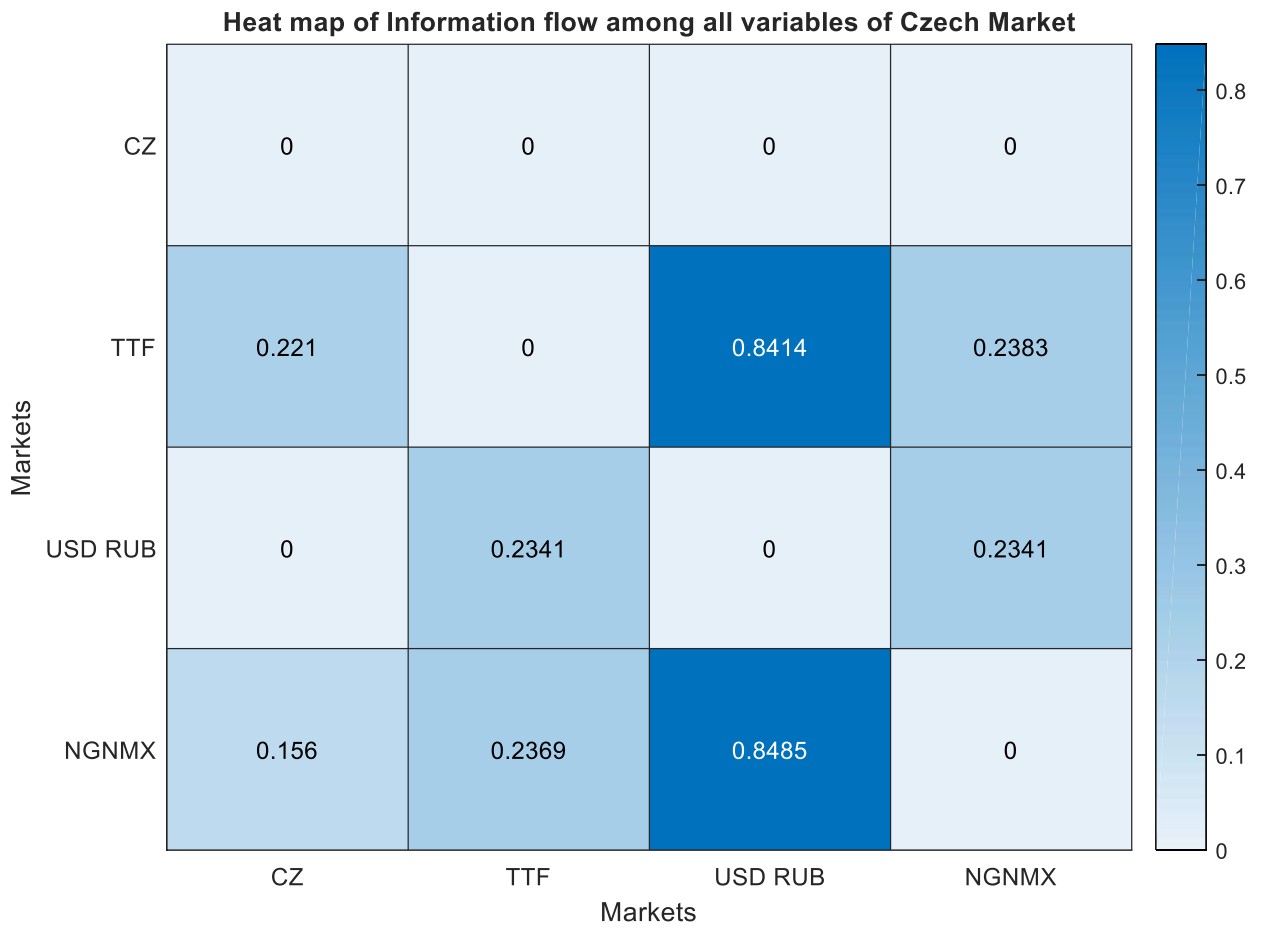
Heat map of Information flow among all variables of Czech Market
Markets
CZ
TTF
USD RUB
NGNMX
0
0.221
0.156
0.2341
0.2369
0.8414
0.8485
0.2383
0.8
0.7
0.6
0.5
0.4
0.3
0.2
0.1

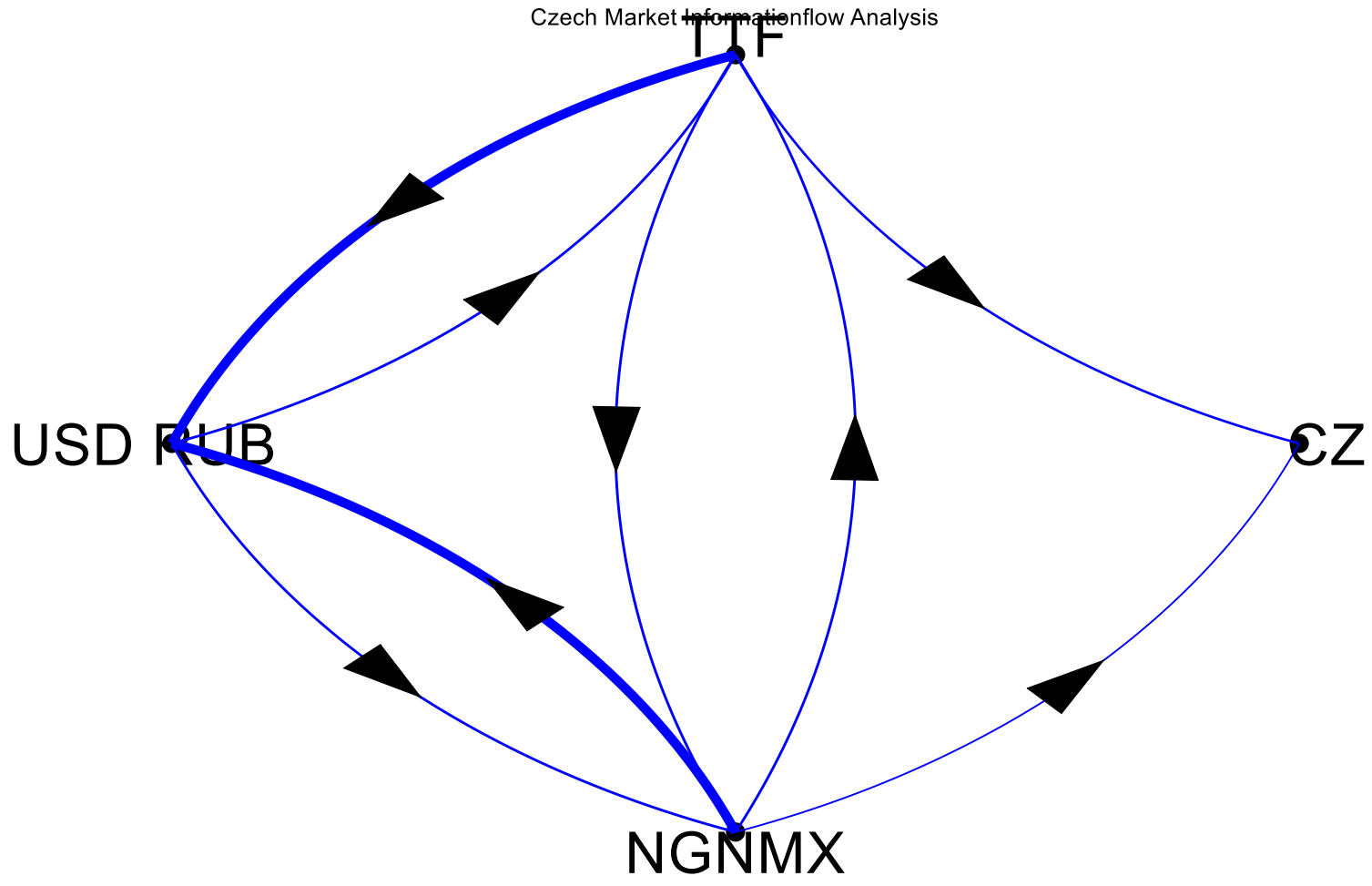
Czech Market Informationflow Analysis
TTF
USD RUB
CZ
NGNMX

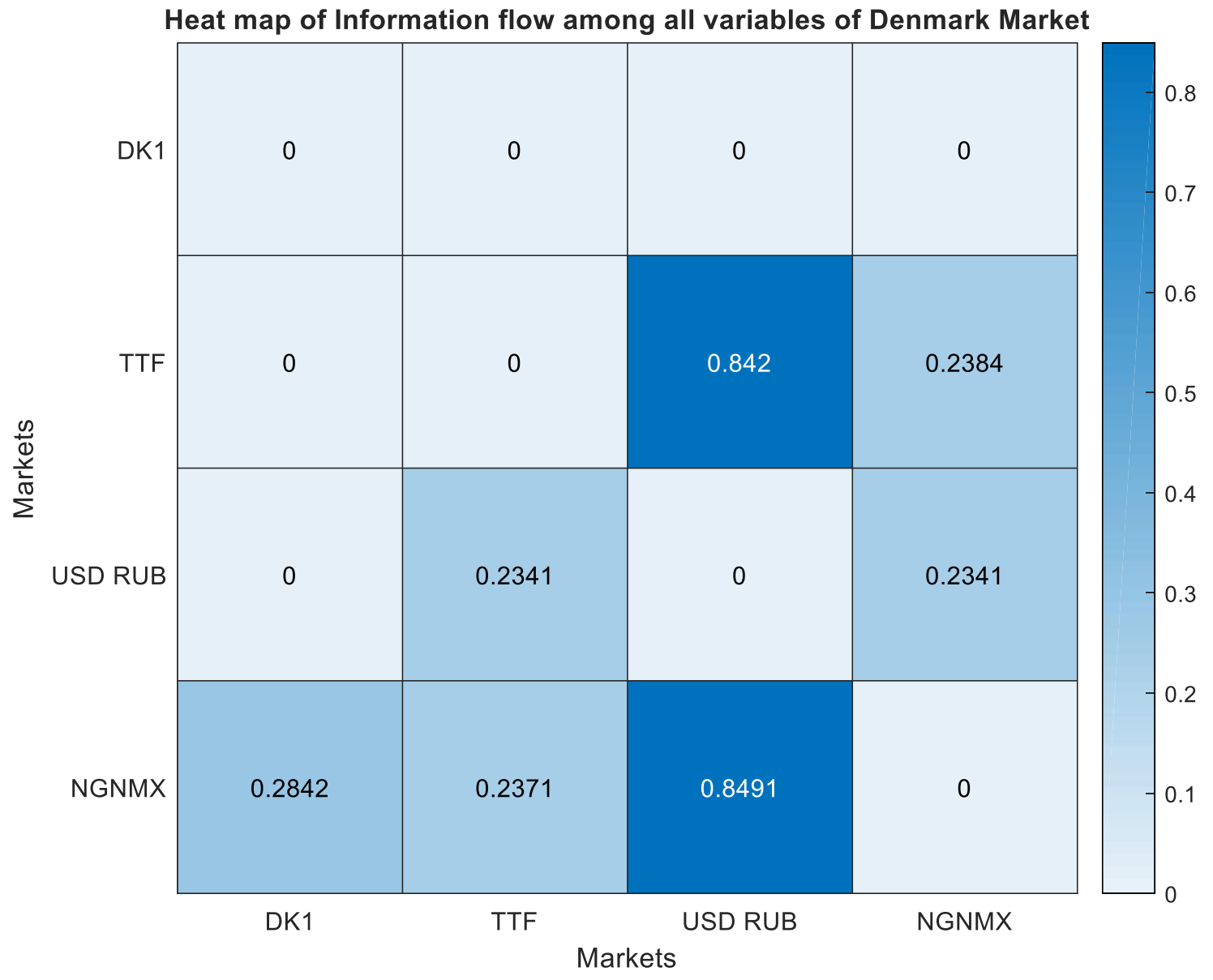
Heat map of Information flow among all variables of Denmark Market
DK1
TTF
USD RUB
NGNMX
Markets
0
0
0
0
0
0
0.842
0.2384
0
0.2341
0
0.2341
0.2842
0.2371
0.8491
0
DK1
TTF
USD RUB
NGNMX
Markets
0.8
0.7
0.6
0.5
0.4
0.3
0.2
0.1
0

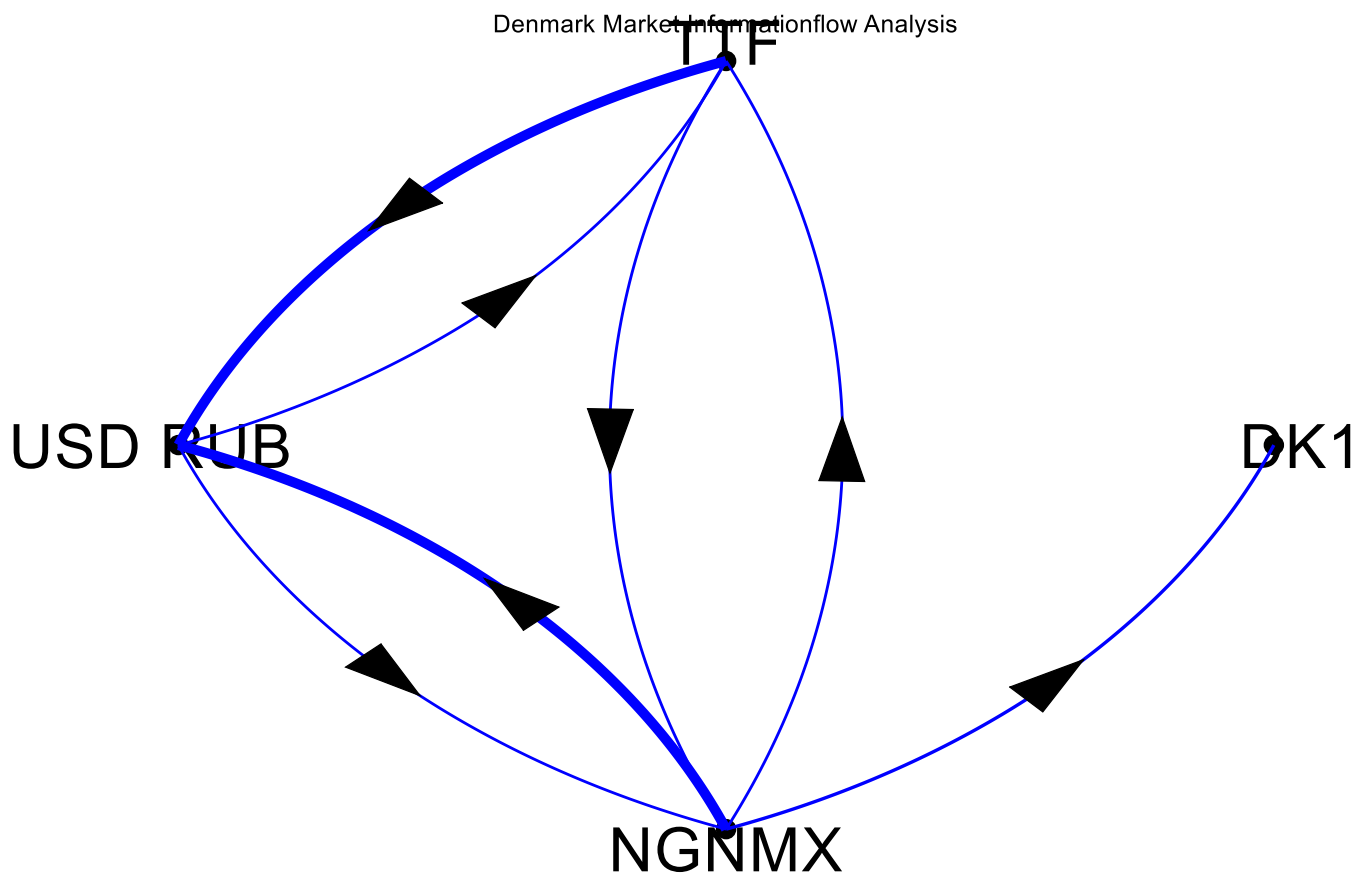
Denmark Market Informationflow Analysis
TTF
USD RUB
DK1
NGNMX

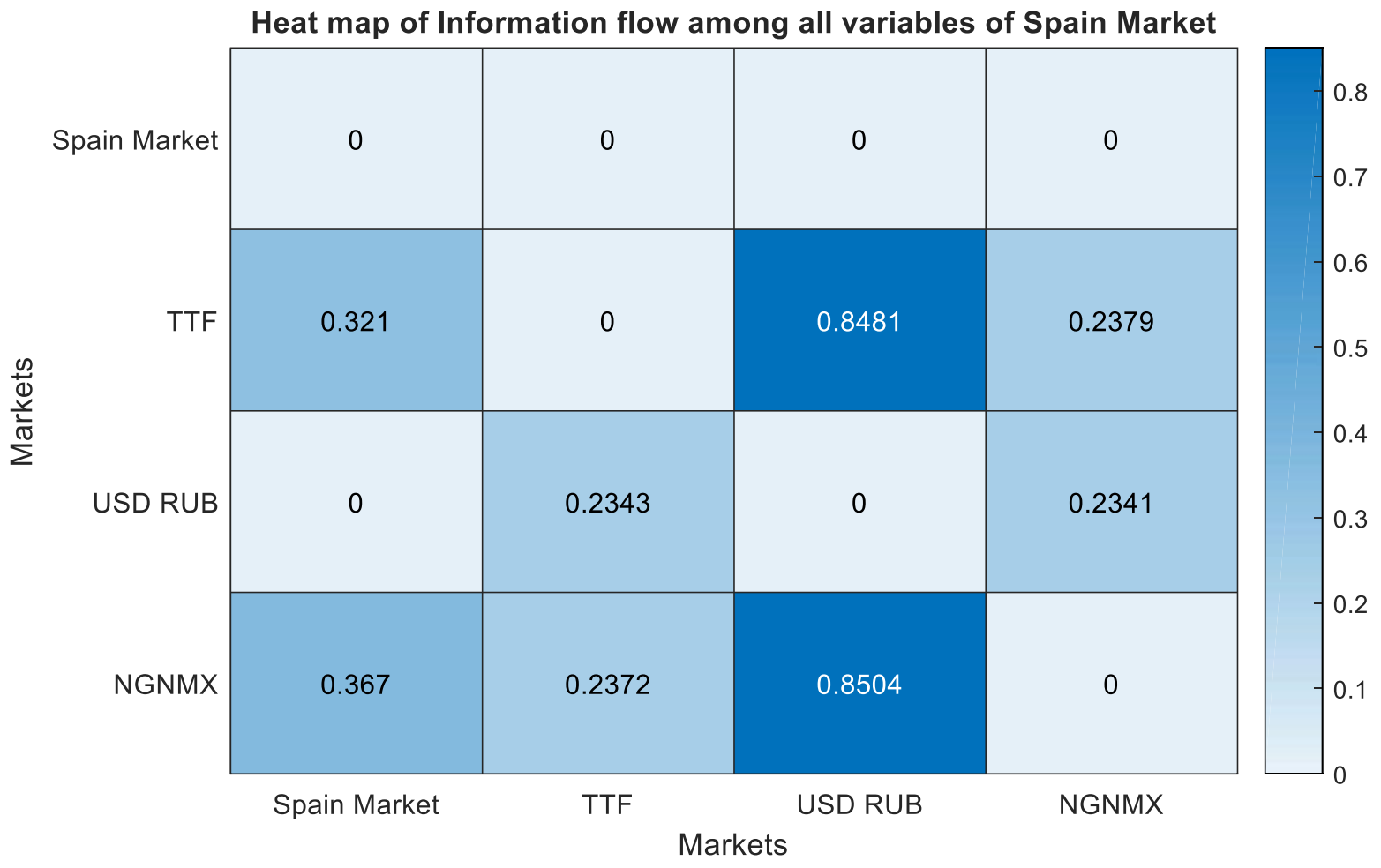
Heat map of Information flow among all variables of Spain Market
Markets
Spain Market
TTF
USD RUB
NGNMX
0 0 0 0
0.321 0 0.8481 0.2379
0 0.2343 0 0.2341
0.367 0.2372 0.8504 0
Spain Market
TTF
USD RUB
NGNMX
Markets
0.8
0.7
0.6
0.5
0.4
0.3
0.2
0.1
0

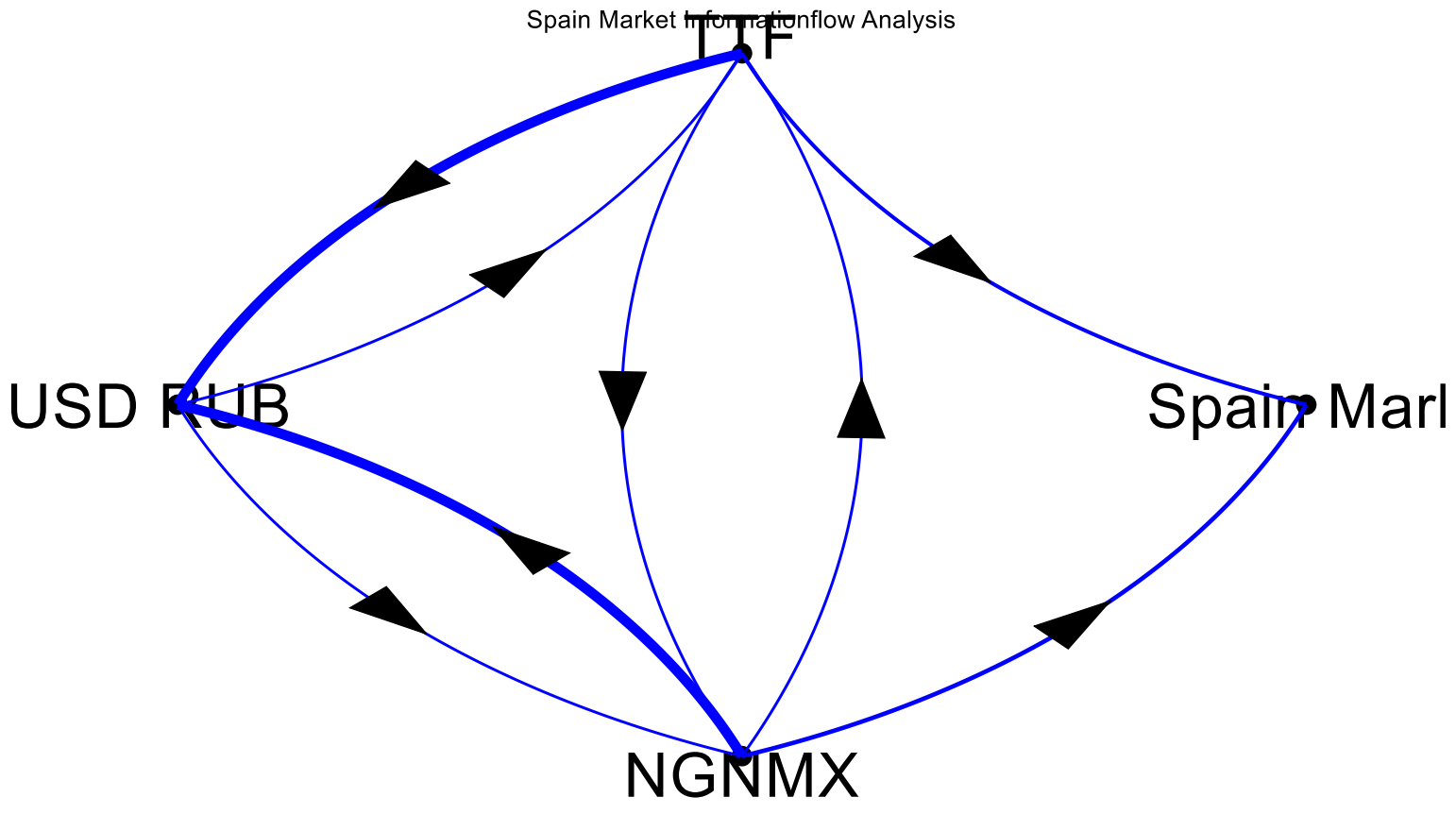
Spain Market Informationflow Analysis
TTF
USD RUB
Spain Marl
NGNMX

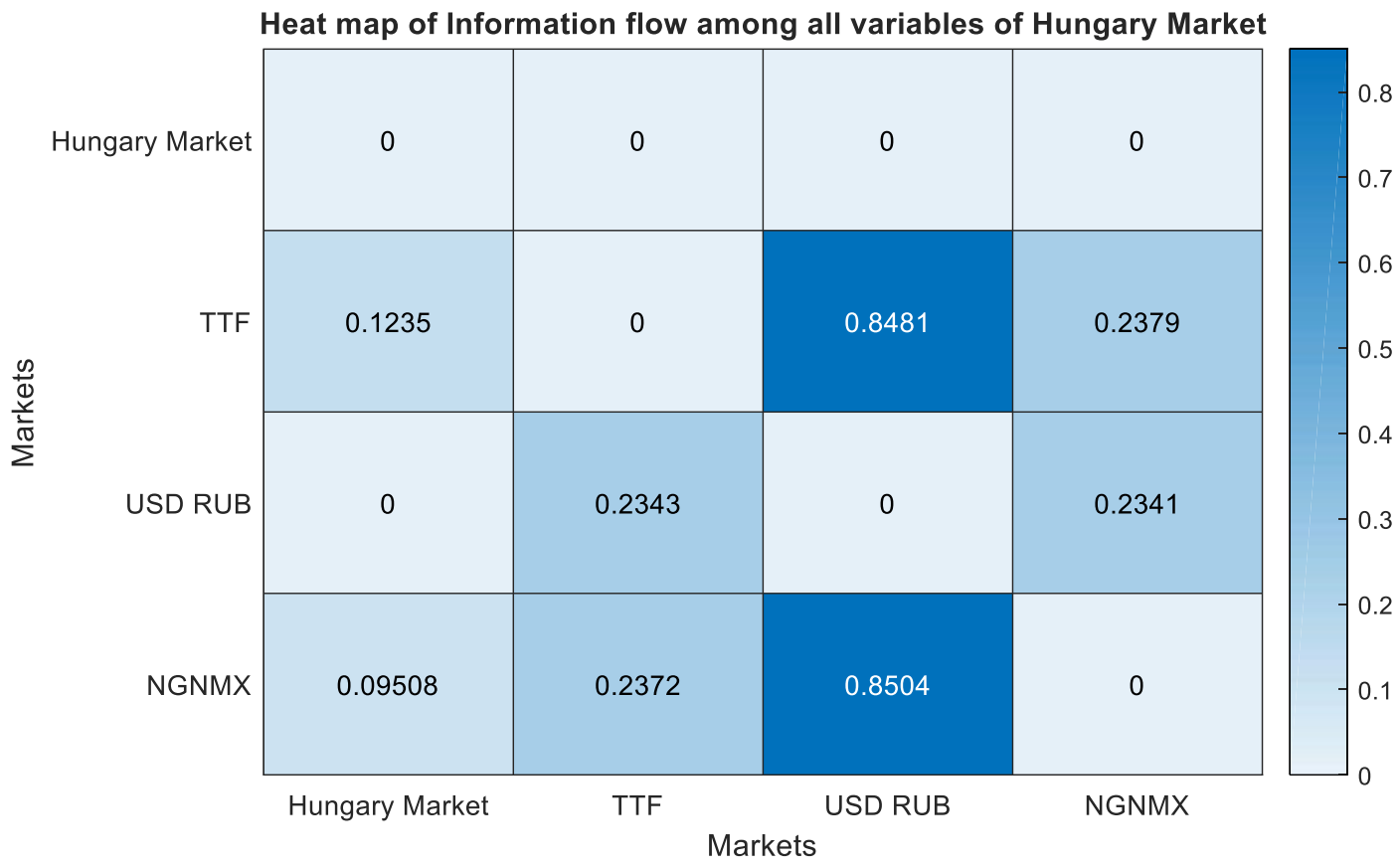
Heat map of Information flow among all variables of Hungary Market
Hungary Market
TTF
USD RUB
NGNMX
Markets
0
0
0
0
0.1235
0
0.8481
0.2379
0
0.2343
0
0.2341
0.09508
0.2372
0.8504
0
Hungary Market
TTF
USD RUB
NGNMX
Markets
0.8
0.7
0.6
0.5
0.4
0.3
0.2
0.1
0

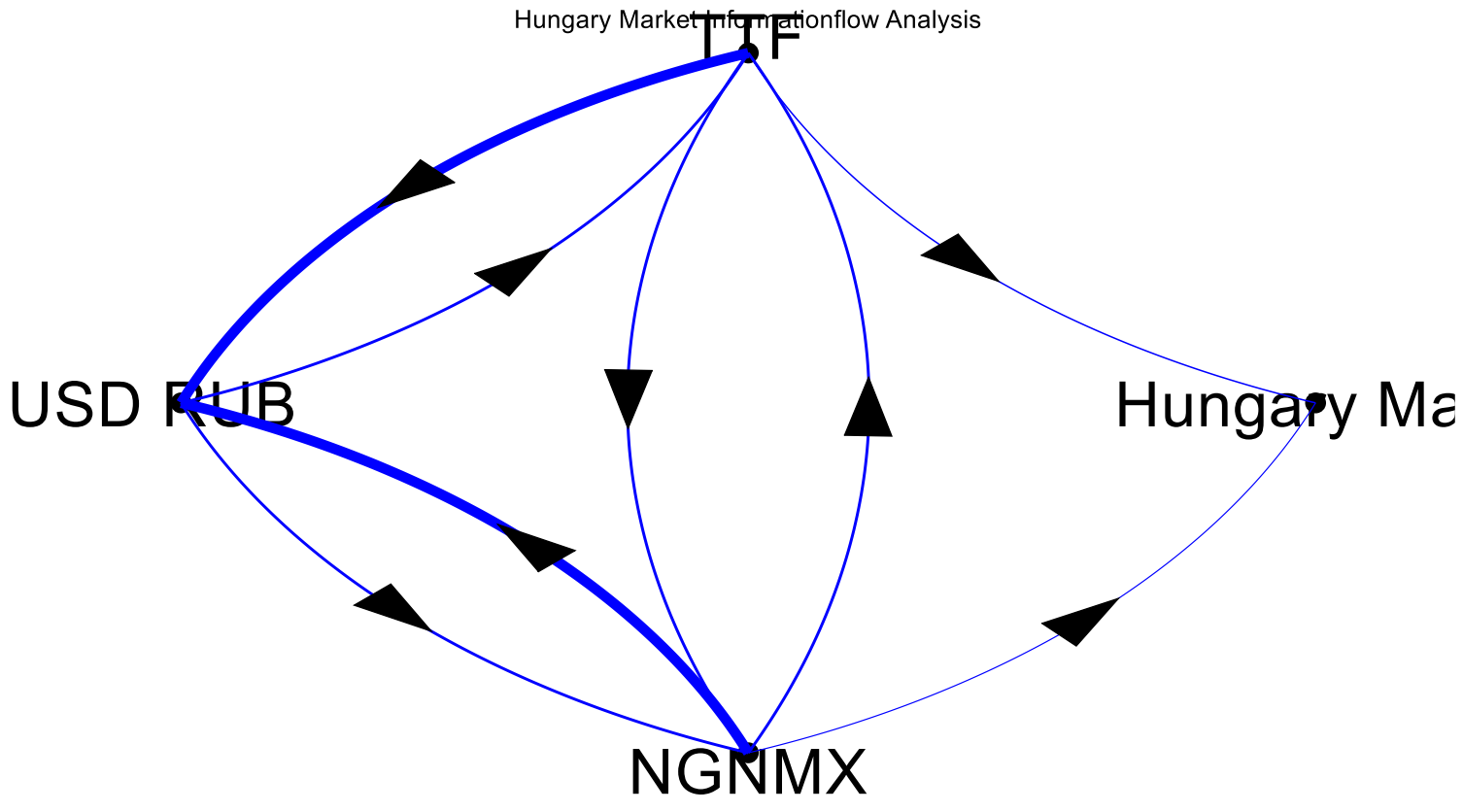
Hungary Market Informationflow Analysis
TTF
USD RUB
Hungary Ma
NGNMX

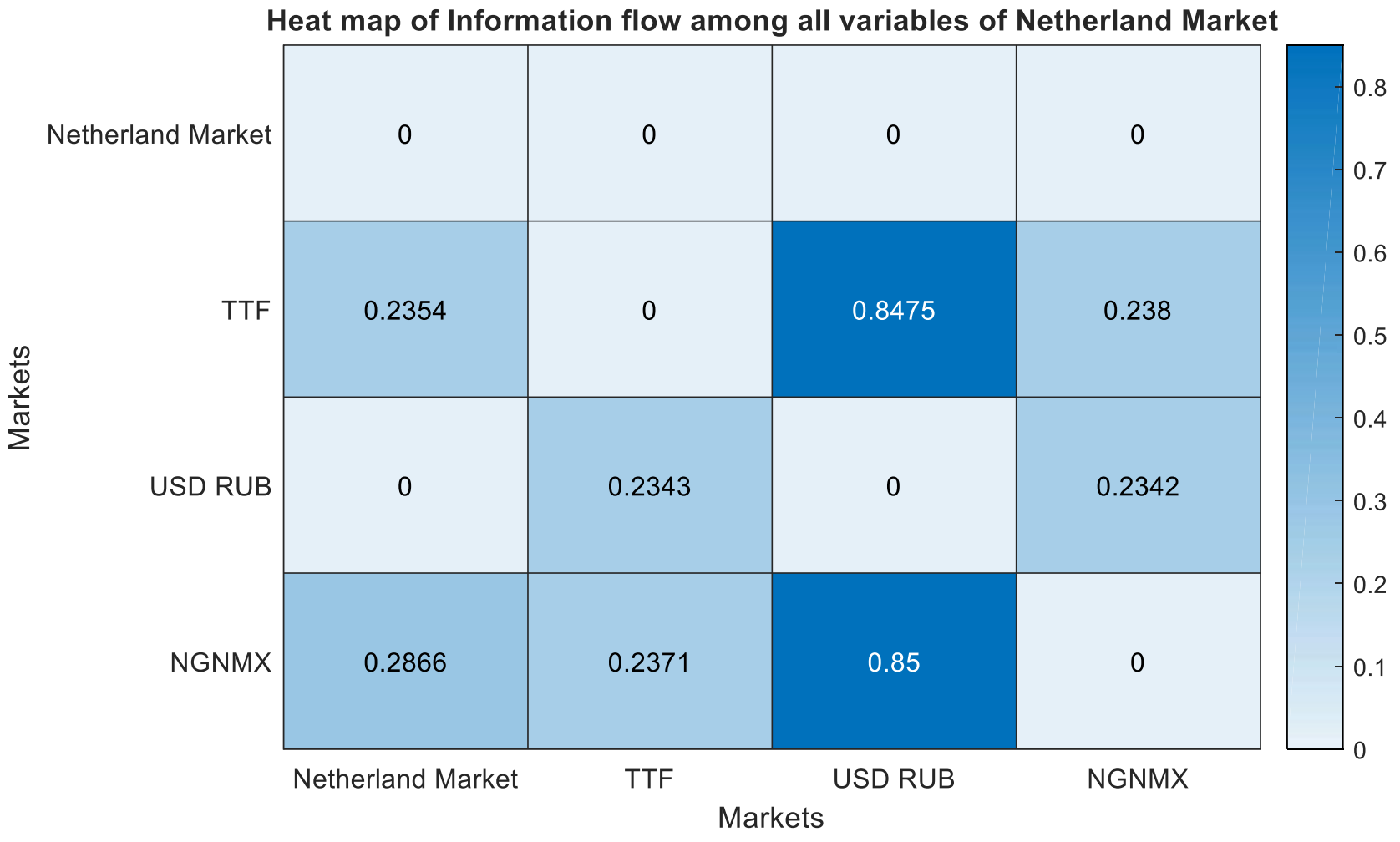
Heat map of Information flow among all variables of Netherland Market
Markets
Netherland Market
TTF
USD RUB
NGNMX
0
0
0
0
0.2354
0
0.8475
0.238
0
0.2343
0
0.2342
0.2866
0.2371
0.85
0
Netherland Market
TTF
USD RUB
NGNMX
Markets
0.8
0.7
0.6
0.5
0.4
0.3
0.2
0.1
0

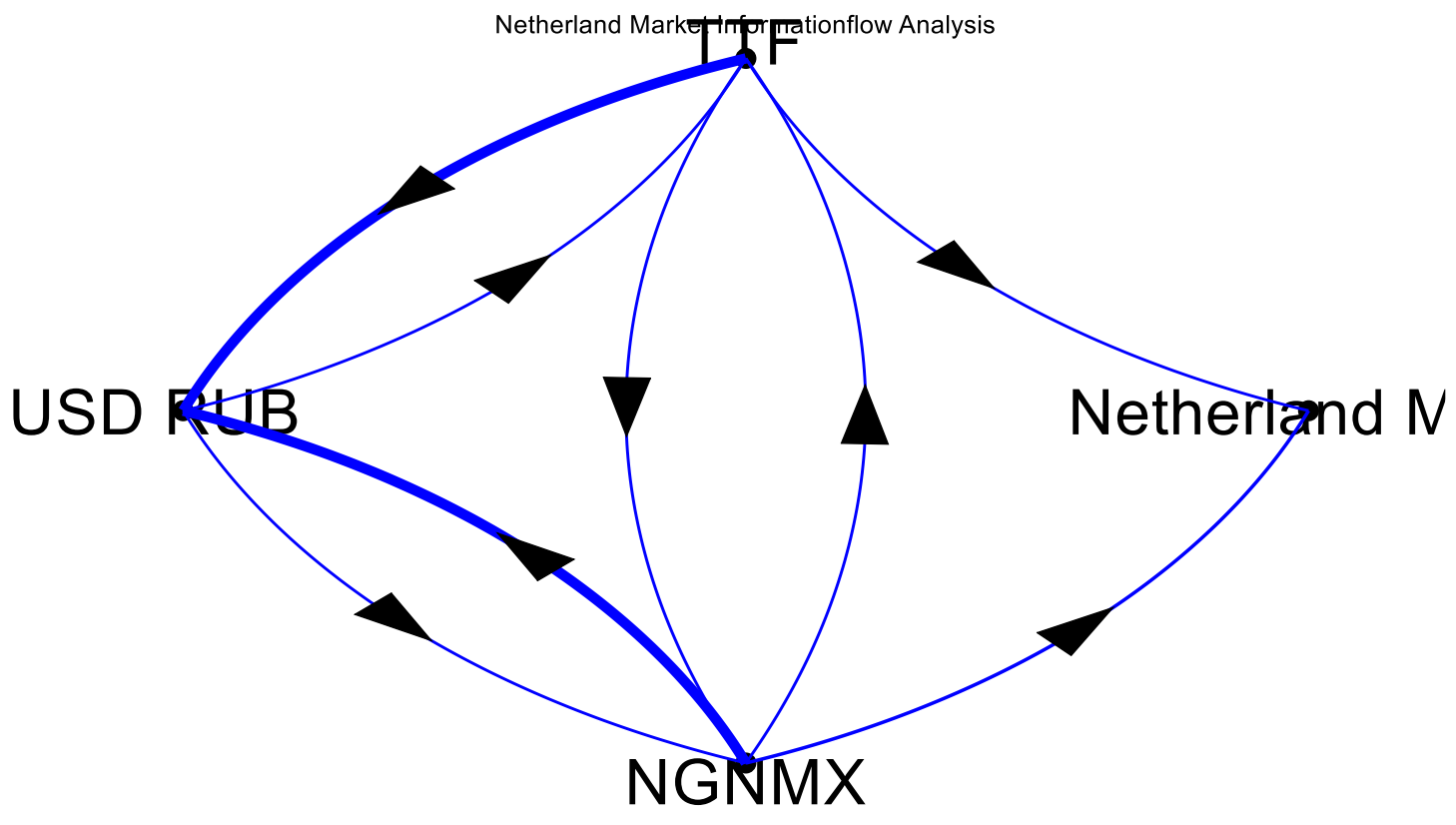
Netherland Market Informationflow Analysis
TTF
USD RUB
Netherland M
NGNMX

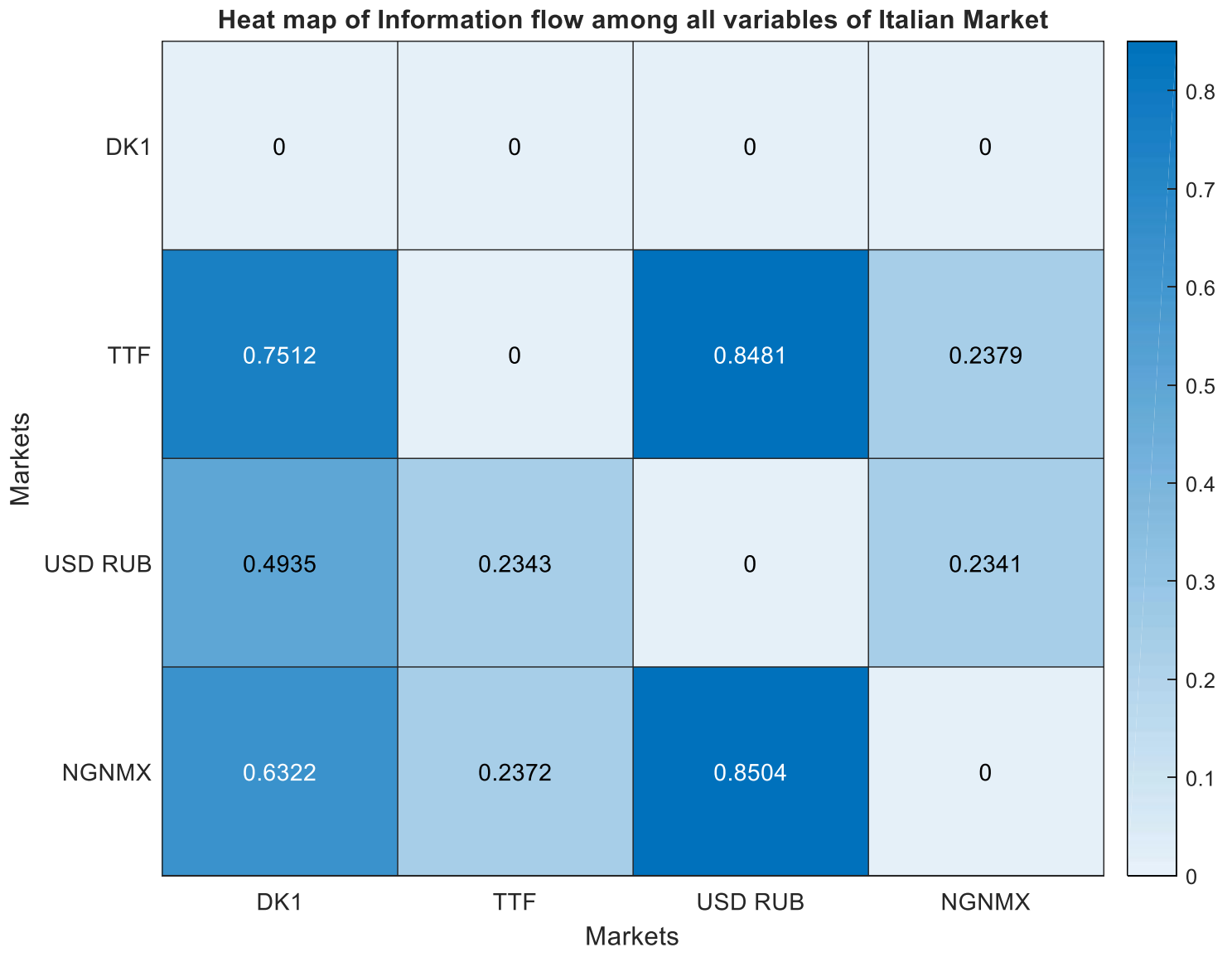
Heat map of Information flow among all variables of Italian Market
Markets
DK1
TTF
USD RUB
NGNMX
0
0
0
0
0.7512
0
0.8481
0.2379
0.4935
0.2343
0
0.2341
0.6322
0.2372
0.8504
0
DK1
TTF
USD RUB
NGNMX
Markets
0.8
0.7
0.6
0.5
0.4
0.3
0.2
0.1
0

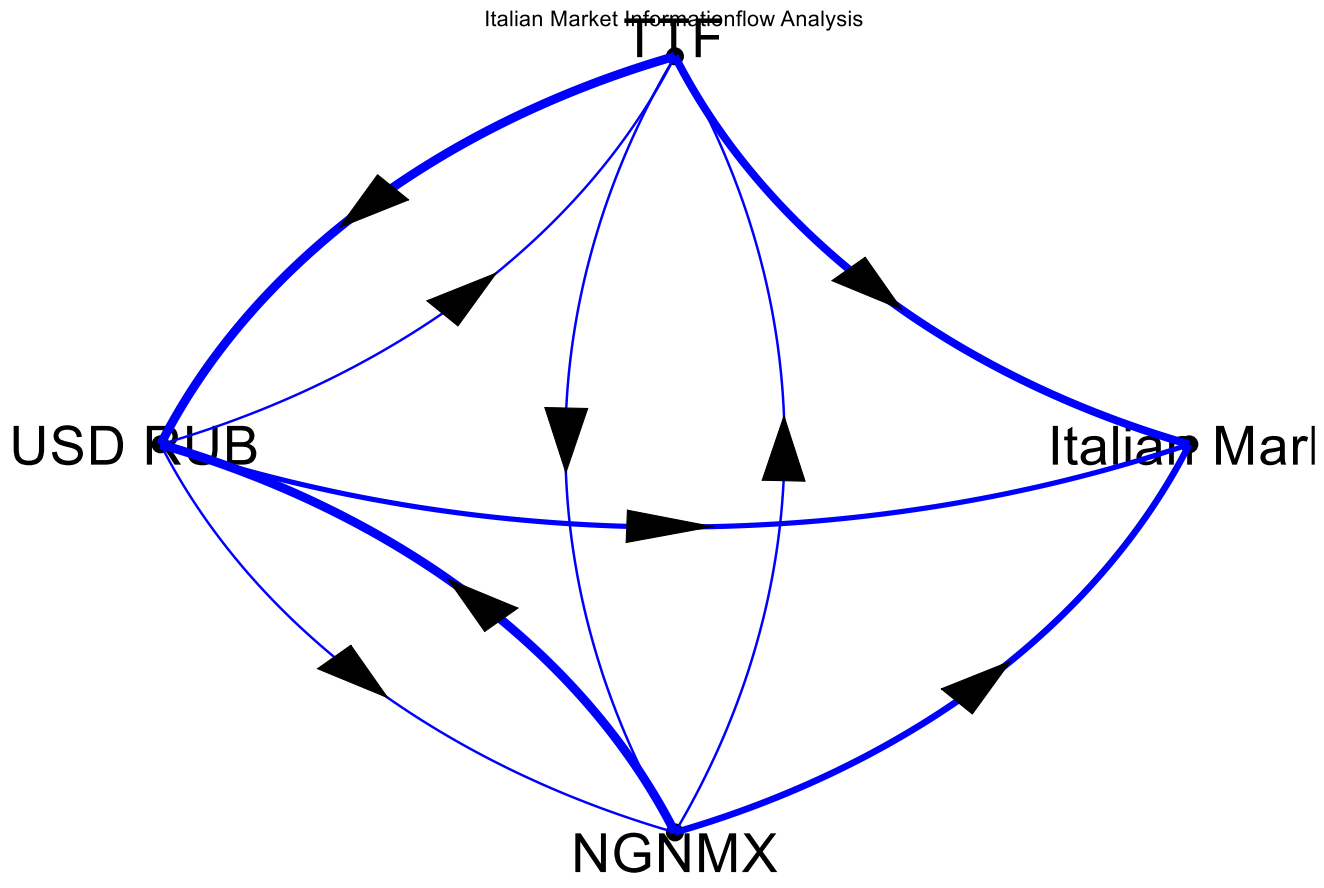
Italian Market Informationflow Analysis
TTF
USD RUB
Italian Marl
NGNMX

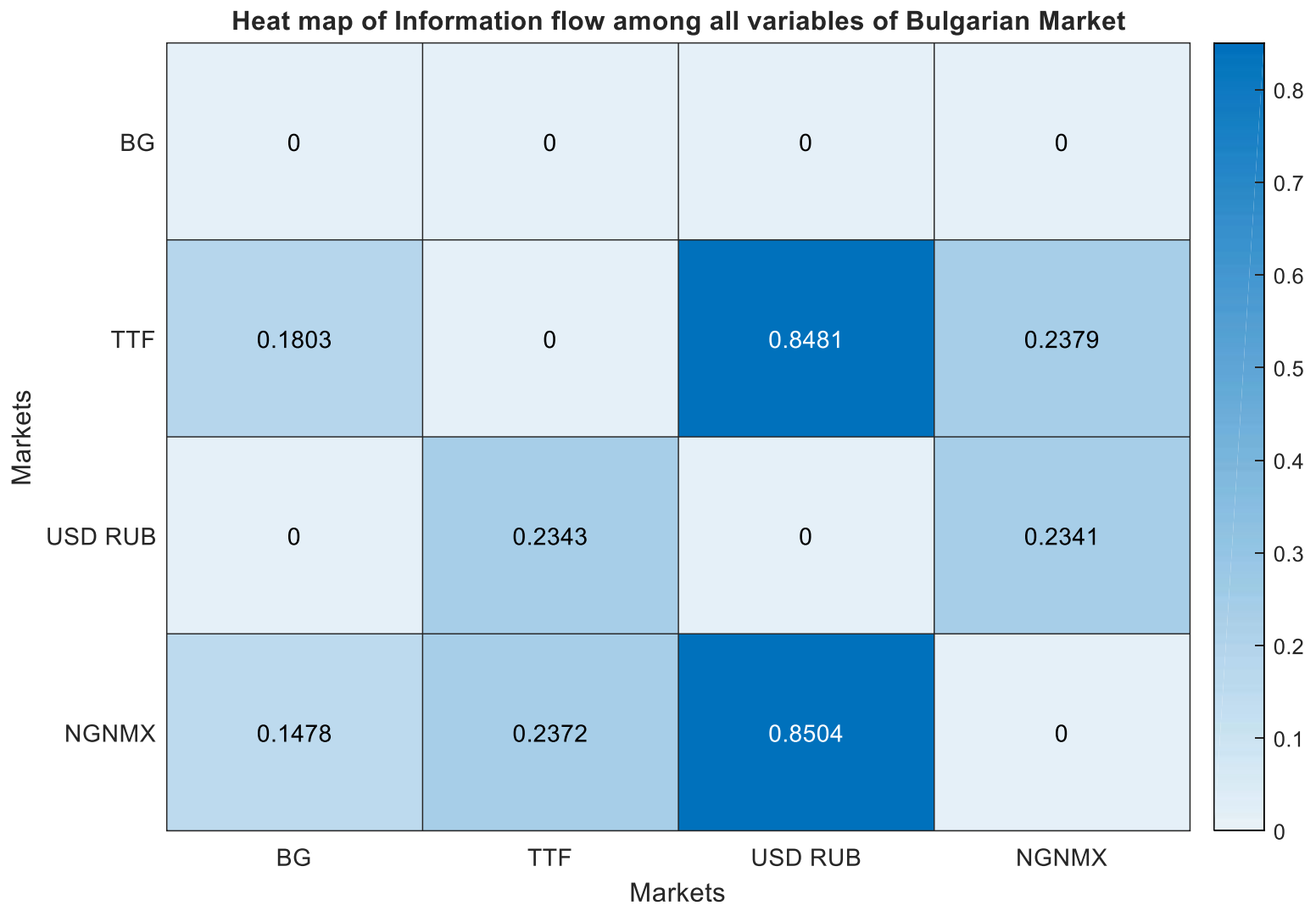
Heat map of Information flow among all variables of Bulgarian Market
BG
TTF
USD RUB
NGNMX
Markets
0
0.1803
0.1478
0.2343
0.2372
0.8481
0.8504
0.2379
0.2341
Markets
0.1
0.2
0.3
0.4
0.5
0.6
0.7
0.8

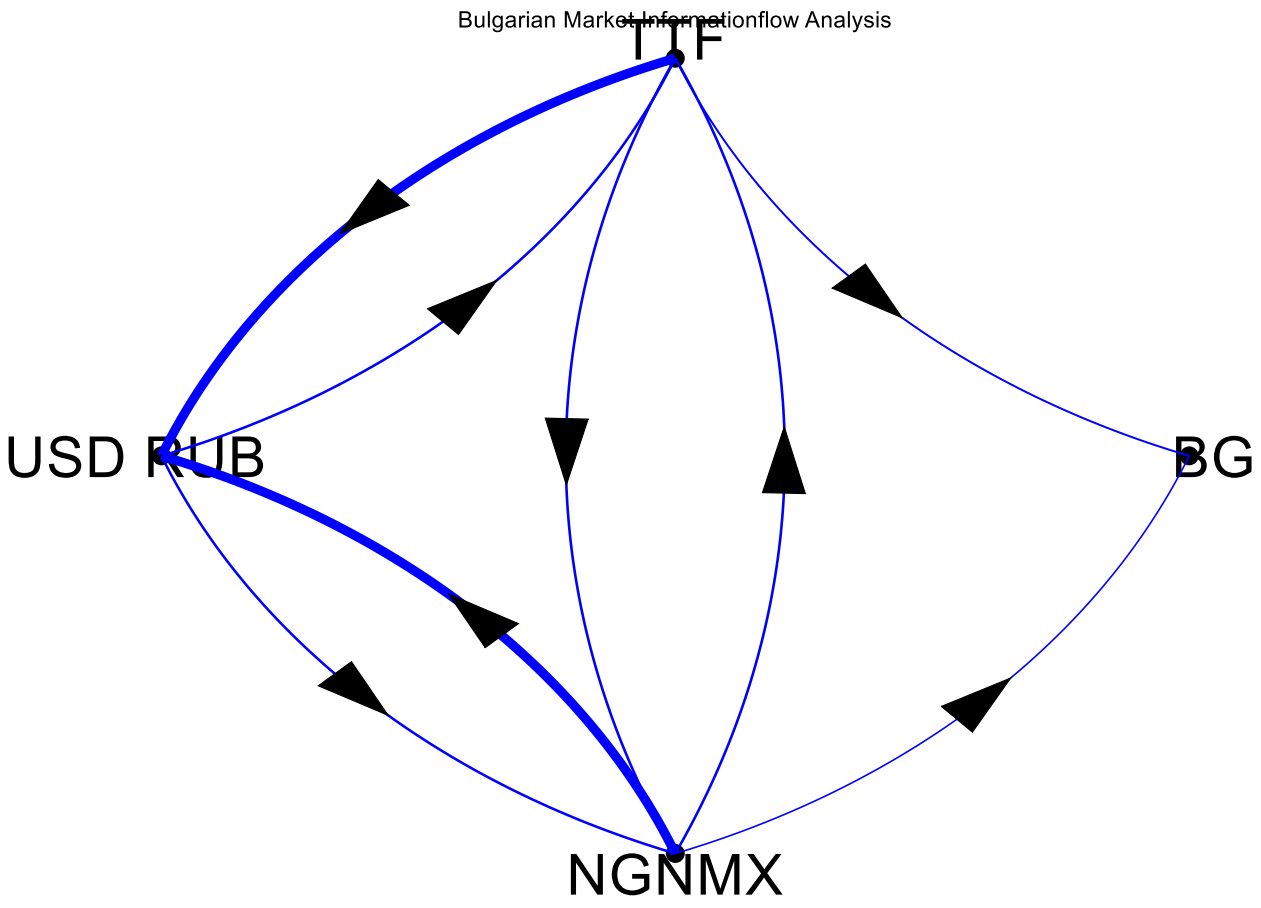
Bulgarian Market Informationflow Analysis
TTF
USD RUB
BG
NGNMX

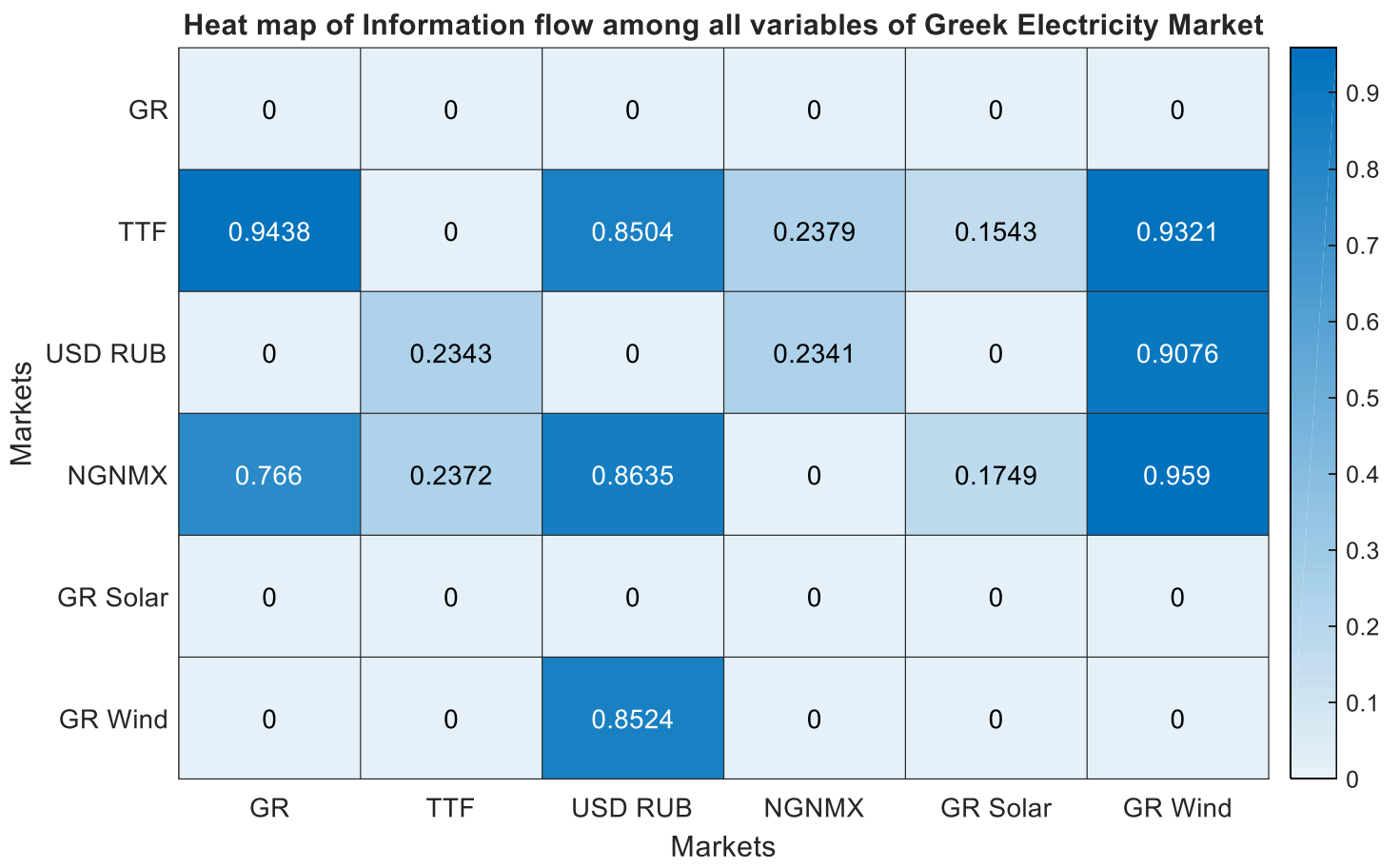
Heat map of Information flow among all variables of Greek Electricity Market
GR
TTF
USD RUB
NGNMX
GR Solar
GR Wind
Markets
0 0 0 0 0 0
0.9438 0 0.8504 0.2379 0.1543 0.9321
0 0.2343 0 0.2341 0 0.9076
0.766 0.2372 0.8635 0 0.1749 0.959
0 0 0 0 0 0
0 0 0.8524 0 0 0
GR
TTF
USD RUB
NGNMX
GR Solar
GR Wind
Markets
0.9
0.8
0.7
0.6
0.5
0.4
0.3
0.2
0.1
0

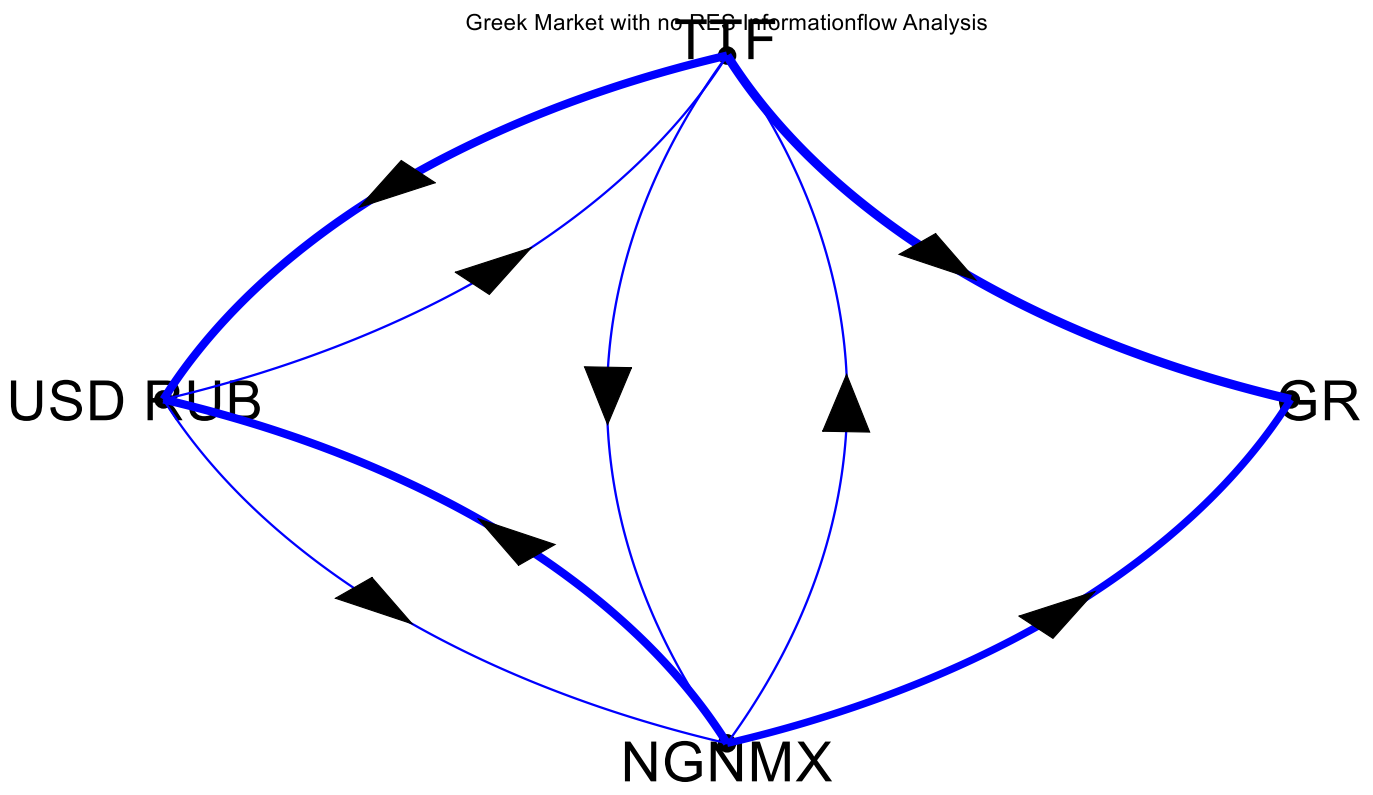
Greek Market with no RES Informationflow Analysis
TTF
USD RUB
GR
NGNMX